\documentclass{article}

\usepackage{PRIMEarxiv}

\usepackage[utf8]{inputenc} 
\usepackage[T1]{fontenc}    
\usepackage{hyperref}       
\usepackage{url}            
\usepackage{booktabs}       
\usepackage{amsfonts}       
\usepackage{nicefrac}       
\usepackage{microtype}      
\usepackage{lipsum}
\usepackage{fancyhdr}       
\usepackage{graphicx}       
\graphicspath{{media/}}     

\usepackage{cite}
\usepackage{amsmath,amsfonts}
\usepackage{algorithmic}
\usepackage{graphicx}
\usepackage{textcomp}

\usepackage{amssymb}
\usepackage{soul}
\usepackage{enumitem}
\usepackage{multirow}
\usepackage{booktabs,hhline}
\usepackage{longtable}
\usepackage{arydshln}
\usepackage{threeparttable}
\usepackage{fontawesome}

\usepackage{adjustbox}

\usepackage{hyperref}

\usepackage{diagbox}

\usepackage{soul}
\usepackage{float}
\usepackage{siunitx}
\usepackage[usestackEOL]{stackengine}

\usepackage{supertabular}

\usepackage{colortbl}
\usepackage[table]{xcolor}
\definecolor{Mycolor2}{HTML}{00F9DE}
\definecolor{B7D1E9}{HTML}{B7D1E9}
\definecolor{DDEBF7}{HTML}{DDEBF7}
\definecolor{5690C4}{HTML}{5690C4}
\definecolor{A3C4E1}{HTML}{A3C4E1}
\definecolor{699DCB}{HTML}{699DCB}
\definecolor{689CCB}{HTML}{689CCB}
\definecolor{2F75B5}{HTML}{2F75B5}
\definecolor{D2E4F3}{HTML}{D2E4F3}
\definecolor{D6E7F5}{HTML}{D6E7F5}
\definecolor{4F8BC1}{HTML}{4F8BC1}
\definecolor{C0D8EC}{HTML}{C0D8EC}
\definecolor{8EB5D9}{HTML}{8EB5D9}
\definecolor{A2C3E1}{HTML}{A2C3E1}
\definecolor{DBEAF7}{HTML}{DBEAF7}
\definecolor{3F80BB}{HTML}{3F80BB}
\definecolor{DCEBF7}{HTML}{DCEBF7}
\definecolor{DCEAF7}{HTML}{DCEAF7}
\definecolor{699CCB}{HTML}{699CCB}
\definecolor{6FA0CD}{HTML}{6FA0CD}
\definecolor{D1E3F3}{HTML}{D1E3F3}
\definecolor{D7E7F5}{HTML}{D7E7F5}
\definecolor{B1CDE7}{HTML}{B1CDE7}
\definecolor{BED6EB}{HTML}{BED6EB}
\definecolor{7FABD3}{HTML}{7FABD3}
\definecolor{CEE1F1}{HTML}{CEE1F1}
\definecolor{CDE0F1}{HTML}{CDE0F1}
\definecolor{AECBE5}{HTML}{AECBE5}
\definecolor{9FC1E0}{HTML}{9FC1E0}
\definecolor{72A3CF}{HTML}{72A3CF}
\definecolor{DAE9F6}{HTML}{DAE9F6}
\definecolor{8EB6D9}{HTML}{8EB6D9}
\definecolor{9EC1DF}{HTML}{9EC1DF}
\definecolor{D5E6F4}{HTML}{D5E6F4}
\definecolor{BBD4EB}{HTML}{BBD4EB}
\definecolor{6EA0CD}{HTML}{6EA0CD}
\definecolor{D4E5F4}{HTML}{D4E5F4}
\definecolor{6B9ECC}{HTML}{6B9ECC}
\definecolor{83AED5}{HTML}{83AED5}
\definecolor{C3DAED}{HTML}{C3DAED}
\definecolor{D3E4F3}{HTML}{D3E4F3}
\definecolor{CCE0F1}{HTML}{CCE0F1}
\definecolor{C3DAEE}{HTML}{C3DAEE}
\definecolor{98BCDD}{HTML}{98BCDD}
\definecolor{BBD4EA}{HTML}{BBD4EA}
\definecolor{B3CFE8}{HTML}{B3CFE8}
\definecolor{7DAAD3}{HTML}{7DAAD3}
\definecolor{9BBFDE}{HTML}{9BBFDE}
\definecolor{C5DBEE}{HTML}{C5DBEE}
\definecolor{CBDFF1}{HTML}{CBDFF1}
\definecolor{C8DDEF}{HTML}{C8DDEF}
\definecolor{B2CEE7}{HTML}{B2CEE7}
\definecolor{9CBFDF}{HTML}{9CBFDF}
\definecolor{86B0D6}{HTML}{86B0D6}
\definecolor{B5D0E8}{HTML}{B5D0E8}
\definecolor{4A88C0}{HTML}{4A88C0}
\definecolor{C3D9ED}{HTML}{C3D9ED}
\definecolor{D0E2F2}{HTML}{D0E2F2}
\definecolor{A8C7E3}{HTML}{A8C7E3}
\definecolor{80ACD4}{HTML}{80ACD4}
\definecolor{9BBEDE}{HTML}{9BBEDE}
\definecolor{95BADC}{HTML}{95BADC}
\definecolor{4785BE}{HTML}{4785BE}
\definecolor{CFE2F2}{HTML}{CFE2F2}
\definecolor{7FACD4}{HTML}{7FACD4}
\definecolor{B9D3EA}{HTML}{B9D3EA}
\definecolor{CADEF0}{HTML}{CADEF0}
\definecolor{AAC9E4}{HTML}{AAC9E4}
\definecolor{D3E5F4}{HTML}{D3E5F4}
\definecolor{BFD7EC}{HTML}{BFD7EC}
\definecolor{C9DEF0}{HTML}{C9DEF0}
\definecolor{5891C5}{HTML}{5891C5}
\definecolor{6398C9}{HTML}{6398C9}
\definecolor{D8E8F5}{HTML}{D8E8F5}
\definecolor{84AFD5}{HTML}{84AFD5}
\definecolor{79A8D1}{HTML}{79A8D1}
\definecolor{A4C4E2}{HTML}{A4C4E2}
\definecolor{C1D8ED}{HTML}{C1D8ED}
\definecolor{79A7D1}{HTML}{79A7D1}
\definecolor{A1C3E1}{HTML}{A1C3E1}
\definecolor{5B93C6}{HTML}{5B93C6}
\definecolor{78A7D1}{HTML}{78A7D1}

\makeatletter
\g@addto@macro\GTS@PredefinedLeftCmds{%
  \GTS@TestLeft\enit@align\GTS@Cdr 
}
\g@addto@macro\GTS@DisablePredefinedCmds{%
  \let\enit@align\@empty 
}
\makeatother

\usepackage{tabularx}
\newcolumntype{P}[1]{>{\RaggedRight\hspace{0pt}}p{#1}}

\newcommand*\rotN{\rotatebox{90}}

\def\tsc#1{\csdef{#1}{\textsc{\lowercase{#1}}\xspace}}
\tsc{WGM}
\tsc{QE}
\def\BibTeX{{\rm B\kern-.05em{\sc i\kern-.025em b}\kern-.08em
    T\kern-.1667em\lower.7ex\hbox{E}\kern-.125emX}}

\title{Systematic Literature Review on Application of Learning-based Approaches in Continuous Integration
\thanks{\textit{\underline{Citation}}: 
\textbf{Arani, A. K., M., Le, T. H., M., Zahedi \& Babar, M. A. (2023, May). Systematic Literature Review on Application of Learning-based Approaches in Continuous Integration. arXiv preprint (2024)}} 
}

\author{Ali Kazemi Arani \\
  CREST-The Centre for Research\\on Engineering Software Technologies \\
  University of Adelaide \\
  Adelaide, SA 5005, Australia\\
  \texttt{Ali.KazemiArani@adelaide.edu.au} \\
   \And
  Triet Huynh Minh Le \\
  CREST-The Centre for Research\\on Engineering Software Technologies \\
  University of Adelaide \\
  Adelaide, SA 5005, Australia\\
  \texttt{triet.h.le@adelaide.edu.au} \\
   \And
  Mansooreh Zahedi \\
  School of Computing and Information Systems\\
  University of Melbourne \\
  Melbourne, VIC 3010, Australia\\
  \texttt{mansooreh.zahedi@unimelb.edu.au} \\
  \And
  M. Ali Babar \\
  CREST-The Centre for Research\\on Engineering Software Technologies \\
  University of Adelaide \\
  Adelaide, SA 5005, Australia\\
  \texttt{ali.babar@adelaide.edu.au} \\
}

\begin{document}
\maketitle

\begin{abstract}
\textbf{Context:} Machine learning (ML) and deep learning (DL) analyze raw data to extract valuable insights in specific phases. The rise of continuous practices in software projects emphasizes automating Continuous Integration (CI) with these learning-based methods, while the growing adoption of such approaches underscores the need for systematizing knowledge. \textbf{Objective:} Our objective is to comprehensively review and analyze existing literature concerning learning-based methods within the CI domain. We endeavour to identify and analyse various techniques documented in the literature, emphasizing the fundamental attributes of training phases within learning-based solutions in the context of CI. \textbf{Method:} We conducted a Systematic Literature Review (SLR) involving 52 primary studies. Through statistical and thematic analyses, we explored the correlations between CI tasks and the training phases of learning-based methodologies across the selected studies, encompassing a spectrum from data engineering techniques to evaluation metrics. \textbf{Results:} This paper presents an analysis of the automation of CI tasks utilizing learning-based methods. We identify and analyze nine types of data sources, four steps in data preparation, four feature types, nine subsets of data features, five approaches for hyperparameter selection and tuning, and fifteen evaluation metrics. Furthermore, we discuss the latest techniques employed, existing gaps in CI task automation, and the characteristics of the utilized learning-based techniques. \textbf{Conclusion:} This study provides a comprehensive overview of learning-based methods in CI, offering valuable insights for researchers and practitioners developing CI task automation. It also highlights the need for further research to advance these methods in CI.
\end{abstract}


\section{Introduction}
\par Continuous Integration (CI) refers to the software development practice of automatically integrating code changes through frequent automated build processes, has gained popularity for enhancing software delivery speed and reliability through early issue detection~\cite{shahin2017continuous}. CI enables rapid testing, building, and software preparation at any time~\cite{jin2020cost}. This facilitates bug detection and fixation, resulting in faster delivery and improved software quality. This approach reduces the development costs and enhances customer satisfaction~\cite{shahin2017continuous}.

\par Frequent changes in CI environments produce large amounts of data. It can be difficult and demanding in terms of time and resources to extract and analyze these data, particularly in large-scale projects~\cite{xie2018cutting}. CI imposes considerable expenses, significantly impacting software development expenses~\cite{mamata2022failure}. Notably, Google and Mozilla reported monthly CI process costs in millions~\cite{jin2020cost, hilton2016usage}. The CI phase consumes more than half of the resources in software development~\cite{abdelkarim2022tcp}, posing a barrier for smaller companies. Thus, enhancing CI pipeline performance and reducing associated costs are crucial for software development.

\par CI's popularity, complex data, and costs drive the adoption of learning-based methods including Machine Learning (ML) and Deep Learning (DL) techniques for analysis, aiding software engineers with valuable insights. These insights enhance the CI feedback loop, analyzing development data, test logs, CI phase outcomes, and operational environment data~\cite{figalist2019supporting}.

\par Learning-based methods use mathematical models to acquire knowledge, make decisions, or improve performance based on data and experience~\cite{shafiq2021literature}. These techniques can efficiently predict complex task outcomes. In Continuous Integration (CI) environments, learning-based methods can enhance task performance by predicting software defects without executing the current version. They provide benefits such as accurately predicting outcomes of unit tests~\cite{lee2019classifying} and regression tests~\cite{yang2021sparse}. They also aid software engineers in timely decision-making, such as task assignment to different profiles~\cite{brandtner2015sqa}.

\par Given the broad scope of learning-based methods and their diverse applications in CI, understanding their impact on CI processes is crucial. This includes discerning the decision-making steps in developing these methods, termed Machine Learning for Continuous Integration (ML4CI). An exploration of this area offers cross-disciplinary insights, emphasizing the necessity of a systematic literature review in ML4CI.

\par This study endeavours to address the existing gap by conducting a comprehensive Systematic Literature Review (SLR) that concentrates on learning-based methodologies within CI development phases. Through the analysis of 52 carefully selected papers spanning from 2000 to August 2023, the objective is to elucidate the progress made in the automation of CI tasks through the application of learning-based techniques. The developmental steps involve data collection/preparation, feature engineering, training, tuning, and evaluation of these methods.

\par Comprehensively analyzing data from all CI tasks offers researchers and practitioners a synthesized overview of information sources and applied techniques. This equips them to leverage these resources in CI pipelines, potentially automating processes and guiding future endeavours.

\par Insights from this review benefit researchers and practitioners, offering a foundational platform to advance CI phases. By reusing or refining learning-based methods, more efficient practices can emerge. This research is significant for those employing existing approaches or developing new solutions for CI, considering identified gaps. Additionally, our discussion addresses opportunities to optimize current solutions or address neglected areas in real-world CI settings.

\par In summary, the main \textbf{contributions} of this paper are threefold:

\begin{itemize}
    \item First, the paper provides a clear and comprehensive understanding of the six identified phases and ten tasks in CI that can be automated by using learning-based methods, along with their sequence and relationships in the CI pipeline.

    \item Second, the paper analyses the data and techniques used for data preparation, feature engineering, training, tuning and evaluation of the learning-based methods concerning the CI phases and tasks.

    \item Finally, the paper discusses the future directions in applying learning-based solutions to CI and the existing inconsistencies, which can serve as a guide for researchers and practitioners in the field. 

\end{itemize}

\par In previous research, our focus was on using ML methods to extract automated CI tasks. We specifically looked at the training approaches used in state-of-the-art (SOTA) learning-based methods for each CI task~\cite{arani2023sok}. This paper extends our previous work by updating the list of papers through searches in two additional indexing databases and reviewing earlier publications. Furthermore, we conduct an analysis specifically focusing on each training phase of ML methods. Additionally, we employ thematic analysis to systematically extract information from the published works and conduct deeper analysis. The objective is to offer a more comprehensive and detailed account, providing additional information and insights into learning-based methods within the context of CI. 
\par The rest of this paper is organized as follows: Section~\ref{sec:BG} provides background and related works, followed by the research methodology in Section~\ref{Scopus}. Section~\ref{sec:Find} presents synthesized data based on the proposed research questions (RQs). Section~\ref{sec:limit} outlines the study's limitations. Finally, Section~\ref{sec:Disc} discusses the findings, and Section~\ref{sec:conc} concludes the study.

\par Henceforth, we will use the term \textbf{`ML'} as an abbreviation for the combined term \textbf{`Learning-based'}.


\section{Background and Related Works}
\label{sec:BG}
\par In this section, we will introduce the essential concepts of CI and the applications of ML in software engineering, which are the primary focus of this SLR.

\subsection{Overview of Continuous Integration}

\par CI encompasses the processes of building, testing, validating software, and managing commit-related actions, including addressing reported bugs and messages from developers, before advancing to the deployment phase~\cite{fitzgerald2017continuous}. A development team integrates and merges their source code frequently, often multiple times a day, to build the software~\cite{shahin2017continuous}. Automated testing and quick feedback are crucial to preventing issues from propagating to the delivery phase or affecting the development process of other team members~\cite{fitzgerald2017continuous}. The validation phase in CI provides feedback to developers on detected bugs or performance issues, and they also monitor deployed software and the related materials in operations and version control systems (VCS) to maintain its performance~\cite{debois2011devops}.

\begin{figure*}
    \centering
    \includegraphics[width=0.75\linewidth]{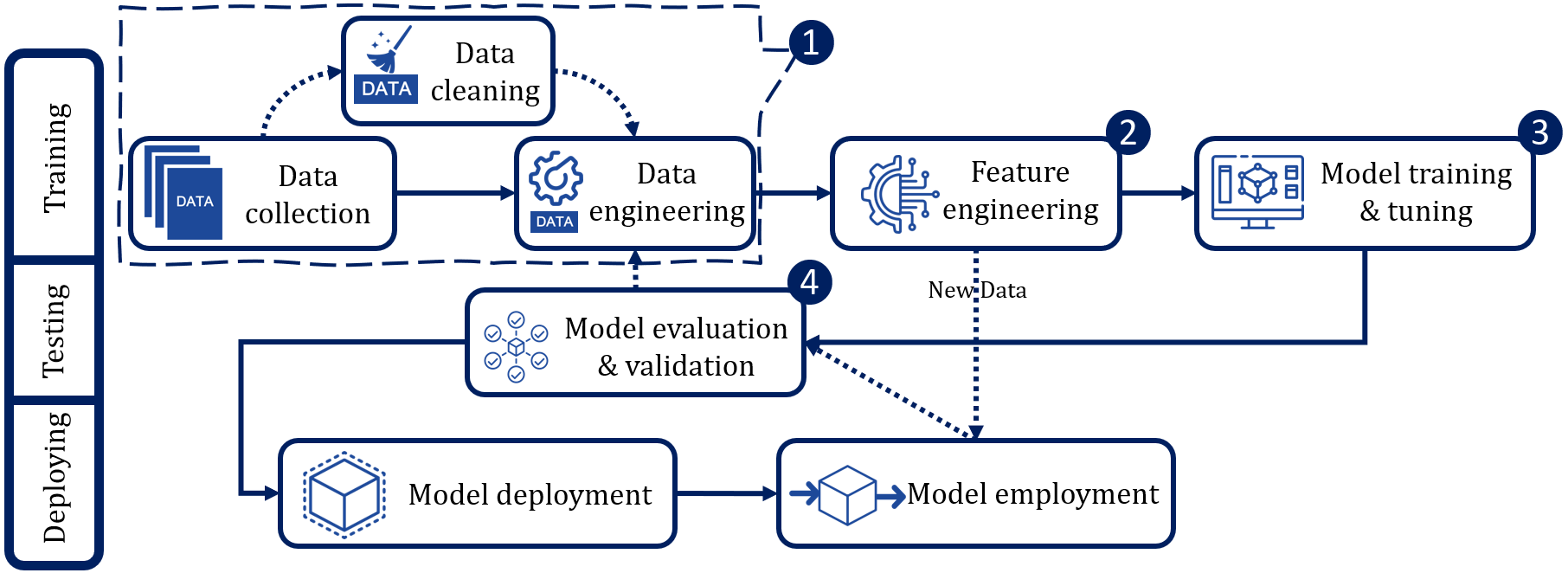}
    \caption{The four phases of ML life cycle. \textbf{Note:} the required steps for training an ML model are distinguished by numbers}
    \label{Fig:MLOverview}
\end{figure*}

\subsection{Machine Learning in Software Engineering}
\label{Section:MLforCI}

\par ML methods enhance software product quality by automating tasks through pattern analysis and learning from historical data~\cite{mitchell2006discipline}. During training, models use a subset of input data to recognize patterns, making predictions or classifications for new data based on their training methodology, which includes supervised, unsupervised~\cite{bacstanlar2014introduction}, semi-supervised~\cite{russell2009artificial}, and Reinforcement Learning (RL)~\cite{andrew1998reinforcement}.

\par For instance, a supervised ML model predicts build outcomes with features like changed classes and committer experience~\cite{hassan2017change}. Unsupervised learning clusters data based on natural language features to assign tasks to related profiles~\cite{brandtner2015sqa}. Section~\ref{Sec:Models} further explores these learning types, highlighting their common development process.


\label{Sec:RelatedWorks}

\par Previous studies have explored ML techniques in software engineering, with a notable focus on ML methods exclusively in Test Case Selection and Prioritization (TSP)~\cite{pan2022test}. This systematic review included 29 studies but did not examine data-related concepts. Their primary focus was on comparisons between studies and considerations related to reproducibility and repeatability in the TSP domain.

\par In a broader perspective, a systematic mapping conducted by Durelli et al.~\cite{durelli2019machine} explored the application of machine learning methods in software testing, emphasizing their scalability and effectiveness in addressing complex issues within software testing systems.

\par Unlike the mentioned studies concentrating on the testing phase, Shafiq et al.~\cite{shafiq2021literature} conducted a comprehensive literature review covering various aspects of ML methods in software engineering, including software requirements, design, construction, quality, and maintenance. They highlighted demographic data and identified challenges, such as the uncertain nature of ML techniques, data availability, and the increasing complexity of software products.

\par Zhang, and Tsai~\cite{zhang2003machine} compiled a comprehensive list of SE problems addressed with ML techniques, discussing their application and the pros and cons of using ML models in SE. Ali and Gravino~\cite{ali2019systematic} conducted a systematic review of 75 studies to assess commonly used ML models, datasets, and accuracy measurement methods in software development effort estimation (SDEE).

\par Our study exclusively explores the application of ML within the CI domain. Unlike other studies that address broader areas of software engineering, our research offers a detailed analysis of ML-based solutions developed to be employed in CI.

\par The literature proposes four primary steps in ML model development for CI~\cite{gada2021automated}: Data Collection and Data Engineering (Step 1), Feature Engineering (Step 2), Training and Hyper-parameter Tuning (Step 3), and Model Evaluation (Step 4). The representation in Figure~\ref{Fig:MLOverview} corresponds to these phases as outlined in reference~\cite{gada2021automated}.

\par During the initial Data Collection and Engineering step, data can be gathered from various sources, including raw data from CI tools and monitoring sensors~\cite{zhang2003data}. Post-collection, it is crucial to extract relevant features and adapt them for the subsequent training phase~\cite{khurana2016automating}. For effective ML model training, selecting an appropriate algorithm and adjusting hyperparameters for performance optimization is essential. Finally, thorough evaluation and validation are necessary to ensure the models' effectiveness for real-world applications~\cite{alpaydin2020introduction}. Regularly repeating these steps is important to keep models updated with new data.


\section{Research Methodology}
\label{Scopus}
\begin{figure*}
    \centering
    \includegraphics[width=0.78\linewidth]{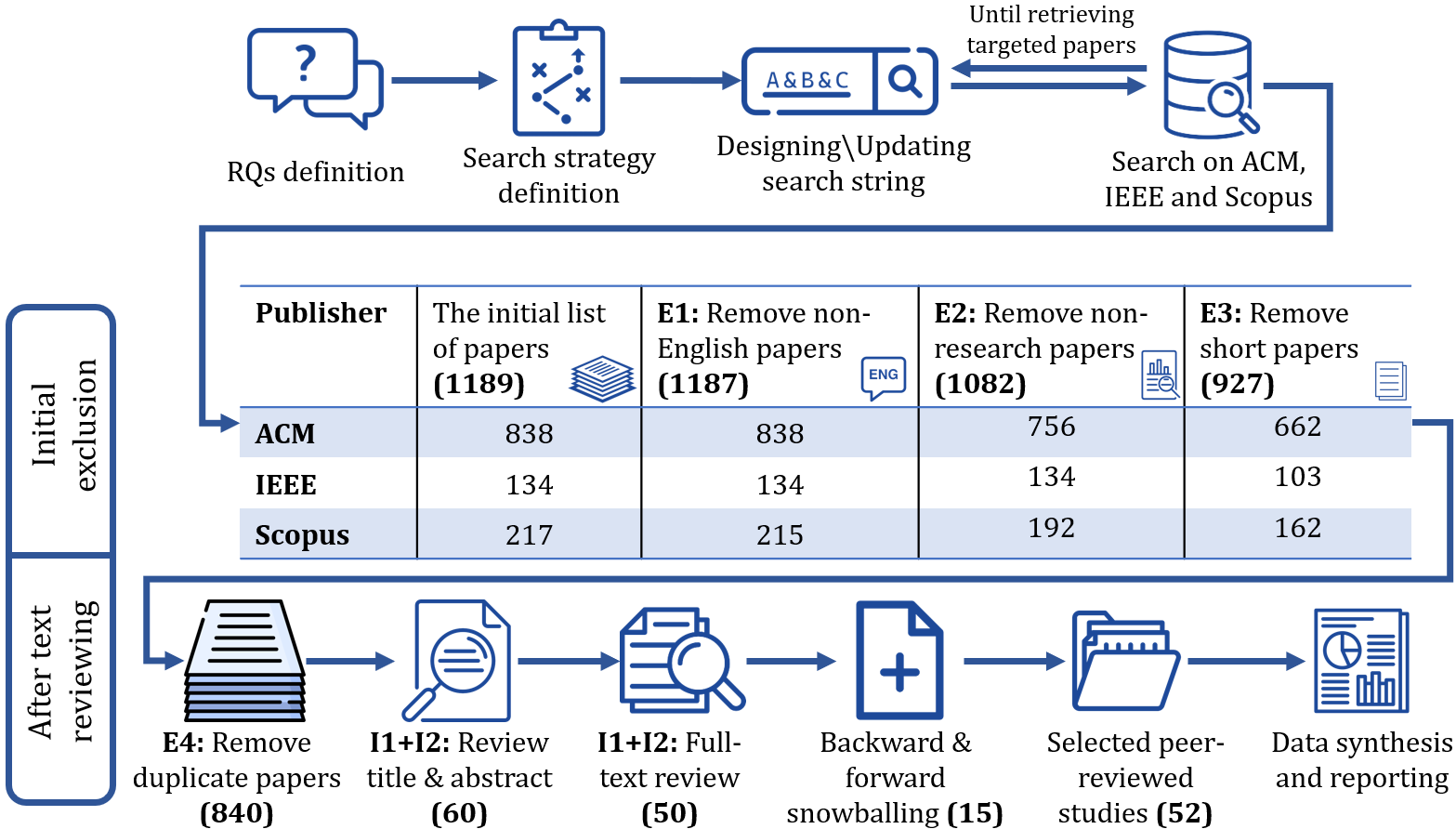}
    \caption{Overview of the research methodology. \textbf{Note:} Two arrows with opposite directions present the iterative actions. The \textbf{``I''} and \textbf{``E''} stand for the Inclusion and Exclusion criteria according to Table~\ref{Table:InclusionExclusion}, respectively, and numbers in parenthesis present the total number of selected papers in each step.}
    \label{Fig:StudySelection}
\end{figure*}
\par To gain insights into the developmental stages of ML-based methods in CI, this SLR followed Kitchenham's guidelines~\cite{kitchenham2007guidelines}. Utilizing the ACM\footnote{\underline{https://dl.acm.org/}}, IEEE\footnote{\underline{https://www.ieee.org/}}, and Scopus\footnote{\underline{https://www.scopus.com/}} indexing systems with a specified search string method facilitated the retrieval of relevant studies. These databases were chosen for their extensive coverage of journals and conferences in Software Engineering and Computer Science domains~\cite{kitchenham2010systematic}.

\par Previous studies affirm the effectiveness of this search strategy in collecting high-quality papers~\cite{kitchenham2010systematic}. A summary of the research methodology and the number of studies in each phase is presented in Figure~\ref{Fig:StudySelection}. Further details are provided in the subsequent section.

\subsection{Research Questions}

\begin{table*}
    \centering
    \caption{Research questions and motivations.}
    \label{Table_old:RQs}
    \begin{tabular}{p{0.25\linewidth}| p{0.70\linewidth}}
    \toprule
        \textbf{Research Questions} & \textbf{Motivation} \\
    \toprule
        \textbf{RQ1:} What CI tasks can be automated using ML-based approaches? &
         In this first research question, we seek to explain the specific CI tasks and phases that ML methods can effectively automate. Also, we present the output and input of these CI phases. 
         This insight not only enhances our understanding of the integration process but also sheds light on underexplored CI phases and tasks that practitioners can further investigate.
         \\\hline
         \textbf{RQ2:} What datasets and data preparation techniques are used in automating CI tasks using ML? &
         Given the data-driven nature of ML models, the characteristics and preparation of input datasets significantly influence the efficacy of ML-based solutions in CI. This research question investigates the frequently employed datasets and associated data engineering methodologies, including approaches to address challenges like class imbalance. 
         \\\hline
        \textbf{RQ3:} What feature extraction strategies are utilized to train ML models for automating the CI tasks?  &
        Beyond raw data, the selection and engineering of features play a pivotal role in shaping the effectiveness of ML models. By identifying the array of feature types and techniques for transforming raw data into machine-readable inputs, we contribute to a repository of knowledge that can be used in future studies. This knowledge empowers researchers and practitioners to optimize the design of ML models for automating CI tasks. \\\hline
        \textbf{RQ4:} What ML modeling and tuning techniques are used to automate CI tasks? &
         This research question focuses on the connection between ML model types and CI tasks. By uncovering the relationship between these two elements, we aim to pinpoint potential gaps in the application of ML models for automating CI tasks. Moreover, we extract the common approaches for tuning these models, shedding light on hyper-parameter tuning methods that enhance their performance. \\\hline
        \textbf{RQ5:} How the ML models are evaluated in case of automating the CI tasks? &
        In the final research question, we focus on the evaluation of ML model performance within the realm of CI. Our objective is to categorize the commonly employed evaluation metrics and techniques, elucidating their correlation with the four distinct ML algorithm categories outlined in Section~\ref{Section:MLforCI}. This categorization enables researchers and practitioners to conduct informed comparisons of various ML methodologies for automating the CI tasks fostering a good perspective based on existing literature.  \\
    \bottomrule
    \end{tabular}

\end{table*}

\par 
To initiate this SLR, five research questions (RQs) were formulated to identify the phases of CI and the ML model development techniques used for these tasks. Section~\ref{Sec:CITasks} provides an overview of the ML models and CI tasks identified based on these RQs. You can find a detailed list of these RQs and their motivations in Table~\ref{Table_old:RQs}.


\subsection{Search Strategy}

\par The next step in this SLR is designing an appropriate search strategy for finding relevant studies~\cite{kitchenham2007guidelines}. The search strategy configuration for extracting studies from databases is detailed below.

\begin{table*}
    \footnotesize
    \centering
    \caption{Search string used to extract relevant papers on the application of ML methods in CI. The search string is composed of two segments, including ``Machine Learning'' and its associated synonyms and ``Continuous Integration''.}
    \label{SearchString}
    \begin{tabular}{c |p{0.9\linewidth}}
    \toprule
        \textbf{A} & \textbf{TITLE-ABS-KEY} ((``Machine Learning'' \textbf{OR} ``Deep Learning'' \textbf{OR} ``Reinforcement Learning'' \textbf{OR} ``Supervised Learning'' \textbf{OR} ``Unsupervised Learning'' \textbf{OR} ``Semi-supervised Learning'' \textbf{OR} ``Data Mining'' \textbf{OR} ``Text Mining'' \textbf{OR} ``Natural Language Processing'' ) \\
        \cdashline{1-2}
        
        \textbf{B} & \textbf{AND} (``Continuous Integration'') \\

    \bottomrule
    \end{tabular}
    
\end{table*}


\subsubsection{Search String}
\label{SSScetion}

\par Our search string consists of two segments: \textbf{A) ``Machine Learning''} and related synonyms, and \textbf{B) ``Continuous Integration''}. These choices were informed by a review of previous studies and SLRs~\cite{shahin2017continuous, van2020crop, wang2022machine, malhotra2015systematic}, with the complete search string detailed in Table~\ref{SearchString}. The search string was executed on 26th July 2023, and snowballing was conducted on 7th August 2023.
\par Before implementing our search string to identify relevant literature, the first and third authors conducted a preliminary search on Google Scholar using keywords such as `machine learning,' `Continuous Integration', and related terms. This search aimed to identify initial relevant studies on the application of machine learning (ML) methods in Continuous Integration (CI). This preliminary search served as a validation set of papers. Subsequently, we iteratively refined our search string to retrieve all the papers included in our validation set. However, due to its tendency to yield numerous irrelevant results~\cite{shahin2017continuous}, Google Scholar was not used for our primary search.
\par To refine our search string, we iteratively ran it on ACM, IEEE, and Scopus databases, identifying and adding missing search terms for comprehensive coverage in the initial phase. The refined search string was then executed on these three databases, resulting in 1189 hits.


\subsubsection{Inclusion and Exclusion Criteria}
\label{Sec:InclusionExclusion}

\begin{table*}
    \centering
    \caption{Inclusion and exclusion criteria.}
    \label{Table:InclusionExclusion}
    \begin{tabular}{ l l } 
    \toprule
    \textbf{ID} &\textbf{Inclusion criteria}  \\ 
    \bottomrule
    I1 & Studies that employed ML-based methods in the CI environment. \\ 
    I2 & Related to software development and engineering. \\
    \toprule
    \textbf{ID} & \textbf{Exclusion criteria}  \\ 
    \bottomrule
    E1 & The paper is tagged as non-English language. \\
    E2 & Non-research papers including conference reviews and reports, notes, and short surveys. \\
    E3 & Short papers (i.e., less than five pages) \\
    E4 & Duplicate papers \\
    \bottomrule
    \end{tabular}
    
\end{table*}

\par For collecting a list of high-quality studies, we established both inclusion and exclusion criteria. You can find the selection process and these criteria in Figure~\ref{Fig:StudySelection} and Table~\ref{Table:InclusionExclusion} respectively. We applied these criteria to all 1189 studies from the three databases, involving several steps in the paper selection process. Initially, non-English papers were excluded. Non-research documents, such as reports and notes, were disregarded. Additionally, papers with fewer than five pages were excluded for comprehensive method explanations. To prevent redundancy, duplicated papers across all databases were removed before reviewing the content.

\par In subsequent stages, the first and second authors reviewed paper titles and abstracts, categorizing them into two lists and curating a selection of potential papers relevant to ML in CI. A comprehensive examination of the full text followed, ensuring relevance and content quality, with the cooperation of the first and second authors and continuous input from the third and fourth authors. According to Wohlin's guidelines~\cite{wohlin2014guidelines}, the selected papers must be published in peer-reviewed venues.

\par To ensure comprehensive coverage of relevant literature, we employed snowballing on the initial set of 50 papers~\cite{wohlin2014guidelines}, with the assistance of researchers whose names are presented in the acknowledgment section. Backward and forward snowballing involved reviewing references in selected papers and papers citing them, respectively. We repeated this process and reviewed steps until no more related papers could be identified.

\par In backward snowballing, 11 papers were identified, and in forward snowballing, four potential papers were found before full-text review. After reviewing these 15 potential papers, we filtered out 13 based on inclusion and exclusion criteria. In the end, 52 papers on the application of ML in CI were selected. The list and corresponding IDs are presented in the appendices, specifically in Table~\ref{Table_old:PaprID}.


\subsection{Data Extraction and Data Synthesis}
\label{Sec:Synth}
\par \textit{\textbf{Data extraction:}} Following Kitchenham's guidelines~\cite{kitchenham2007guidelines}, we designed a form to systematically collect information, including demographic data, from selected studies. The first author organized the extracted data in a Microsoft Excel spreadsheet according to the extraction form provided in Table~\ref{Table:ExtractedData}, with random checks performed by the second author.


\par \textit{\textbf{Data synthesis:}} This study aims to comprehensively analyze, classify, and report both qualitative and quantitative data. Thematic analysis was utilized for synthesizing qualitative data~\cite{braun2006using} a method recognized for its applicability in the software engineering domain ~\cite{cruzes2011recommended}. Quantitative data were presented in a raw format, including demographic information, or were derived from the synthesized qualitative data. The process of qualitative thematic analysis comprised six steps, as outlined below. The initial execution of these steps was undertaken by the first author and thoroughly discussed with the second author for potential modifications. The third and fourth authors oversaw this process and offered feedback for necessary adjustments.
\begin{enumerate}[noitemsep]
    
    \item \emph{Familiarizing with data:} We reviewed and annotated each field of the extracted data.
    \item \emph{Generating initial codes:} At this stage, we broke down the data into smaller parts and assigned codes, meaningful words or phrases acting as labels. This step involved iterative merging and revision of codes.
    \item \emph{Searching for themes:} After finalizing the codes, we reorganized and gathered relevant data for each code, defining potential themes.
    \item \emph{Reviewing themes:} We reviewed potential themes for relevance with the extracted data and codes for each research question.
    \item \emph{Defining and naming themes:} In the last step, we defined coherent and precise names for each theme, along with clear definitions.
    \item \emph{Reporting:} Our analysis and findings were mapped to each research question and reported in section~\ref{sec:Find}
\end{enumerate}

\section{Findings}
\label{sec:Find}
\par In this section, we present the findings derived from our comprehensive analysis of demographic data and synthesized data, and we will provide answers to our five research questions as outlined in Table~\ref{Table_old:RQs}.


\subsection{Demographic Data}

\par This section presents the demographic data concerning the four research design attributes, namely \textbf{(1)} publication year, \textbf{(2)} publisher venue and type of the venues: conference, journal or workshop, \textbf{(3)} study context and research methods, and \textbf{(4)} types of learning algorithms. The demographic information is valuable for researchers who wish to conduct research in this area, as it provides insights into where relevant studies can be found and their trends over the years, as well as critical areas by reviewing the list of keywords. Furthermore, it offers an overview of the study context and research methods employed in conducted studies~\cite{shahin2017continuous}.
\subsubsection{Distribution of Studies}
\begin{figure}
    \centering
    \includegraphics[width=0.7\linewidth]{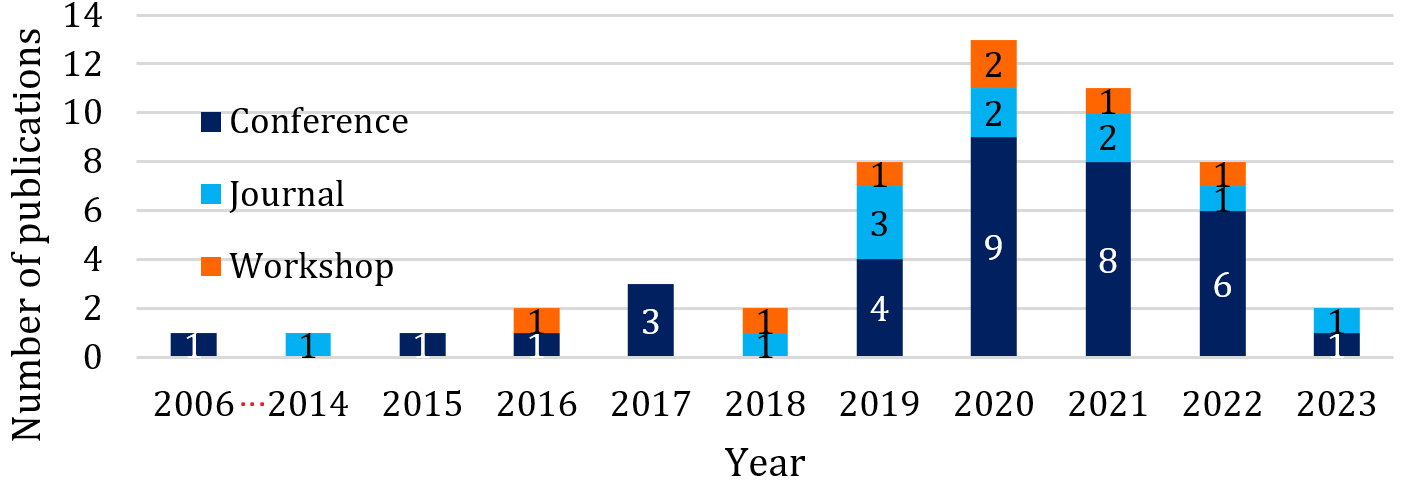}
    \caption{Number of selected studies published per year and their distribution over publication venues. \textbf{Note:} No paper was published between 2006 and 2014 --- Due to running the search string on July 2023, and snowballing on August 2023, the list of published papers in 2023 is incomplete.}
    \label{Fig:YearVenue}
\end{figure}
\par Since we ran the search string on July 26th 2023 and conducted the snowballing on August 7th 2023, we can say that a total of 52 studies were published between 2000 and August 2023. As shown in Figure~\ref{Fig:YearVenue}, the first study on the application of ML techniques in CI was published in 2006~\cite{hassan2006using}, which focused on predicting build validation in the CI phase. Notably, demographic data reveals that no paper was published on this topic between 2006 and 2014. However, the application of ML methods in CI has gradually gained more attention from researchers, and the number of studies started increasing since 2014. Additionally, the results indicate that a majority of papers (65.4\%) were published at conferences. It is worth mentioning that from 2021 all the studies focused only on enhancing the regression testing and predicting the build validations. This point presents the importance of these two phases in the CI pipeline.

\subsubsection{Distribution of Keywords}
\begin{figure}
    \centering
    \includegraphics[width=0.5\linewidth]{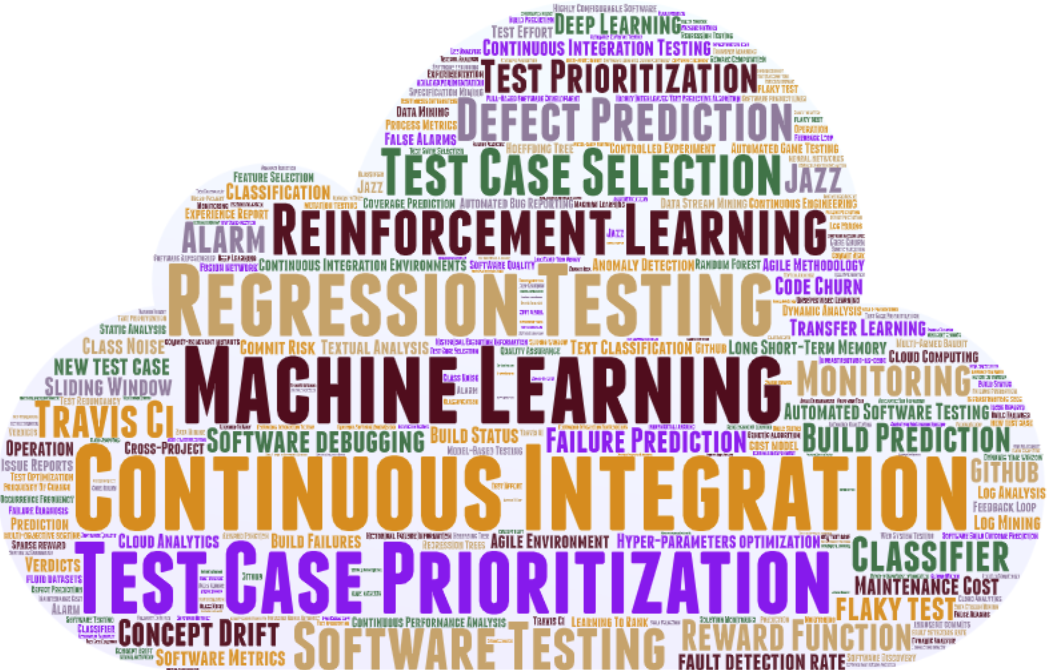}
    \caption{Word cloud of the keywords in selected primary studies in the CSE area.}
    \label{Fig:WordCloud}
\end{figure}

\par The concept of CI encompasses various areas and tools. To illustrate the areas that researchers have focused on, we can examine the keywords used in the selected studies. Figure~\ref{Fig:WordCloud} displays the frequency of each keyword in the primary studies and highlights the most frequent ones. The domains covered in this SLR are diverse, including Software Engineering concepts (e.g. Continuous Integration, Test Case Prioritization, Monitoring), software tools (e.g. Travis CI), and ML concepts (e.g. Classifier, Reward Function). The most common keywords in the selected studies are Continuous Integration, Machine Learning, Regression Testing, and Test Case Prioritization. This point depicts the importance of the regression testing phase of CI and especially its test case prioritization task.

\begin{table*}

\caption{Distribution of the 52 selected primary studies on publication venues. The color intensity in this table corresponds to the number of studies, with lighter shades indicating lower counts and darker blues representing higher counts.}
\centering
\label{Table:PubVenue}
\resizebox{\textwidth}{!}{%
\begin{tabular}{l|l|l}
\hline
\textbf{Publication Venue} & \textbf{\#} & \textbf{\%} \\ \hline

\cellcolor{B7D1E9}IEEE/ACM International Conference on Automated Software Engineering (ASE) & \cellcolor{B7D1E9} \textbf{3} & \cellcolor{B7D1E9} \textbf{5.8} \\
\cellcolor{B7D1E9}International Conference on Software Engineering (ICSE) & \cellcolor{B7D1E9} \textbf{3} & \cellcolor{B7D1E9} \textbf{5.8} \\
\cellcolor{B7D1E9}International Conference on Software Testing, Verification and Validation Workshops (ICSTW) & \cellcolor{B7D1E9} \textbf{3} & \cellcolor{B7D1E9} \textbf{5.8} \\
\cellcolor{B7D1E9}Proceedings of the ACM SIGSOFT International Symposium on Software Testing and Analysis (STA) & \cellcolor{B7D1E9} \textbf{3} & \cellcolor{B7D1E9} \textbf{5.8} \\
\cellcolor{B7D1E9}IEEE Transactions on Software Engineering (TSE) & \cellcolor{B7D1E9} \textbf{3} & \cellcolor{B7D1E9} \textbf{5.8} \\
\cellcolor{DDEBF7}Proceedings of the Asia-Pacific Symposium on Internetware (APSI) & \cellcolor{DDEBF7} \textbf{2} & \cellcolor{DDEBF7} \textbf{3.8} \\
\cellcolor{DDEBF7}CEUR Workshop Proceedings (CEUR) & \cellcolor{DDEBF7} \textbf{2} & \cellcolor{DDEBF7} \textbf{3.8} \\
\cellcolor{DDEBF7}Proceedings of the Annual International Conference on Computer Science and Software Engineering (CSSE) & \cellcolor{DDEBF7} \textbf{2} & \cellcolor{DDEBF7} \textbf{3.8} \\
\cellcolor{DDEBF7}IEEE Conference on Software Testing, Validation and Verification (ICST) & \cellcolor{DDEBF7} \textbf{2} & \cellcolor{DDEBF7} \textbf{3.8} \\
\cellcolor{DDEBF7}International Conference on Industrial, Engineering and Other Applications of Applied Intelligent Systems (IEA/AIE) & \cellcolor{DDEBF7} \textbf{2} & \cellcolor{DDEBF7} \textbf{3.8} \\
\cellcolor{DDEBF7}Journal of Software: Evolution and Process (JSEP) & \cellcolor{DDEBF7} \textbf{2} & \cellcolor{DDEBF7} \textbf{3.8} \\
\cellcolor{DDEBF7}IEEE International Conference on Software Analysis, Evolution and Re-engineering (SANER)  & \cellcolor{DDEBF7} \textbf{2} & \cellcolor{DDEBF7} \textbf{3.8} \\
\cellcolor{699DCB}Others (These venues only published one study) & \cellcolor{699DCB}\textbf{23} & \cellcolor{699DCB}\textbf{44.2} \\ \hline
\end{tabular}%
}

\end{table*}

\par Table~\ref{Table:PubVenue} displays the distribution of published papers on the application of ML in CI across 35 venues, with 12 venues publishing more than one study. Among these 12 venues, IEEE/ACM International Conference on Automated Software Engineering (ASE), International Conference on Software Engineering (ICSE), International Conference on Software Testing, Verification and Validation Workshops (ICSTW), Proceedings of the ACM SIGSOFT International Symposium on Software Testing and Analysis (STA), and IEEE Transactions on Software Engineering (TSE) stand out as the top venues for publishing research on the topic of application of ML in CI, with each venue publishing three papers.
It is worth noting that most of the papers (44.2\%) were published in 23 different venues, with Software Engineering venues being responsible for the majority of papers (65.2\%). The other domains of venues include Computer Science, Artificial Intelligence, Web and databases.

\subsubsection{Study Context and Research Methods}
\label{sec:datatype}

\begin{table}
\centering
\caption{The number of papers about each research and data analysis type. The color intensity in this table corresponds to the number of studies, with lighter shades indicating lower counts and darker blues representing higher counts. The bolded numbers represent the count of studies and their percentage in total.}
\label{Table:context}
\begin{tabular}{c|ccc} \toprule 
 & \textbf{Industry} & \textbf{Open-Source} & \textbf{Simulation}  \\ \toprule
\Centerstack{\textbf{Study IDs}\\\textbf{(Number of}\\\textbf{studies,}\\\textbf{percentage)}}
& \cellcolor{A3C4E1} \Centerstack{S1, S4, S8, S9, S13, S18, S21,\\S24, S26, S27, S28, S29, S30,\\S37, S39, S42, S43, S44, S46,\\S47, S49 \textbf{(21, 40.4\%)}}
& \cellcolor{699DCB} \Centerstack{S2, S3, S5, S6, S7, S10, S11, S14, S15, S16,\\S17, S20, S22, S23, S25, S26, S31, S32, S33,\\S34, S35, S36, S37, S38, S40, S41, S45,\\S48, S50, S51, S52 \textbf{(31, 59.6\%)}}
& \cellcolor{DDEBF7} \Centerstack{S12, S19\\\textbf{(2, 3.8\%)}}  \\ \bottomrule
\end{tabular}
\end{table}

\par Table~\ref{Table:context} provides a classification of the study context of the reviewed papers into three groups:``Industry'', ``Open-Source'', and ``Non-industry (simulation)''. Industrial studies were conducted with real-world data sets from software companies, such as projects in Microsoft company (S1)~\cite{philip2019fastlane}, to verify the applicability of the proposed methods in a practical closed-source software environment. On the other hand, studies in the ``Open-Source'' category validated their methods on real-world data projects from industrial open-source software, such as the Apache projects (S10)~\cite{brandtner2015sqa}. The distribution of the studies in Industrial and Open-Source categories presents the applicability of the ML methods in different CI environments. 

\par Based on Table~\ref{Table:context}, it can be observed that only two studies (S12, S19) conducted research in simulated environments, while the majority of studies (96.2\%) were situated in the Industry and Open-source categories. Simulation-based data sets refer to studies that implemented and evaluated their methods in simulated environments. In S19, the authors used the Gym library to simulate a CI environment by using execution logs of test cases and training a Reinforcement Learning (RL) agent. The simulator recreates a similar environment as a real environment for testing execution history. In S12, the authors simulated a cloud environment by installing packages manually and validated their ML method for discovering the installed software on containers and Virtual Machines (VMs) in testing the whole system. This highlights the usefulness of simulation-based data sets in evaluating and testing ML-based methods in a controlled and reproducible environment.

\par Moreover, we conducted a clustering analysis of the primary studies based on the types of research methods used. Typically, studies are categorized into two groups, namely quantitative and qualitative research, and we assigned the studies to each group based on their presented results and conclusions. Quantitative research methods involve the collection of empirical data through measurements and procedures, while qualitative studies are more descriptive and leave more room for interpretation~\cite{firestone1987meaning}.

\par In evaluating the effectiveness of quantitative research methods, researchers often rely on metrics such as the accuracy and performance of ML models, whereas qualitative methods may involve surveys, interviews, and other techniques to gather feedback from users of the trained ML models~\cite{dixon2005synthesising}. The results presented that none of the studies in this area employed qualitative methods for evaluating their solutions. Hence, a significant gap exists in the case of using qualitative analysis in the area of ML for CI. Furthermore, in the SLR paper, we provide additional information on the evaluation methods of quantitative research, which can assist researchers and practitioners in understanding how quantitative results can be validated and what are the most commonly used evaluation metrics in the area of ML for CI.

\vspace{3pt}
\noindent\fcolorbox{black}{blue!10}{%
\begin{minipage}{\columnwidth}
\vspace{1pt}
\textbf{Summary:}
\vspace{1pt}
\par $\bullet$ The number of published studies related to the application of ML in CI has been on the rise, and gradually narrowed down to only regression testing and build validation over time.
\par $\bullet$ All of the studies included in this SLR utilized quantitative research methods. However, a lack of qualitative assessment of the results is visible in the literature.
\end{minipage}}

\begin{figure*}[!h]
    \centering
    \includegraphics[width=\textwidth]{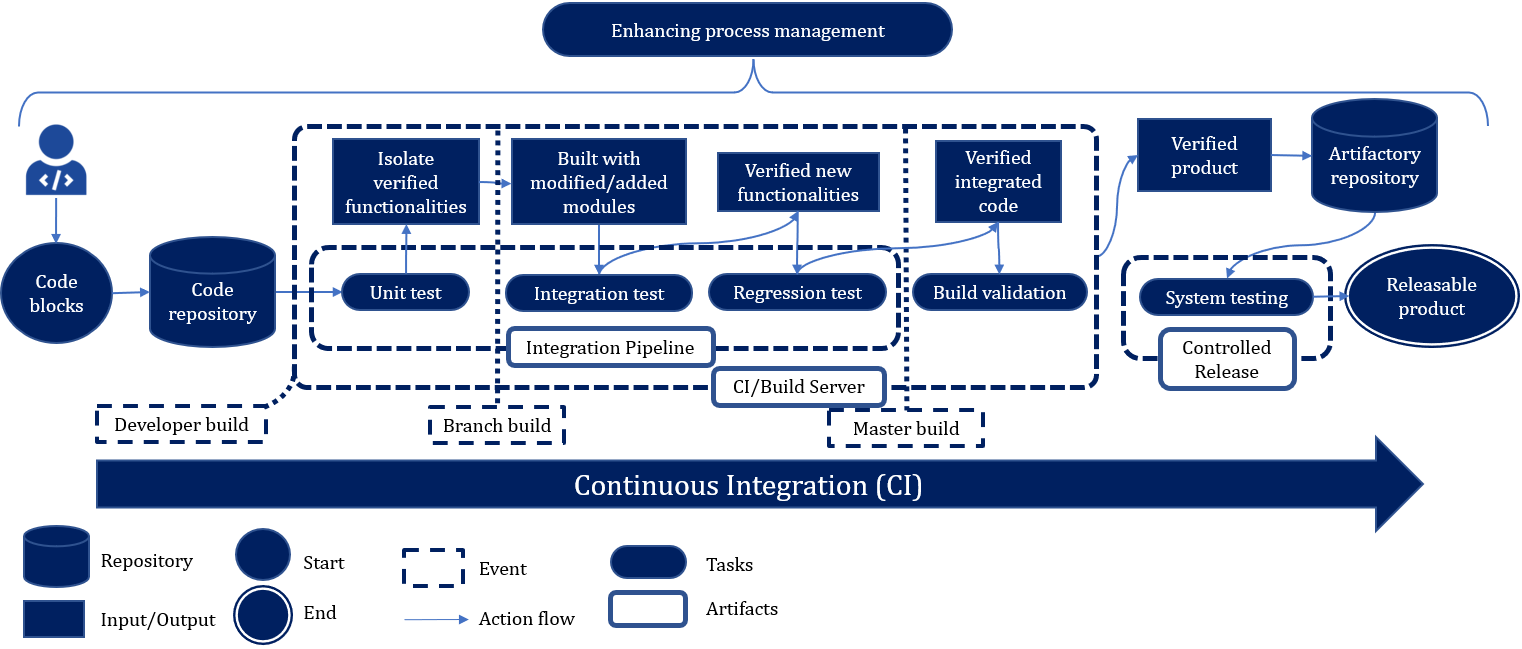}
    \caption{Overview of connection between six CI phases and their in/output.}
    \label{Fig:CIPhases}
\end{figure*}

\subsection{RQ1: Continuous Integration Phases and Tasks}
\label{Sec:CITasks}
\par Our analysis uncovered six distinct ML-enhanced phases within CI, interconnected as illustrated in Figure~\ref{Fig:CIPhases}. These phases are presented in the order of their sequence within the CI pipeline. To highlight automated tasks within each CI phase, we have identified them in an underlined format, briefly explained in the description of each phase. Table~\ref{Table:StudiesCItasks} lists ten identified automated CI tasks across these six phases. It is worth mentioning that, due to the limited number of studies and the use of different evaluation metrics, we did not compare state-of-the-art methods in each of the other tasks except for the Test Optimization and Build Prediction tasks.

\begin{table*}
\caption{List of papers in each CI phase with related tasks and total count in parentheses. The table's color intensity corresponds to the number of studies, with lighter shades indicating lower counts and darker blues representing higher counts. The bolded numbers represent the count of studies and their percentage in total.}
\label{Table:StudiesCItasks}
\resizebox{\textwidth}{!}{%
\begin{tabular}{c|c|cc|ccc|c|cc|c}
\toprule
\Centerstack{\textbf{CI}\\\textbf{Tasks}} &   \rotN{\Centerstack[l]{\textbf{Unit Test}\\\textbf{Prediction}}} &
  \rotN{\Centerstack[l]{\textbf{Branch Coverage}\\\textbf{Prediction}}} &
  \rotN{\Centerstack[l]{\textbf{Integration Test}\\\textbf{Prediction}}} &
  \rotN{\textbf{Test Optimization}} &
  \rotN{\textbf{Defect Prediction}} &
  \rotN{\Centerstack[l]{\textbf{Flaky Test}\\\textbf{Detection}}} &
  \rotN{\textbf{Build Prediction}} &
  \rotN{\Centerstack[l]{\textbf{Installed Software}\\\textbf{Discovery}}} &
  \rotN{\Centerstack[l]{\textbf{Performance Test}\\\textbf{Optimization}}} &
  \rotN{\Centerstack[l]{\textbf{Activity}\\\textbf{Management}}} \\ 
  \hline
 \Centerstack{\textbf{CI}\\\textbf{Phase}} &\multicolumn{1}{c|}{\Centerstack{\textbf{Unit}\\\textbf{Test}}}&\multicolumn{2}{c|}{\Centerstack{\textbf{Integration}\\ \textbf{Test}}}&\multicolumn{3}{c|}{\textbf{Regression Test}}&\multicolumn{1}{c|}{\Centerstack{\textbf{Build}\\\textbf{Validation}}}&\multicolumn{2}{c|}{\Centerstack{\textbf{System}\\\textbf{Test}}}&\multicolumn{1}{c}{\Centerstack{\textbf{Process}\\\textbf{Management}}} \\ \hline
\textbf{Studies} &
  \cellcolor{DDEBF7}\Centerstack{S2\\\textbf{(1, 1.9\%)}} &
  \cellcolor{D6E7F5}\Centerstack{S3, S15\\\textbf{(2, 3.9\%)}} &
  \cellcolor{D6E7F5}\Centerstack{S18, S23\\\textbf{(2, 3.9\%))}}&
  \cellcolor{699DCB}\Centerstack{S7, S8, S11, S14,\\S16, S19, S20, S22,\\S26, S28, S29, S30,\\S31, S33, S34, S35,\\S36, S37, S39, S41,\\S45, S46, S47, S48,\\S50, S52 \textbf{(26, 50\%)}} &
  \cellcolor{C0D8EC}\Centerstack{S9, S13,\\S17, S32\\\textbf{(4, 7.7\%)}} &
  \cellcolor{DDEBF7}\Centerstack{S40\\\textbf{(1, 1.9\%)}}&
  \cellcolor{8EB5D9}\Centerstack{S1, S5, S6,\\S21, S24, S25,\\S27, S38, S42,\\S43, S44, S51\\\textbf{(12, 23.1\%)}} &
  \cellcolor{DDEBF7}\Centerstack{S12\\\textbf{(1, 1.9\%)}}&
  \cellcolor{DDEBF7}\Centerstack{S4\\\textbf{(1, 1.9\%)}}&
  \cellcolor{D6E7F5}\Centerstack{S10, S17\\\textbf{(2, 3.9\%)}} \\ \hline

\end{tabular}
}

\end{table*}

\par \textit{\textbf{Unit Test (UT):}} According to Stolberg's definition~\cite{stolberg2009enabling}, the Unit Test phase serves as the initial step in the CI pipeline. This phase involves validating newly developed code within an isolated environment. It occurs whenever a developer commits new or modified compiled code. The code built during this step is referred to as the ``Developer build'', as depicted in Figure~\ref{Fig:CIPhases}. 
\par In CI, the master branch is typically kept free from failures, with code changes made on a separate ``developer branch'' throughout this study. Among the 52 studies, only S2 focused on this phase, conducting \underline{Unit Test Prediction}. The authors enhanced static code checkers' accuracy using ML models to predict false positives (wrongly predicting safe code) (S2), reducing the risk of releasing buggy software.

\par \textit{\textbf{Integration Test (IT):}} After validating units with Unit Test in the developer branch, the next step is to integrate the new module with other developed modules on the same branch~\cite{al2019predicting}. This phase involves testing various modules and new functionalities in combination with the entire system or sub-systems. ML-based methods in this CI phase aim to predict test outcomes (S18) and reduce the computational load of integration tests by skipping safe commits (S23), executing fewer test cases, and reducing overall costs (S3, S15) through ML-based \underline{Integration Test Prediction} or \underline{Branch Coverage Prediction}.

\begin{table*}
\caption{Comparison of the three SOTA methods in optimizing the regression testing. \textbf{Note:} TCP is Test Case Prioritization and TCS is Test Case Selection Tasks; RL is Reinforcement Learning, NN is Neural Networks, and MAB is Multi-Armed Bandit.}
\label{table:TestOpt_SOTA}
\resizebox{\textwidth}{!}{%
\begin{tabular}{l|ccccll}
\hline
\textbf{SID} &
  \textbf{Method Name} &
  \textbf{Strategy} &
  \textbf{Model Properties} &
  \textbf{Minimum Required Data} &
  \Centerstack{\textbf{Advantages}} &
  \Centerstack{\textbf{Limitations}} \\ \hline
\textbf{S7} &
  RETECS &
  TCP \& TCS &
  RL (NN-based agent) &
  60 test cycles &
  \Centerstack[l]{$\bullet$ Language-Free\\ $\bullet$ Not required source code} &
  \Centerstack[l]{$\bullet$ Not appropriate for\\~~large scale datasets}
  \\ \cdashline{1-7}
\textbf{S20} &
  COLEMAN &
  TCP &
  RL (MAB-based agent) &
  100 test cycles &
  \Centerstack[l]{$\bullet$ Language-Free\\ $\bullet$ Lightweight} &
  \Centerstack[l]{$\bullet$ Cold start}
  \\\cdashline{1-7}
\textbf{S43} &
  DeepOrder &
  TCP &
  DL &
  4 test cycles & 
  \Centerstack[l]{$\bullet$ Language-Free\\ $\bullet$ Efficient for large datasets\\ $\bullet$ Time-efficient}    &
  \Centerstack[l]{$\bullet$ Non-interpratable}\\
  \hline
\end{tabular}%
}
\end{table*}

\par \textit{\textbf{Regression Test (RT):}} Following integrated code validation and testing, the software product's current version undergoes comprehensive testing based on previously designed test cases~\cite{ali2019enhanced}. Regression tests encompass various types, such as structural and functional testing~\cite{elsner2021empirically}. Given the time and resource intensity of regression testing~\cite{pan2020dynamic}, the majority of studies (31 out of 52) focus on this area. The main ML-based solutions explored by 26 studies are Test Case Prioritization (TCP) and Test Case Selection (TCS), contributing to \underline{Test Optimization}. In TCP, ML-based solutions expedite the detection of software defects by ordering test cases based on the likelihood of revealing faults more quickly~\cite{lima2020learning}. TCS focuses on selecting test cases that assess code changes while consuming fewer resources and less time~\cite{martins2021supervised}. Additionally, selected studies introduced other ML-based solutions, including \underline{Defect Prediction} (S9, S13, S17, and S32) and \underline{Flaky Test Detection} (S40).

\par Among the 26 papers primarily focused on Test Optimization, three stand out in terms of the accuracy of the trained ML models selected as State-Of-The-Art (SOTA) approaches: S7, S20, and S43. These papers introduced the Reinforced Test Case Prioritization and Selection (RETECS), Combinatorial VOlatiLE Multi-Armed BANdit (COLEMAN), and Deep learning for test case prioritization (DeepOrder) methods, respectively. According to the authors, both COLEMAN and DeepOrder outperformed the RETECS method, previously recognized as the latest SOTA approach for test optimization in regression testing. Notably, upon reviewing these COLEMAN and DeepOrder studies, it was observed that both employed the IOF/ROL dataset. In this dataset, DeepOrder yielded slightly better Normalized Average Percentage of Fault Detection (NAPFD) (see section~\ref{Sec:MethodEval}) results than the COLEMAN method. However, it is essential to emphasize that we cannot conclusively assert that DeepOrder universally outperforms the COLEMAN method, as further examinations are required.

\par For a comprehensive overview of these three SOTA methods and their respective advantages, please refer to Table~\ref{table:TestOpt_SOTA}. Importantly, all these SOTA methods share a common characteristic: they do not require access to the source code for test case prioritization, making them language-independent. Their sole prerequisite is the availability of test case metadata, reducing training time and facilitating the use of larger datasets.

\par Regarding other tasks in the RT phase, only one study focuses on Flaky Test Detection, while four studies address Defect Prediction CI tasks. Our comparative analysis will exclusively concentrate on the body of published works concerning Defect Prediction tasks within the RT phase.

\par However, it is imperative to acknowledge that these studies have diverse objectives and methodological properties. The wide range of research goals and methodologies within this limited corpus prevents direct comparisons. Here, we present the objectives of these studies to afford a comprehensive understanding of their respective aims and contributions.

\par \emph{Performance Enhancement for Low Defect Percentage Datasets \textbf{(S9)}:} This study proposes methods with good performance on datasets characterized by a low defect percentage.
\par \emph{Impact of Feature Extraction Methods \textbf{(S13)}:} Investigations in this category assess the influence of feature extraction techniques, such as bag-of-words and word embedding, on ML method effectiveness.
\par \emph{Transfer Learning for New Project Implementation \textbf{(S17)}:} Studies in this category explore transfer learning techniques for using trained ML models in new projects to predict defects.
\par \emph{Continuous Defect Prediction Based on Code Change Features \textbf{(S32)}:} Research efforts in this context introduce ML methods specifically designed to continuously predict defects by leveraging code change features.

\begin{table*}
\caption{Comparison of the three SOTA methods in build outcome validation. \textbf{Note:} DNN is Deep Neural Network, LSTM is Long Short Term Memory.
}
\label{Table:BuildValid_SOTA}
\resizebox{\textwidth}{!}{%
\begin{tabular}{l|ccccc}
\hline
\textbf{SID} &
  \textbf{Method Name} &
  \textbf{Strategy} &
  \textbf{Model Properties} &
  \textbf{Results} &
  \textbf{Advantages} \\ \hline
\textbf{S21} &
  SmartBuildSkip &
  Predicting pass builds &
  Random Forest &
  \Centerstack[l]{$\bullet$ Outperform SOTA (S24)\\ $\bullet$ Save 30\% on building time~~~~~~~~~~~~~~~~~~~~~~~~~~~~~~~~~} &
  $\bullet$ Lightweight~~~~~~~~~~~~~~~~~~~~~~~~~~~~~~~ \\ \cdashline{1-6}
\textbf{S25} &
  BuildFast &
  Predicting fail builds &
  XGBoost &
  \Centerstack[l]{$\bullet$ Outperform SOTA (S6 and S24)\\$\bullet$  47.5\% improved F1 score of SOTA method~~~~~~~} &
  \Centerstack[l]{$\bullet$ Chronological~~~~~~~~~~~~~~~~~~~~~~~~~~~~~\\$\bullet$ Lightweight} \\ \cdashline{1-6}
\textbf{S38} &
  DL-CIBuild &
  Predicting build outcome &
  DNN (LSTM) &
  $\bullet$ Improved F1-score of best ML methods by 10\% &
  \Centerstack[l]{$\bullet$ Chronological\\$\bullet$ No feature engineering required\\$\bullet$ Works within and cross projects} \\ \hline
\end{tabular}%
}
\end{table*}

\par \textit{\textbf{Build Validation (BV):}} Following the preceding CI phases, developers generally have confidence in the functionalities and performance of developed software units. At this stage, the software version is ready for merging into the master branch, representing the final product for building based on the developed units within this branch. The Build Validation phase focuses on ensuring the stability of integrated codes before releasing the software for system testing~\cite{finlay2014data}. Given the significant computational cost associated with building software products~\cite{xie2018cutting}, ML-based solutions in this phase primarily target \underline{Build Prediction} and reduce building efforts. 
\par Among the 12 studies concentrating on the BV task, eleven predicted the build outcome using data from the same project (within-project). In contrast, study S5 used building information from various projects to train a cross-project model, allowing build prediction for projects lacking sufficient data~\cite{xia2017empirical}. Given the substantial focus on build validation, we reviewed published studies in this area and identified three SOTA methods in terms of the top three accurate ML models, namely S21, S25, and S38, each employing distinct strategies.

\par \textbf{S21} introduced a method to predict passing builds and save time by skipping unnecessary building tests (SmartBuildSkip) using a Random Forest method. Their approach reduced the time spent on building tests by 30\%, outperforming the reviewed ML method in predicting passing builds. It is worth noting that S21 considered a build as likely to fail if it had failed in the previous execution, excluding failed builds from evaluations.

\par In contrast, \textbf{S25} focused on predicting failing builds, utilizing a decision-tree-based method (XGBoost) for their \textit{BuildFast} approach. Results showed a remarkable 47.5\% improvement in the F1-score for predicting failed builds compared to existing SOTA methods. An advantage of BuildFast over SmartBuildSkip is its consideration of the chronological order of input data during model training, enabling its use in real-world environments.

\par The latest SOTA method in build validation, \textbf{S38}, employed Long Short-Term Memory (LSTM), a deep learning-based approach, known as \textit{DL-CIBuild}. DL-CIBuild, a chronological method, outperformed existing SOTA ML-based methods in F1-score by 10\%. DL-CIBuild has advantages over SmartBuildSkip and BuildFast: it does not require feature engineering and can be applied to both intra-project and cross-project build validation tasks. Table~\ref{Table:BuildValid_SOTA} offers a concise overview of these three SOTA methods in build validation in CI.

\par \textit{\textbf{System Test (ST):}} In the final step, System Test, the complete software system undergoes testing to ensure all aspects, including performance, functionality, and compatibility, are correct. ML-based solutions in this phase are limited, mainly focusing on \underline{Installed Software Discovery} for compliance, security, and efficiency (S12) and generating test cases for \underline{Performance Test Optimization} to detect defects in the system (S4).

\par \textit{\textbf{Process Management (PM):}} This step involves communication among developers and the production of numerous documents, ML methods are applied for \underline{Activity} \underline{Management}. S10 classifies CI environment information for project management committee members, while S17 proposes an ML-based approach for automatically labeling reported issues, aiding software engineers in assigning the correct issues to developers. Notably, this phase primarily relies on natural language-based information, and the limited use of Large Language Models (LLM) is noteworthy. 

\par 
In industrial projects, the sequence of Integration and Regression tests may be altered or combined based on project requirements and priorities, deviating from the sequence in Figure~\ref{Fig:CIPhases}. Selected studies for each CI phase and task are listed in Table~\ref{Table:StudiesCItasks}.


\vspace{1pt}
\noindent\fcolorbox{black}{blue!10}{%
\begin{minipage}{\columnwidth}
\vspace{1pt}
\textbf{Summary:}
\vspace{1pt}

\par $\bullet$ Six phases and ten tasks are identified in CI pipelines: Unit Test, Integration Test, Regression Test, Build Validation, System Test, and Process Management. Detailed descriptions of their sequence and input/output are provided.
\vspace{1pt}

\par $\bullet$ Out of the 52 selected studies, 31 focused on Regression Testing and 12 on Build Validation due to their high costs, emphasizing their critical roles in ensuring software quality. Other CI phases had nine studies collectively, highlighting the significance of RT and BV.
\vspace{1pt}


\par $\bullet$ Key tasks include Test Optimization in Regression Testing and Build Prediction in Build Validation. SOTA methods for these tasks are RETECS, COLEMAN, DeepOrder, SmartBuildSkip, BuildFast, and DL-CIBuild, demonstrating advancements in CI.
\end{minipage}}


\subsection{RQ2: Data Sets and Data Engineering Methods}
\label{Sec:DataEng}
\par Input data selection, data engineering, and model training are crucial factors influencing ML model performance~\cite{schwabacher2005survey}. The nature and availability of input data vary across CI tasks, emphasizing the importance of understanding data sources and engineering methods.

\par To facilitate informed decisions, this section offers insights into data sources, types, and engineering techniques employed in selected studies. Analyzing this information aids researchers and practitioners in understanding diverse approaches in ML4CI and discovering potential research directions.

\subsubsection{Data Sources} In the reviewed literature, 67 diverse data sources were utilized across the 52 selected ML-based CI studies. These sources differ in characteristics like lines of code (LOC), number of builds, and project specificity. For instance, study S3 used Google Dagger (848 LOC), contrasting with S2 employing a Samsung dataset with over 27 million LOC. Similarly, builds in different studies varied significantly, e.g., the Jazz project in S44 had 199 builds, while S5 used the TravisTorrent dataset with over 300,000 builds. These variations emphasize the need to comprehend data source characteristics in ML-based CI research.

\begin{table*}
\caption{This table summarizes frequently used study data sources, their correlation with identified CI tasks in each CI phase, and study IDs. color intensity in the table shows numerical values, with lighter shades indicating lower and darker blues representing higher values. The bolded numbers represent the count of studies and their percentage in total. \textbf{Acronyms:} \textit{\textbf{UT:}} Unit Test, \textit{\textbf{IT:}} Integration Test, \textit{\textbf{RT:}} Regression Test, \textit{\textbf{BV:}} Build Validation, \textit{\textbf{ST:}} System Test and \textit{\textbf{PM:}} Process Management.}
\label{Table:DataSourceCI}
\resizebox{\textwidth}{!}{%
\begin{tabular}{cccc|c|cccccccccc}
\hline


  \rotN{\textbf{\# Test Cases}} &
  \rotN{\textbf{\# CI Cycles}} &
  \rotN{\textbf{\# Test Outcomes}} &
  \rotN{\textbf{Fail Test Rate}} &
  
 &
\rotN{\Centerstack[l]{\textbf{Unit Test}\\\textbf{Prediction}}} &
  \rotN{\Centerstack[l]{\textbf{Branch Coverage}\\\textbf{Prediction}}} &
  \rotN{\Centerstack[l]{\textbf{Integration Test}\\\textbf{Prediction}}} &
  \rotN{\Centerstack[l]{\textbf{Test}\\\textbf{Optimization}}} &
  \rotN{\Centerstack[l]{\textbf{Defect}\\\textbf{Prediction}}} &
  \rotN{\Centerstack[l]{\textbf{Flaky Test}\\\textbf{Detection}}} &
  \rotN{\Centerstack[l]{\textbf{Build Prediction}}} &
  \rotN{\Centerstack[l]{\textbf{Installed Software}\\\textbf{Discovery}}} &
  \rotN{\Centerstack[l]{\textbf{Performance Test}\\\textbf{Optimization}}} &
  \rotN{\Centerstack[l]{\textbf{Activity}\\\textbf{Management}}} \\ 
  \hline
\multicolumn{4}{c|}{\textbf{Data Properties}} & 
 \textbf{CI Datasets} &
\multicolumn{1}{c|}{\textbf{UT}}&
\multicolumn{2}{c|}{\textbf{IT}}&
\multicolumn{3}{c|}{\textbf{RT}}&
\multicolumn{1}{c|}{\textbf{BV}}&
\multicolumn{2}{c|}{\textbf{ST}}&
\multicolumn{1}{c}{\textbf{PM}}
\\  
\hline

\cellcolor{A2C3E1}1941   & \cellcolor{DDEBF7}320   & \cellcolor{DBEAF7}32260   & \cellcolor{3F80BB}28.79\% & \textbf{IOF/ROL}        &  \cellcolor{DDEBF7}-- & \cellcolor{DDEBF7}--      &\cellcolor{DDEBF7}-- & \cellcolor{3F80BB}\Centerstack{S7, S11, S14, S19,\\S20, S22, S28, S31,\\S33, S35, S37, S48\\ \textbf{(12, 23.1\%)}}   &\cellcolor{DDEBF7}-- &\cellcolor{DDEBF7}-- &\cellcolor{DDEBF7}--                              &\cellcolor{DDEBF7}-- &\cellcolor{DDEBF7}-- &\cellcolor{DDEBF7}--  \\ \cdashline{1-15}
 \cellcolor{DCEBF7}89     & \cellcolor{DCEAF7}352   & \cellcolor{DCEAF7}25594   & \cellcolor{699CCB}19.36\% & \textbf{Paint Control}  &\cellcolor{DDEBF7}-- &\cellcolor{DDEBF7}--      &\cellcolor{DDEBF7}-- & \cellcolor{6FA0CD}\Centerstack{S7, S11, S14, S19,\\S20, S22, S31, S33,\\S35, S37, S48\\ \textbf{ (11, 21.2\%)}}        &\cellcolor{DDEBF7}-- &\cellcolor{DDEBF7}-- &\cellcolor{DDEBF7}--                              &\cellcolor{DDEBF7}-- &\cellcolor{DDEBF7}-- &\cellcolor{DDEBF7}--  \\\cdashline{1-15}
 \cellcolor{D1E3F3}457 & \cellcolor{D7E7F5}434  & \cellcolor{B1CDE7}332650 & \cellcolor{DDEBF7}0.02\%  & \textbf{Apache Commons} &\cellcolor{DDEBF7}-- &\cellcolor{BED6EB}\Centerstack{S3, S15\\ \textbf{(2, 3.8\%)} } &\cellcolor{DDEBF7}-- & \cellcolor{7FABD3}\Centerstack{S19, S20, S22, S33,\\ S35, S48 \textbf{(6, 11.6\%)}}                             &\cellcolor{DDEBF7}--  &\cellcolor{DDEBF7}-- &\cellcolor{DDEBF7}--                              &\cellcolor{DDEBF7}-- &\cellcolor{DDEBF7}-- &\cellcolor{CEE1F1}\Centerstack{S10\\ \textbf{(1, 1.9\%)} }\\\cdashline{1-15}
 \cellcolor{3F80BB}5555   & \cellcolor{DDEBF7}336   & \cellcolor{3F80BB}1260618 & \cellcolor{DCEBF7}0.25\%  &\textbf{GSDTSR}         &\cellcolor{DDEBF7}-- &\cellcolor{DDEBF7}--      &\cellcolor{DDEBF7}-- & \cellcolor{4F8BC1}\Centerstack{S7, S14, S16, S22,\\S31, S33, S35, S37,\\ S48 \textbf{(9, 17.3\%)}}              &\cellcolor{DDEBF7}-- &\cellcolor{DDEBF7}-- &\cellcolor{DDEBF7}--                              &\cellcolor{DDEBF7}-- &\cellcolor{DDEBF7}-- &\cellcolor{DDEBF7}--  \\\cdashline{1-15}
  \cellcolor{CDE0F1}568    & \cellcolor{A3C4E1}1312  & \cellcolor{7FABD3}694395 & \cellcolor{DDEBF7}0.03\%   &\textbf{Google Guava}   &\cellcolor{DDEBF7}-- &\cellcolor{BED6EB}\Centerstack{S3, S15\\ \textbf{(2, 3.8\%)} } &\cellcolor{DDEBF7}-- & \cellcolor{AECBE5}\Centerstack{S20, S28, S48 \\ \textbf{(3, 5.8\%)} }                                                          &\cellcolor{DDEBF7}--  &\cellcolor{DDEBF7}-- &\cellcolor{DDEBF7}--                              &\cellcolor{DDEBF7}-- &\cellcolor{DDEBF7}-- &\cellcolor{DDEBF7}--  \\\cdashline{1-15}
 \cellcolor{9FC1E0}2010   & \cellcolor{2F75B5}3263  & \cellcolor{72A3CF}781273  & \cellcolor{DAE9F6}0.62\%  &\textbf{Rails}          &\cellcolor{DDEBF7}-- &\cellcolor{DDEBF7}--      &\cellcolor{DDEBF7}-- & \cellcolor{8EB6D9}\Centerstack{S20, S22, S33, S35,\\S48 \textbf{(5, 9.7\%)}}                                  &\cellcolor{DDEBF7}--  &\cellcolor{DDEBF7}-- &\cellcolor{DDEBF7}--                              &\cellcolor{DDEBF7}-- &\cellcolor{DDEBF7}-- &\cellcolor{DDEBF7}--  \\\cdashline{1-15}
 N/A  & N/A  & N/A     & N/A     &\textbf{TravisTorrent}  &\cellcolor{DDEBF7}-- &\cellcolor{DDEBF7}--      &\cellcolor{DDEBF7}-- &\cellcolor{DDEBF7}--                                                                      &\cellcolor{DDEBF7}--  &\cellcolor{DDEBF7}-- & \cellcolor{9EC1DF}\Centerstack{S5, S6, S21,\\S29 \textbf{(4, 7.7\%)}} &\cellcolor{DDEBF7}-- &\cellcolor{DDEBF7}-- &\cellcolor{DDEBF7}--  \\\cdashline{1-15}
 \cellcolor{D5E6F4}303    & \cellcolor{BBD4EB}988   & \cellcolor{6EA0CD}815598  & \cellcolor{DDEBF7}0.15\%  &\textbf{MyBatis}        &\cellcolor{DDEBF7}-- &\cellcolor{DDEBF7}--      &\cellcolor{DDEBF7}-- &\cellcolor{AECBE5}\Centerstack{ S20, S33, S48 \\\textbf{(3, 5.8\%)}}                                                           & \cellcolor{CEE1F1}\Centerstack{ S32 \\\textbf{(1, 1.9\%)}} &\cellcolor{DDEBF7}-- &\cellcolor{DDEBF7}--                              &\cellcolor{DDEBF7}-- &\cellcolor{DDEBF7}-- &\cellcolor{DDEBF7}--  \\\cdashline{1-15}
 \cellcolor{D4E5F4}360 & \cellcolor{6B9ECC}2257 & \cellcolor{83AED5}663470 & \cellcolor{DDEBF7}0.07\%  &\textbf{Google Closure} &\cellcolor{DDEBF7}-- &\cellcolor{DDEBF7}--      &\cellcolor{DDEBF7}-- &\cellcolor{AECBE5}\Centerstack{ S20, S33, S48 \\\textbf{(3, 5.8\%)}}                                                           &\cellcolor{DDEBF7}--  &\cellcolor{DDEBF7}-- &\cellcolor{DDEBF7}--                              &\cellcolor{DDEBF7}-- &\cellcolor{DDEBF7}-- &\cellcolor{DDEBF7}--  \\\cdashline{1-15}
 \cellcolor{DDEBF7}46     & \cellcolor{D1E3F3}638   & \cellcolor{DDEBF7}14601   & \cellcolor{DCEBF7}0.19\%  &\textbf{Google Auto}    &\cellcolor{DDEBF7}-- &\cellcolor{DDEBF7}--      &\cellcolor{DDEBF7}-- &\cellcolor{AECBE5} \Centerstack{ S20, S33, S48 \\\textbf{(3, 5.8\%)}}                                                           &\cellcolor{DDEBF7}--  &\cellcolor{DDEBF7}-- &\cellcolor{DDEBF7}--                              &\cellcolor{DDEBF7}-- &\cellcolor{DDEBF7}-- &\cellcolor{DDEBF7}--  \\\cdashline{1-15}
 \cellcolor{DCEAF7}106    & \cellcolor{3F80BB}3813  & \cellcolor{C3DAED}204161  & \cellcolor{D3E4F3}1.82\%  &\textbf{Dspace}         &\cellcolor{DDEBF7}-- &\cellcolor{DDEBF7}--      &\cellcolor{DDEBF7}-- &\cellcolor{AECBE5} \Centerstack{ S20, S33, S48 \\\textbf{(3, 5.8\%)}}                                                           &\cellcolor{DDEBF7}--  &\cellcolor{DDEBF7}-- &\cellcolor{DDEBF7}--                              &\cellcolor{DDEBF7}-- &\cellcolor{DDEBF7}-- &\cellcolor{DDEBF7}--  \\\cdashline{1-15}
 N/A  & N/A  & N/A     & N/A     &\textbf{Google Dagger}  &\cellcolor{DDEBF7}-- &\cellcolor{BED6EB}\Centerstack{S3, S15\\ \textbf{(2, 3.8\%)} } &\cellcolor{DDEBF7}-- &\cellcolor{DDEBF7}--                                                                      &\cellcolor{DDEBF7}--  &\cellcolor{DDEBF7}-- &\cellcolor{DDEBF7}--                              &\cellcolor{DDEBF7}-- &\cellcolor{DDEBF7}-- &\cellcolor{DDEBF7}--  \\ \hline
\end{tabular}%
}
\end{table*}

\par To highlight frequently used sources, 11 publicly available datasets were employed in more than one study, as indicated by the numbers in parentheses.

\par To underscore the applicability of the data sources, we have presented 11 datasets that have been utilized in multiple studies and are reported as publicly available at the time of the studies' publication. The number in parentheses indicates the frequency with which each dataset has been utilized in these studies.

IOF/ROL~(12),
Paint-Control~(11) 
from ABB Robotics company,
Apache-Commons~(9),
GSDTSR~(9),
Google Guava~(5),
Rails~(5),
TravisTorrent CI projects~(5),
MyBatis~(4),
Google Closure~(3),
Google Auto~(3),
DSpace~(3),
 and 
Google Dagger~(2).

\par Table~\ref{Table:DataSourceCI} provides a summary of the properties of these datasets, with particular emphasis on datasets related to testing data in CI. Additionally, the data properties of Travis Torrent and Google Dagger are presented as N/A due to their unavailability at the time of writing this paper.

\par To assist researchers in selecting suitable datasets based on their defined research problems, this section presents key statistics and summaries of selected datasets. For instance, ABB company, a prominent industrial robot supplier, provides robot software and equipment, with datasets such as Paint Control (PC) and IOF/ROL containing historical information on test results and over 300 CI cycles. CI cycles encompass all development tasks, including coding, building, and testing, occurring continuously to ensure code integration and quality throughout the software development lifecycle before preparing a software product for deployment~\cite{vassallo2019enabling}. PC includes 89 test cases, 352 CI cycles, 25,594 verdicts, and a 19.36\% failure rate, while IOF/ROL consists of 1941 test cases, 320 CI cycles, 32,260 verdicts, and a 28.79\% failure rate. The GSDTSR dataset, an open resource from Google, has 336 CI cycles, with a notably low failure rate of around 0.25\%. Considering these datasets alongside the previously discussed sources enables researchers to make informed choices based on their research objectives and requirements, facilitating evaluations in diverse dataset scenarios.
\par Table~\ref{Table:DataSourceCI} highlights notable characteristics of datasets. The IOF/ROL and PC datasets stand out for their higher failure rates, a rarity in CI, making them popular among researchers. The Apache datasets gain popularity for their good structure and quality~\cite{chen2022focus}. Additionally, the GSDTSR, Google Guava, Rails, MyBatis, and Google Closure datasets find frequent use due to their substantial test outcome volume, as indicated in Table~\ref{Table:DataSourceCI}.

\begin{table*}
\centering
\caption{Definition of the identified data types in the selected studies. \textbf{Acronyms:} \textit{\textbf{MD:}} Meta Data.}
\label{Table:DataTypes}
\begin{tabular}{p{0.12\linewidth}|p{0.4\linewidth}|p{0.4\linewidth}}
\toprule
         \Centerstack{\textbf{Data Type}}    & \Centerstack{\textbf{Description}} & \Centerstack{\textbf{ML-based Example}}  \\ \toprule
\textbf{Source Code} & Actual code written by developers that forms the basis of a software application. Usually, it requires preprocessing to be understandable for ML methods. & Detecting text similarity or source-code coverage by tests to predict the outcome of a test through tokenizing it or making the Abstract Syntax Trees (AST). \\ \cdashline{1-3}
\textbf{Code MD} &  This data type extracts more information than the Source Code data type by considering specific characteristics of the source code.
& Predicting test outcomes by calculating the changes and complexity of the code (e.g. number of changed LOC, or the depth changed module in the inheritance tree) \\\cdashline{1-3}
\textbf{Test Case} & This data type covers the textual information of test cases, including names and test codes. & Prioritizing the execution of tests based on analyzing the test codes.    \\\cdashline{1-3}
\textbf{Test MD} & This data type is related to the result of test executions and is gathered by analyzing test logs. & Extracting metadata information such as the time and duration of the test, result, and code coverage of each test case, and selecting a portion of them for execution.    \\\cdashline{1-3}
\textbf{Commit MD} & This data type refers to all development changes to the system under test (SUT) except the text of the submitted codes. & Developing a model by using the number of commits, commit time, the branch of code, and the committer's experience as input data.    \\\cdashline{1-3}
\textbf{Build Logs} & These logs present historical information about the outcome of previous builds. & Training an ML model for validating future builds based on analyzing the text of build configurations and their outcomes.    \\\cdashline{1-3}
\textbf{Project MD} & This data type depicts a holistic view of the project and is usually used in combination with other data types for training ML models. & The size of the development team and the age of the project are examples of this data type.   \\\cdashline{1-3}
\textbf{Texts} &  This data type includes all texts except source codes, logs, and test codes and requires text processing techniques & Documents, user stories, reported issues, and commit messages are some examples of this data source.  \\\cdashline{1-3}
\textbf{System Logs} & System logs are files that show the behavior of the system in different situations and the impact of taking actions in the system. & Discovering the installed software on cloud systems by analyzing the tree of files and paths \\

\bottomrule
\end{tabular}
\end{table*}

\par TravisTorrent and Apache datasets are extensively employed in CI research. TravisTorrent, spanning 1,359 projects (402 Java, 898 Ruby, 59 in other languages), encompasses 2,640,825 builds. The Apache dataset comprises diverse projects like Cassandra, Ivy, Lang, Drill, and Math, featuring CI cycles ranging from 55 to 438. Researchers favor these datasets for their versatility, enabling evaluations across a broad spectrum of scenarios.

\par Table~\ref{Table:DataSourceCI} illustrates the correlation between frequently used data sources and identified CI tasks in selected studies. Apache datasets prove versatile and employed across various CI phases due to project diversity. Conversely, TravisTorrent is notably applied in Build Prediction, and Google Dagger in predicting branch coverage during the Integration Test phase.

\subsubsection{Data Types} This section categorizes and explains data types within CI pipelines. Nine categories, including Source Code, Code Meta-Data (MD), Test Case, Test MD, Commit MD, Build Logs, Project MD, Textual Data or Texts, and System Logs, are identified. Table~\ref{Table:DataTypes} provides a detailed overview of these data types, offering valuable insights for research purposes. Integrating innovative data types could unlock further exploration opportunities.

\par Here we aim to elucidate the correlation between data types and CI tasks, providing valuable insights into how different data types can enhance the efficiency of each task. This understanding informs the development and deployment of CI strategies, guiding the selection of data types for future research in automating CI tasks.

\begin{table*}
\centering
\caption{Relation between data types and CI tasks. color intensity in the table shows numerical values, with lighter shades indicating lower and darker blues representing higher values. The bolded numbers represent the count of studies and their percentage in total. \textbf{Acronyms:} \textit{\textbf{MD:}} Meta Data,  \textit{\textbf{UT:}} Unit Test, \textit{\textbf{IT:}} Integration Test, \textit{\textbf{RT:}} Regression Test, \textit{\textbf{BV:}} Build Validation, \textit{\textbf{ST:}} System Test and \textit{\textbf{PM:}} Process Management.}
\label{Table_old:DataCI}
\resizebox{\textwidth}{!}{%
\begin{tabular}{c|l|ccccccccc}
\hline
\textbf{CI Phases} &
\textbf{CI Tasks} &
  \textbf{Source Code} &
  \textbf{Code MD} &
  \textbf{Test Case} &
  \textbf{Test MD} &
  \textbf{Commit MD} &
  \multicolumn{1}{l}{\textbf{Build Logs}} &
  \multicolumn{1}{l}{\textbf{Project MD}} &
  \multicolumn{1}{l}{\textbf{Texts}} &
  \multicolumn{1}{l}{\textbf{System Logs}} \\ \hline
  \textbf{UT} &
\Centerstack[l]{\textbf{Unit Test}\\\textbf{Prediction}} &
  \cellcolor{D5E6F4}\Centerstack{S2 \\ \textbf{(1, 1.9\%)}} &
  \cellcolor{DDEBF7}-- &
  \cellcolor{DDEBF7}-- &
  \cellcolor{DDEBF7}-- &
  \cellcolor{DDEBF7}-- &
  \cellcolor{DDEBF7}-- &
  \cellcolor{DDEBF7}-- &
  \cellcolor{DDEBF7}-- &
  \cellcolor{DDEBF7}-- \\ \cdashline{1-11}
\multirow{2}{*}{\textbf{IT}} &
\Centerstack[l]{\textbf{Branch Coverage}\\\textbf{Prediction}} &
  \cellcolor{D5E6F4}\Centerstack{S15\\\textbf{(1, 1.9\%)}} &
  \cellcolor{CCE0F1}\Centerstack{S3, S15\\\textbf{(2, 3.8\%)}} &
  \cellcolor{DDEBF7}-- &
  \cellcolor{DDEBF7}-- &
  \cellcolor{DDEBF7}-- &
  \cellcolor{DDEBF7}-- &
  \cellcolor{DDEBF7}-- &
  \cellcolor{DDEBF7}-- &
  \cellcolor{DDEBF7}-- \\ \cdashline{2-11}
&
\Centerstack[l]{\textbf{Integration Test}\\\textbf{Prediction}} &
  \cellcolor{D5E6F4}\Centerstack{S18\\\textbf{(1, 1.9\%)}} &
  \cellcolor{D5E6F4}\Centerstack{S18\\\textbf{(1, 1.9\%)}} &
  \cellcolor{DDEBF7}-- &
  \cellcolor{D5E6F4}\Centerstack{S18\\\textbf{(1, 1.9\%)}} &
  \cellcolor{D5E6F4}\Centerstack{S23\\\textbf{(1, 1.9\%)}} &
  \cellcolor{DDEBF7}-- &
  \cellcolor{DDEBF7}-- &
  \cellcolor{D5E6F4}\Centerstack{S23\\\textbf{(1, 1.9\%)}} &
  \cellcolor{DDEBF7}-- \\ \cdashline{1-11}
\multirow{3}{*}{\textbf{RT}} &
\Centerstack[l]{\textbf{Test}\\\textbf{Optimization}} &
  \cellcolor{C3DAEE}\Centerstack{S8, S39, S47\\\textbf{(3, 5.8\%)}} &
  \cellcolor{C3DAEE}\Centerstack{S19, S36, S52\\\textbf{(3, 5.8\%)}} &
  \cellcolor{CCE0F1}\Centerstack{S36, S39\\\textbf{(2, 3.8\%)}} &
  \cellcolor{4A88C0}\Centerstack{S7, S8, S11, S16,\\ S19, S22, S26,\\ S28, S30, S31,\\ S33, S35, S37,\\ S39, S41, S45,\\ S46, S48, S50,\\ S52 \textbf{(20, 38.5\%)}} &
  \cellcolor{D5E6F4}\Centerstack{S28\\\textbf{(1, 1.9\%)}} &
  \cellcolor{D5E6F4}\Centerstack{S36\\\textbf{(1, 1.9\%)}} &
  \cellcolor{CCE0F1}\Centerstack{S30, S31\\\textbf{(2, 3.8\%)}} &
  \cellcolor{D5E6F4}\Centerstack{S45\\\textbf{(1, 1.9\%)}} &
  \cellcolor{DDEBF7}-- \\ \cdashline{2-11}
 &
\Centerstack[l]{\textbf{Defect}\\\textbf{Prediction}} &
  \cellcolor{D5E6F4}\Centerstack{S13\\\textbf{(1, 1.9\%)}} &
  \cellcolor{C3DAEE}\Centerstack{S9, S17, S32\\\textbf{(3, 5.8\%)}} &
  \cellcolor{DDEBF7}-- &
  \cellcolor{C3DAEE}\Centerstack{S13, S14, S32\\\textbf{(3, 5.8\%)}} &
  \cellcolor{D5E6F4}\Centerstack{S9\\\textbf{(1, 1.9\%)}} &
  \cellcolor{DDEBF7}-- &
  \cellcolor{DDEBF7}-- &
  \cellcolor{DDEBF7}-- &
  \cellcolor{DDEBF7}-- \\ \cdashline{2-11}
 &
\Centerstack[l]{\textbf{Flaky Test}\\\textbf{Detection}} &
  \cellcolor{DDEBF7}-- &
  \cellcolor{DDEBF7}-- &
  \cellcolor{D5E6F4}\Centerstack{S40\\\textbf{(1, 1.9\%)}} &
  \cellcolor{D5E6F4}\Centerstack{S40\\\textbf{(1, 1.9\%)}} &
  \cellcolor{DDEBF7}-- &
  \cellcolor{DDEBF7}-- &
  \cellcolor{DDEBF7}-- &
  \cellcolor{DDEBF7}-- &
  \cellcolor{DDEBF7}-- \\ \cdashline{1-11}
\textbf{BV} &
\Centerstack[l]{\textbf{Build}\\\textbf{Prediction}} &
  \cellcolor{D5E6F4}\Centerstack{S5\\\textbf{(1, 1.9\%)}} &
  \cellcolor{98BCDD}\Centerstack{S6, S21, S24,\\ S25, S38, S42,\\ S44, S51\\ \textbf{(8, 15.4\%)}} &
  \cellcolor{DDEBF7}-- &
  \cellcolor{BBD4EA}\Centerstack{S1, S5,\\S25, S38\\\textbf{(4, 7,7\%)}} &
  \cellcolor{98BCDD}\Centerstack{S1, S5, S6, S21,\\ S24, S25, S43,\\ S51 \textbf{(8, 15.4\%)}} &
  \cellcolor{BBD4EA}\Centerstack{S24, S25,\\ S27, S43\\\textbf{(4, 7.7\%)}} &
  \cellcolor{D5E6F4}\Centerstack{S21\\\textbf{(1, 1.9\%)}} &
  \cellcolor{DDEBF7}-- &
  \cellcolor{DDEBF7}-- \\ \cdashline{1-11}
\multirow{2}{*}{\textbf{ST}} &
\Centerstack[l]{\textbf{Installed Software}\\\textbf{Discovery}} &
  \cellcolor{DDEBF7}-- &
  \cellcolor{DDEBF7}-- &
  \cellcolor{DDEBF7}-- &
  \cellcolor{DDEBF7}-- &
  \cellcolor{DDEBF7}-- &
  \cellcolor{DDEBF7}-- &
  \cellcolor{DDEBF7}-- &
  \cellcolor{DDEBF7}-- &
  \cellcolor{D5E6F4}\Centerstack{S12\\\textbf{(1, 1.9\%)}} \\ \cdashline{2-11}
 &
\Centerstack[l]{\textbf{Performance Test}\\\textbf{Optimization}} &
  \cellcolor{DDEBF7}-- &
  \cellcolor{DDEBF7}-- &
  \cellcolor{DDEBF7}-- &
  \cellcolor{D5E6F4}\Centerstack{S4\\\textbf{(1, 1.9\%)}} &
  \cellcolor{DDEBF7}-- &
  \cellcolor{DDEBF7}-- &
  \cellcolor{DDEBF7}-- &
  \cellcolor{DDEBF7}-- &
  \cellcolor{DDEBF7}-- \\ \cdashline{1-11}
\textbf{PM} &
\Centerstack[l]{\textbf{Activity} \\\textbf{Management}} &
  \cellcolor{DDEBF7}-- &
  \cellcolor{DDEBF7}-- &
  \cellcolor{DDEBF7}-- &
  \cellcolor{DDEBF7}-- &
  \cellcolor{D5E6F4}\Centerstack{S10\\\textbf{(1, 1.9\%)}} &
  \cellcolor{DDEBF7}-- &
  \cellcolor{DDEBF7}-- &
  \cellcolor{CCE0F1}\Centerstack{S10, S49\\\textbf{(2, 3.8\%)}} &
  \cellcolor{DDEBF7}-- \\ \hline
\multicolumn{2}{l|}{
\textbf{Total usage in studies}} &
  \cellcolor{B3CFE8}\textbf{8, 15.4\%} &
  \cellcolor{7DAAD3}\textbf{17, 32.7\%} &
  \cellcolor{D1E3F3}\textbf{3, 5.8\%} &
  \cellcolor{4A88C0}\textbf{30, 57.7\%} &
  \cellcolor{9BBFDE}\textbf{12, 23.1\%} &
  \cellcolor{C5DBEE}\textbf{5, 9.6\%} &
  \cellcolor{D1E3F3}\textbf{3, 5.8\%} &
  \cellcolor{CBDFF1}\textbf{4, 7.7\%} &
  \cellcolor{D5E6F4}\textbf{1, 1.9\%} \\ \hline
\end{tabular}%
}

\end{table*}

\par Table~\ref{Table_old:DataCI} highlights the prevalent use of source code metadata over raw source code data in the selected studies. This tendency is likely attributed to the growing availability of code analysis tools like the CK tool\footnote{\underline{https://github.com/mauricioaniche/ck}}, which automates the computation of established metrics such as Chidamber and Kemerer (CK). These tools have significantly streamlined the extraction of source code metadata for researchers. Notably, S3, S9, and S15 explicitly mentioned utilizing these tools for extracting CK metrics.

\par S15, S40, and S44 utilized Halstead metrics with available tools to delve into source code attributes, enriching the development of ML-based CI solutions. Extracting source code metadata from source code is cost-effective and straightforward, involving metrics like comments, methods, lines of code, and parameters. These metrics are easily obtainable and do not demand extensive computational resources. Moreover, these data types are well-suited for ML models as they can be readily converted into numerical values. 
\par Table~\ref{Table_old:DataCI} reveals the prevalent use of source code metadata in studies related to Build Prediction, Test Optimization, and Defect Prediction in Regression Testing, as well as in Integration Testing within the CI pipeline. However, it is notably absent in the context of Unit Test Prediction, signifying a research gap in this area.

\par Table~\ref{Table_old:DataCI} underscores the limited utilization of data types such as test case codes, build logs, project metadata, textual data, and system logs in automating CI tasks. This underscores the potential benefits of exploring combinations of various data types for each CI task. For example, combining system logs with code metadata could enhance the accuracy of the Build Prediction task in CI. Furthermore, it is noteworthy that test metadata and commit metadata are predominantly employed in the Build Prediction task, indicating areas for further research in ML-based CI approaches.

\subsubsection{Data Preparation} In this section of our literature we mainly focus on data-related techniques used to modify and prepare raw data for ML models. Data cleaning, a fundamental aspect, was not explicitly examined separately due to limited information in the studies. However, data cleaning can be considered a filtering technique within the data preparation process.

\par Table~\ref{Table:DataPreparation} shows that the selected studies utilized ten distinct data preparation techniques. The choice of these techniques and the data's characteristics depends on the overall strategy for addressing the research problem. The table summarizes the employed techniques and their subgroups. The next section provides a comprehensive description and examples for each data preparation group and subgroup.

\begin{table*}
\caption{Correlation between the most commonly used datasets and data preparation methods. color intensity in the table shows numerical values, with lighter shades indicating lower and darker blues representing higher values. The bolded numbers represent the count of studies and their percentage in total.}

\label{Table:DataPreparation}
\resizebox{\textwidth}{!}{%
\begin{tabular}{l|ccc|cc|ccc|cc}
\hline
  \multirow{2}{*}{\textbf{Dataset Names}}
  &
  \multicolumn{3}{c|}{\textbf{Conditioning}} &
  \multicolumn{2}{c|}{\textbf{Building}} &
  \multicolumn{3}{c|}{\textbf{Balancing}} &
  \multicolumn{2}{c}{\textbf{Filtering}} \\ \cline{2-11}
  &
  \multicolumn{1}{l}{\Centerstack{\textbf{Manual Data}\\\textbf{Division}}} &
  \multicolumn{1}{l}{\textbf{Clustering}} &
  \multicolumn{1}{l|}{\Centerstack{\textbf{Objective Data}\\\textbf{Selection}}} &
  \multicolumn{1}{l}{\Centerstack{\textbf{Data}\\\textbf{Augmentation}}} &
  \multicolumn{1}{l|}{\Centerstack{\textbf{Data}\\\textbf{Manipulation}}} &
  \multicolumn{1}{l}{\textbf{Re-sampling}} &
  \multicolumn{1}{l}{\textbf{Oversampling}} &
  \multicolumn{1}{l|}{\textbf{Undersampling}} &
  \multicolumn{1}{l}{\Centerstack{\textbf{Data}\\\textbf{Pruning}}} &
  \multicolumn{1}{l}{\Centerstack{\textbf{Selective Data}\\\textbf{Filtering}}} \\ \hline
\textbf{IOF/ROL} &
  \cellcolor{DDEBF7}-- &
  \cellcolor{DDEBF7}-- &
  \cellcolor{DDEBF7}-- &
  \cellcolor{C8DDEF}\Centerstack{S35\\\textbf{(1, 1.9\%)}} &
  \cellcolor{DDEBF7}-- &
  \cellcolor{C8DDEF}\Centerstack{S35\\\textbf{(1, 1.9\%)}} &
  \cellcolor{B2CEE7}\Centerstack{S28, S37\\\textbf{(2, 3.8\%)}} &
  \cellcolor{C8DDEF}\Centerstack{S37\\\textbf{(1, 1.9\%)}}  &
  \cellcolor{B2CEE7}\Centerstack{S11, S14\\\textbf{(2, 3.8\%)}} &
  \cellcolor{C8DDEF}\Centerstack{S20\\\textbf{(1, 1.9\%)}} \\
\textbf{Paint Control} &
  \cellcolor{DDEBF7}-- &
  \cellcolor{DDEBF7}-- &
  \cellcolor{DDEBF7}-- &
  \cellcolor{C8DDEF}\Centerstack{S35\\\textbf{(1, 1.9\%)}} &
  \cellcolor{DDEBF7}-- &
  \cellcolor{C8DDEF}\Centerstack{S35\\\textbf{(1, 1.9\%)}} &
  \cellcolor{C8DDEF}\Centerstack{S37\\\textbf{(1, 1.9\%)}} &
  \cellcolor{C8DDEF}\Centerstack{S37\\\textbf{(1, 1.9\%)}} &
  \cellcolor{B2CEE7}\Centerstack{S11, S14\\\textbf{(2, 3.8\%)}} &
  \cellcolor{C8DDEF}\Centerstack{S20\\\textbf{(1, 1.9\%)}} \\
\textbf{Apache Commons} &
  \cellcolor{DDEBF7}-- &
  \cellcolor{DDEBF7}-- &
  \cellcolor{DDEBF7}-- &
  \cellcolor{B2CEE7}\Centerstack{S15, S35\\\textbf{(2, 3.8\%)}} &
  \cellcolor{C8DDEF}\Centerstack{S10\\\textbf{(1, 1.9\%)}} &
  \cellcolor{C8DDEF}\Centerstack{S35\\\textbf{(1, 1.9\%)}} &
  \cellcolor{DDEBF7}-- &
  \cellcolor{DDEBF7}-- &
  \cellcolor{DDEBF7}-- &
  \cellcolor{C8DDEF}\Centerstack{S20\\\textbf{(1, 1.9\%)}} \\
\textbf{GSDTSR} &
  \cellcolor{DDEBF7}-- &
  \cellcolor{DDEBF7}-- &
  \cellcolor{C8DDEF}\Centerstack{S16\\\textbf{(1, 1.9\%)}} &
  \cellcolor{C8DDEF}\Centerstack{S35\\\textbf{(1, 1.9\%)}} &
  \cellcolor{DDEBF7}-- &
  \cellcolor{C8DDEF}\Centerstack{S35\\\textbf{(1, 1.9\%)}} &
  \cellcolor{C8DDEF}\Centerstack{S37\\\textbf{(1, 1.9\%)}} &
  \cellcolor{C8DDEF}\Centerstack{S37\\\textbf{(1, 1.9\%)}} &
  \cellcolor{C8DDEF}\Centerstack{S14\\\textbf{(1, 1.9\%)}} &
  \cellcolor{DDEBF7}-- \\
\textbf{Google Guava} &
  \cellcolor{DDEBF7}-- &
  \cellcolor{DDEBF7}-- &
  \cellcolor{DDEBF7}-- &
  \cellcolor{C8DDEF}\Centerstack{S15\\\textbf{(1, 1.9\%)}} &
  \cellcolor{DDEBF7}-- &
  \cellcolor{DDEBF7}-- &
  \cellcolor{C8DDEF}\Centerstack{S28\\\textbf{(1, 1.9\%)}} &
  \cellcolor{DDEBF7}-- &
  \cellcolor{DDEBF7}-- &
  \cellcolor{C8DDEF}\Centerstack{S20\\\textbf{(1, 1.9\%)}} \\
\textbf{Rails} &
  \cellcolor{DDEBF7}-- &
  \cellcolor{DDEBF7}-- &
  \cellcolor{DDEBF7}-- &
  \cellcolor{C8DDEF}\Centerstack{S35\\\textbf{(1, 1.9\%)}} &
  \cellcolor{DDEBF7}-- &
  \cellcolor{C8DDEF}\Centerstack{S35\\\textbf{(1, 1.9\%)}} &
  \cellcolor{DDEBF7}-- &
  \cellcolor{DDEBF7}-- &
  \cellcolor{DDEBF7}-- &
  \cellcolor{C8DDEF}\Centerstack{S48\\\textbf{(1, 1.9\%)}} \\
\textbf{TravisTorrent} &
  \cellcolor{B2CEE7}\Centerstack{S5, S21\\\textbf{(2, 3.8\%)}} &
  \cellcolor{DDEBF7}-- &
  \cellcolor{DDEBF7}-- &
  \cellcolor{C8DDEF}\Centerstack{S29\\\textbf{(1, 1.9\%)}} &
  \cellcolor{DDEBF7}-- &
  \cellcolor{DDEBF7}-- &
  \cellcolor{DDEBF7}-- &
  \cellcolor{DDEBF7}-- &
  \cellcolor{DDEBF7}-- &
  \cellcolor{C8DDEF}\Centerstack{S6\\\textbf{(1, 1.9\%)}} \\
\textbf{MyBatis} &
  \cellcolor{DDEBF7}-- &
  \cellcolor{DDEBF7}-- &
  \cellcolor{DDEBF7}-- &
  \cellcolor{DDEBF7}-- &
  \cellcolor{DDEBF7}-- &
  \cellcolor{DDEBF7}-- &
  \cellcolor{DDEBF7}-- &
  \cellcolor{DDEBF7}-- &
  \cellcolor{C8DDEF}\Centerstack{S32\\\textbf{(1, 1.9\%)}} &
  \cellcolor{C8DDEF}\Centerstack{S20\\\textbf{(1, 1.9\%)}} \\
\textbf{Google Closure} &
  \cellcolor{DDEBF7}-- &
  \cellcolor{DDEBF7}-- &
  \cellcolor{DDEBF7}-- &
  \cellcolor{DDEBF7}-- &
  \cellcolor{DDEBF7}-- &
  \cellcolor{DDEBF7}-- &
  \cellcolor{DDEBF7}-- &
  \cellcolor{DDEBF7}-- &
  \cellcolor{DDEBF7}-- &
  \cellcolor{C8DDEF}\Centerstack{S20\\\textbf{(1, 1.9\%)}} \\
\textbf{Google Auto} &
  \cellcolor{DDEBF7}-- &
  \cellcolor{DDEBF7}-- &
  \cellcolor{DDEBF7}-- &
  \cellcolor{DDEBF7}-- &
  \cellcolor{DDEBF7}-- &
  \cellcolor{DDEBF7}-- &
  \cellcolor{DDEBF7}-- &
  \cellcolor{DDEBF7}-- &
  \cellcolor{DDEBF7}-- &
  \cellcolor{C8DDEF}\Centerstack{S20\\\textbf{(1, 1.9\%)}} \\
\textbf{Dspace} &
  \cellcolor{DDEBF7}-- &
  \cellcolor{DDEBF7}-- &
  \cellcolor{DDEBF7}-- &
  \cellcolor{DDEBF7}-- &
  \cellcolor{DDEBF7}-- &
  \cellcolor{DDEBF7}-- &
  \cellcolor{DDEBF7}-- &
  \cellcolor{DDEBF7}-- &
  \cellcolor{DDEBF7}-- &
  \cellcolor{C8DDEF}\Centerstack{S20\\\textbf{(1, 1.9\%)}} \\
\textbf{Google Dagger} &
  \cellcolor{DDEBF7}-- &
  \cellcolor{DDEBF7}-- &
  \cellcolor{DDEBF7}-- &
  \cellcolor{C8DDEF}\Centerstack{S15\\\textbf{(1, 1.9\%)}} &
  \cellcolor{DDEBF7}-- &
  \cellcolor{DDEBF7}-- &
  \cellcolor{DDEBF7}-- &
  \cellcolor{DDEBF7}-- &
  \cellcolor{DDEBF7}-- &
  \cellcolor{DDEBF7}-- \\
\textbf{Other} &
  \cellcolor{C8DDEF}\Centerstack{S42\\\textbf{(1, 1.9\%)}} &
  \cellcolor{9CBFDF}\Centerstack{S1, S24, S45\\\textbf{(3, 5.8\%)}} &
  \cellcolor{B2CEE7}\Centerstack{S2, S40\\\textbf{(2, 3.8\%)}} &
  \cellcolor{5690C4}\Centerstack{S8, S12, S13,\\S26, S36, S49,\\S43, S46\\\textbf{(8, 15.4\%)}} &
  \cellcolor{C8DDEF}\Centerstack{S23\\\textbf{(1, 1.9\%)}} &
  \cellcolor{DDEBF7}-- &
  \cellcolor{B2CEE7}\Centerstack{S38, S44\\\textbf{(2, 3.8\%)}} &
  \cellcolor{C8DDEF}\Centerstack{S30\\\textbf{(1, 1.9\%)}} &
  \cellcolor{86B0D6}\Centerstack{S9, S34,\\S51, S49\\\textbf{(4, 7.7\%)}} &
  \cellcolor{B2CEE7}\Centerstack{S24, S25\\\textbf{(2, 3.8\%)}} \\ \hline
 \textbf{Num of studies} & 
  \cellcolor{B5D0E8}\textbf{3, 5.8\%} &
  \cellcolor{B5D0E8}\textbf{3, 5.8\%} &
  \cellcolor{B5D0E8}\textbf{3, 5.8\%} &
  \cellcolor{4A88C0}\textbf{11, 21.2\%} &
  \cellcolor{C3D9ED}\textbf{2, 3.8\%} &
  \cellcolor{D0E2F2}\textbf{1, 1.9\%} &
  \cellcolor{A8C7E3}\textbf{4, 7.7\%} &
  \cellcolor{C3D9ED}\textbf{2, 3.8\%} &
  \cellcolor{80ACD4}\textbf{7, 13.5\%} &
  \cellcolor{9BBEDE}\textbf{5, 9.6\%} \\
  
  \hline
\end{tabular}%
}

\end{table*}

\par \textbf{DP1) Conditioning:} This technique adjusts the dataset according to its characteristics and the proposed solution. In the case of \underline{manual data division}, as demonstrated in study S5, the TravisTorrent dataset was partitioned using strategies such as the Burak filter~\cite{xia2017empirical}and the Bellwether filter~\cite{xia2017empirical}. In another instance (S21), data was segregated based on previous build results (pass or fail), resulting in separate training of machine learning models for each data partition.

\par Moreover, \underline{clustering} methods like k-means can divide data based on feature values or data distribution, as in studies S24 and S45. \underline{Objective data selection} involves choosing specific data points or segments based on defined objectives or criteria of our solution, such as selecting lines before and after the changed codes (changed codes are the objective of our solution) for inclusion in the dataset.
\par Utilizing conditioning for data preparation presents a clear advantage in reducing the computational overhead during ML model training, especially in CI environments with significant data volumes~\cite{ali2019enhanced}.

\par \textbf{DP2) Building:} This methodology transforms data to enhance its structure for better compatibility with ML techniques, especially in diverse CI environments. It consists of two sub-groups.

\par The first sub-group, \underline{data augmentation}, is the most commonly used method. Researchers employ it to modify data structures or simulate real-world conditions. For instance, in S13, input data is padded with zeros for consistency and easier processing. Techniques like creating graphs or trees based on regression test results (S29, S36, S46) or simulating cloud-based real environments (S12) fall into this sub-group.

\par The second sub-group, \underline{data manipulation}, involves methods such as text processing (S23) and merging distinct parts of a dataset by identifying correlations, such as combining issue tracking and VCS (S10).

\par \textbf{DP3) Balancing:} Many commonly used data sources display imbalances, which can substantially impact the performance and accuracy of classifiers~\cite{kaur2019systematic}. Addressing this issue involves employing \underline{re-sampling} techniques or training separate models for each class of input data. Re-sampling encompasses both \underline{oversampling}, which involves increasing instances in the minority class (e.g., SMOTE), and \underline{under-sampling}, as demonstrated in S30, which reduces the number of instances in the majority class. Additionally, combining both methods, as exemplified in SMOGN in S37~\cite{sharif2021deeporder}, presents another viable approach.

\par \textbf{DP4) Filtering:} \underline{Data pruning}, a form of data cleaning, is vital in CI, addressing high computational workloads in large-scale datasets and improving the performance of the ML models~\cite{xiao2020lstm}. Seven studies used this method, e.g., S11 removed unexecuted tests, S51 eliminated errored builds, and S9 removed incomplete test data. Issues like these arise from evolving software practices~\cite{wu2019time} and human errors~\cite{xiao2020lstm}. Filtering is crucial in build validation studies, tackling interrupted builds~\cite{martins2021supervised}. \underline{Selective data filtering} can enhance the quality of the training model and reduce training costs by prioritizing data segments that closely resemble the practical environment where the trained ML model will be deployed. For example, in study S24, TravisTorrent was constrained to Java-based projects employing Ant, Maven, and Gradle CI build tools, while Ruby-based projects were excluded.


\vspace{1pt}
\noindent\fcolorbox{black}{blue!10}{%
\begin{minipage}{\columnwidth}
\vspace{1pt}
\textbf{Summary:}
\vspace{1pt}

\par $\bullet$ 12 datasets, nine data types, and ten processing techniques are identified in the reviewed studies.
\vspace{1pt}


\par $\bullet$ Clustering datasets can reduce computation overhead and enhance ML model accuracy due to the high volume of CI data.
\vspace{1pt}


\par $\bullet$ Addressing class imbalance is crucial to improve ML model accuracy in CI data.
\vspace{1pt}
\par $\bullet$ Filtering, addressing exceptional cases like cancelled or errored processes, enhances ML model performance and reduces computational overhead.
\end{minipage}}



\subsection{RQ3: Extracted Features}
\label{Sec:ExtFeatures}
\par This section delves into the feature types and feature engineering techniques employed in the reviewed studies. Given that ML models are data-driven, the choice of feature types and engineering methods directly impacts the performance of these models~\cite{schwabacher2005survey}. Through thematic analysis, we have categorized Features (\textbf{FT}) into four primary groups and nine sub-categories, presented as follows:

\textbf{Relational:} Individual and Spatial,
\textbf{Statistical:} Numerical, Components and Context,
\textbf{Lexical:} Content and Syntactical, and
\textbf{Epochal:} Temporal and Narrative.

\par In the following, the details regarding these features are presented. A brief explanation and examples for these features and the relationship between feature types and CI tasks are presented in Table~\ref{Table_old:features} and~\ref{Table:Tasks_FeatureTypes} respectively. Note that five studies lacked comprehensive feature descriptions. Additionally, feature combinations that were not employed in any of the selected studies have been excluded from Table~\ref{Table:Tasks_FeatureTypes}. These omitted combinations encompass the use of Relational features, the utilization of both Relational and Statistical features, the combination of both Relational and Lexical features, as well as the usage of Statistical and Lexical features.

\begin{table*}
\centering
\caption{Employed features in studies and description of them}
\label{Table_old:features}

\begin{tabular}{p{0.1\textwidth}|p{0.1\textwidth}|p{0.45\textwidth}|p{0.25\textwidth}}
\toprule
\multicolumn{1}{c}{\textbf{Feature types}} & \multicolumn{1}{c}{\textbf{Sub groups}} & \multicolumn{1}{c}{\textbf{Description}} & \multicolumn{1}{c}{\textbf{Example}} \\ \toprule
\multirow{2}{*}{\textbf{Relational}} 
 & \textit{Individual} & These features focus on the behaviour and experiences of individuals. 
 & \textbf{(S1)} Percentage of ownership of a developer on a file
 \\\cdashline{2-4}
 & \textit{Spatial} & These features encompass the communication pathways and interdependencies among distinct components within the space of software system.
 & \textbf{(S3)} Source code coverage of a testing code
 \\ \toprule
\multirow{3}{*}{\textbf{Statistical}} & \textit{Components} & These features encapsulate statistical data pertaining to various aspects of software components, offering insights into their quantitative characteristics.
 & \textbf{(S3)} Depth of inheritance tree (DIT)
 \\\cdashline{2-4}
 & \textit{Context} & These features represent the statistical information within a code including the minimum, maximum, and average of operations or the complexity of components.
 & \textbf{(S44)} Number of operands and operators in the committed code
 \\\cdashline{2-4}
 & \textit{Numerical} & These features are based on calculating straightforward metrics not included in the Components and Context feature types, and they are independent of content.
& \textbf{(S26)} Number of commits on a file
 \\ \toprule
\multirow{2}{*}{\textbf{Lexical}} & \textit{Content} & These features denote the information about the terms and tags in texts including source code, log files, and other text files and reports.
& \textbf{(S8)} Determining text similarities in Java codes via TF-IDF
\\\cdashline{2-4}
 & \textit{Syntactical} & These features only focus on information about the specific programming language reserved words. 
 & \textbf{(S15)} Number of each Java reserved word in a source code
 \\ \toprule
\multirow{2}{*}{\textbf{Epochal}} & \textit{Temporal} & These features represent the time-dependent attributes of any software components from last changes until the present time.
& \textbf{(S21)} The time gap since the last build
\\\cdashline{2-4}
 & \textit{Narrative} & These features represent the historical actions taken and their outcomes leading up to a particular event within the software product. 
 & \textbf{(S36)} The durations of the previous executions of a test case 
 \\
 \bottomrule
\end{tabular}

\end{table*}

\par \textbf{FT1) Relational:} This category encompasses features that depict relationships among elements, such as actions, individuals, and components, influencing the outcomes of CI tasks. It consists of two sub-categories:

\par \underline{\emph{Individual:}} These features focus on individual-related factors, such as developers and test designers, including their experience in software development and file ownership percentages~\cite{philip2019fastlane}. Although pivotal in predicting build outcomes (S43) and testing results of code commits (S23), only four studies employed these features.

\par \underline{\emph{Spatial:}} Spatial features relate to code concepts like coverage, coupling, inheritance, and cohesion, extractable using tools like IntelliJ Idea~\cite{vig2018test}, Rational Software Analyzer (RSA)~\cite{finlay2014data}, and Aniche~\cite{grano2018high} with low computational overhead. Notably, three studies utilized the Chidamber and Kemerer (CK) indices~\cite{chidamber1994metrics}. This feature type found diverse applications, including predicting branch coverage (S3, S15), identifying code defects (S9), detecting flaky tests (S40), predicting build outcomes (S44), and optimizing test cases (S36, S39, S45, S46).

\par \textbf{FT2) Statistical:}
\par These features, are computed and analyzed for statistical data. In a broader context, they serve to show the complexity and scope of projects or committed changes. Table~\ref{Table:Tasks_FeatureTypes} highlights the usage of these features in the Build Prediction task, primarily due to their lower computational overhead, especially in the case of the Numerical feature type. Notably, build prediction tasks due to its full coverage of all tests involve huge data from numerous CI cycles. So, making minimizing computational resources is a serious concern in this task.

\par \underline{\emph{Components:}} Statistical insights into software product composition, such as the number of concrete and abstract classes, functions, and software package dependencies, are provided by Components features. They are useful in predicting branch coverage (S3, S15), defect prediction (S32), build outcomes (S42), flaky test detection (S40), and test optimization based on mutation testing (S29).

\par \underline{\emph{Context:}} This category yields statistical properties offering information about the complexity of software and testing source code. It computes attributes from the source code, including Osmax and Osavg (maximum and average operation size), WMC (weighted method complexity), NOAC (number of operations added), Ocmax and Ocavg (maximum and average operation complexity), Opavg (average operation parameters), CSO (class size operations), CSOA (class size operations attributes), CSA (class size attributes), Query (number of queries), NAAC (number of attributes added), NOIC (number of operations inherited), NOOC (number of operations overridden). It also incorporates the well-known Halstead metrics designed to show program complexity by examining its operators and operands (N, E, V, D, B, n)~\cite{halstead1977elements}. These features are prevalently used in predicting build outcomes (S27, S42, S43, S44), test branch coverage (S15), and detecting flaky tests (S40).

\par \underline{\emph{Numerical:}} Known for computational simplicity, Numerical features are utilized in 23 out of 52 studies for tasks like test optimization (S22, S26, S29, S30, S36, S39), build prediction (S1, S5, S6, S21, S24, S25, S42, S43, S44), test outcome prediction (S23), code defect prediction (S9, S17, S32), CI data management (S10), and flaky test detection (S40). These features encompass various attributes like counts of code line changes, files, sub-systems, classes, methods, active authors on the same file, commits, pull requests in VCS, and comments in source codes and VCS.

\begin{table*}
\caption{Correlation between CI tasks and feature types. color intensity in the table shows numerical values, with lighter shades indicating lower and darker blues representing higher values. The bolded numbers represent the count of studies and their percentage in total. The \faCircle~symbol represents which feature types have been used, while the \faCircleO~symbol represents which feature types are not used. \textbf{Acronyms:} \textit{\textbf{UT:}} Unit Test, \textit{\textbf{IT:}} Integration Test, \textit{\textbf{RT:}} Regression Test, \textit{\textbf{BV:}} Build Validation, \textit{\textbf{ST:}} System Test and \textit{\textbf{PM:}} Process Management. }
\label{Table:Tasks_FeatureTypes}
\resizebox{\textwidth}{!}{%
\begin{tabular}{l|l|l|l|cccccccccc|c}
\toprule
  \multicolumn{4}{c|}{\textbf{Feature Types }} &
\multicolumn{1}{c|}{\textbf{UT}}&
\multicolumn{2}{c|}{\textbf{IT}}&
\multicolumn{3}{c|}{\textbf{RT}}&
\multicolumn{1}{c|}{\textbf{BV}}&
\multicolumn{2}{c|}{\textbf{ST}}&
\multicolumn{1}{c|}{\textbf{PM}} & \\ \hline

    \rotN{\Centerstack[l]{Relational}} &
  \rotN{\Centerstack[l]{Statistical}} &
  \rotN{\Centerstack[l]{Lexical}} &
  \rotN{\Centerstack[l]{Epochal}} &
\rotN{\Centerstack[l]{\textbf{Unit Test}\\\textbf{Prediction}}} &
  \rotN{\Centerstack[l]{\textbf{Branch Coverage}\\\textbf{Prediction}}} &
  \rotN{\Centerstack[l]{\textbf{Integration Test}\\\textbf{Prediction}}} &
  \rotN{\textbf{Test Optimization}} &
  \rotN{\textbf{Defect Prediction}} &
  \rotN{\Centerstack[l]{\textbf{Flaky Test}\\\textbf{Detection}}} &
  \rotN{\textbf{Build Prediction}} &
  \rotN{\Centerstack[l]{\textbf{Installed Software}\\\textbf{Discovery}}} &
  \rotN{\Centerstack[l]{\textbf{Performance Test}\\\textbf{Optimization}}} &
  \rotN{\Centerstack[l]{\textbf{Activity}\\\textbf{Management}}}&
  \rotN{\Centerstack[l]{\textbf{Number of}\\\textbf{studies}}} \\ \hline

\faCircleO & \faCircle & \faCircleO & \faCircleO &
  \cellcolor{DDEBF7}-- &
  \cellcolor{DDEBF7}-- &
  \cellcolor{DDEBF7}-- &
  \cellcolor{CEE1F1}\Centerstack{S29 \textbf{(1, 1.9\%)}} &
  \cellcolor{DDEBF7}-- &
  \cellcolor{DDEBF7}-- &
  \cellcolor{9EC1DF}\Centerstack{S5, S25, S42,\\S44 \textbf{(4, 7.7\%)}} &
  \cellcolor{DDEBF7}-- &
  \cellcolor{DDEBF7}-- &
  \cellcolor{CEE1F1}\Centerstack{S10\\\textbf{(1, 1.9\%)}} &
  \cellcolor{95BADC}\Centerstack{\textbf{6}\\\textbf{11.5\%}} \\ \cdashline{1-15}
\faCircleO & \faCircleO & \faCircle & \faCircleO &
  \cellcolor{CEE1F1}\Centerstack{S2\\\textbf{(1, 1.9\%)}} &
  \cellcolor{DDEBF7}-- &
  \cellcolor{CEE1F1}\Centerstack{S18\\\textbf{(1, 1.9\%)}} &
  \cellcolor{CEE1F1}\Centerstack{S47\\\textbf{(1, 1.9\%)}} &
  \cellcolor{CEE1F1}\Centerstack{S13\\\textbf{(1, 1.9\%)}} &
  \cellcolor{DDEBF7}-- &
  \cellcolor{DDEBF7}-- &
  \cellcolor{CEE1F1}\Centerstack{S12\\\textbf{(1, 1.9\%)}} &
  \cellcolor{DDEBF7}-- &
  \cellcolor{CEE1F1}\Centerstack{S49\\\textbf{(1, 1.9\%)}} &
  \cellcolor{95BADC}\Centerstack{\textbf{6}\\\textbf{11.5\%}} \\ \cdashline{1-15}
\faCircleO & \faCircleO & \faCircleO & \faCircle &
  \cellcolor{DDEBF7}-- &
  \cellcolor{DDEBF7}-- &
  \cellcolor{DDEBF7}-- &
  \cellcolor{4A88C0}\Centerstack{S7, S11, S14, S16,\\S19, S28, S31, S33,\\S35, S37, S48, S50\\\textbf{(12, 23.1\%)}} &
  \cellcolor{DDEBF7}-- &
  \cellcolor{DDEBF7}-- &
  \cellcolor{DDEBF7}-- &
  \cellcolor{CEE1F1}\Centerstack{S4\\\textbf{(1, 1.9\%)}} &
  \cellcolor{DDEBF7}-- &
  \cellcolor{DDEBF7}-- &
  \cellcolor{4785BE}\Centerstack{\textbf{13}\\\textbf{25\%}} \\ \cdashline{1-15}
\faCircle & \faCircleO & \faCircleO & \faCircle &
  \cellcolor{DDEBF7}-- &
  \cellcolor{DDEBF7}-- &
  \cellcolor{DDEBF7}-- &
  \cellcolor{BED6EB}\Centerstack{S45, S46\\\textbf{(1, 3.8\%)}} &
  \cellcolor{BED6EB}\Centerstack{S17, S32\\\textbf{(1, 3.8\%)}} &
  \cellcolor{DDEBF7}-- &
  \cellcolor{DDEBF7}-- &
  \cellcolor{DDEBF7}-- &
  \cellcolor{DDEBF7}-- &
  \cellcolor{DDEBF7}-- &
  \cellcolor{B2CEE7}\Centerstack{\textbf{4}\\\textbf{7.7\%}} \\ \cdashline{1-15}
\faCircleO & \faCircle & \faCircleO & \faCircle &
  \cellcolor{DDEBF7}-- &
  \cellcolor{DDEBF7}-- &
  \cellcolor{DDEBF7}-- &
  \cellcolor{BED6EB}\Centerstack{S22, S52\\\textbf{(1, 3.8\%)}} &
  \cellcolor{DDEBF7}-- &
  \cellcolor{DDEBF7}-- &
  \cellcolor{9EC1DF}\Centerstack{S6, S21, S24,\\S27 \textbf{(4, 7.7\%)}} &
  \cellcolor{DDEBF7}-- &
  \cellcolor{DDEBF7}-- &
  \cellcolor{DDEBF7}-- &
  \cellcolor{95BADC}\Centerstack{\textbf{6}\\\textbf{11.5\%}} \\ \cdashline{1-15}
\faCircleO & \faCircleO & \faCircle & \faCircle &
  \cellcolor{DDEBF7}-- &
  \cellcolor{DDEBF7}-- &
  \cellcolor{DDEBF7}-- &
  \cellcolor{CEE1F1}\Centerstack{S8\\\textbf{(1, 1.9\%)}} &
  \cellcolor{DDEBF7}-- &
  \cellcolor{DDEBF7}-- &
  \cellcolor{DDEBF7}-- &
  \cellcolor{DDEBF7}-- &
  \cellcolor{DDEBF7}-- &
  \cellcolor{DDEBF7}-- &
  \cellcolor{DDEBF7}\Centerstack{\textbf{1}\\\textbf{1.9\%}} \\ \cdashline{1-15}
 \faCircle & \faCircle & \faCircle & \faCircleO &
  \cellcolor{DDEBF7}-- &
  \cellcolor{CEE1F1}\Centerstack{S15\\\textbf{(1, 1.9\%)}} &
  \cellcolor{DDEBF7}-- &
  \cellcolor{DDEBF7}-- &
  \cellcolor{DDEBF7}-- &
  \cellcolor{DDEBF7}-- &
  \cellcolor{DDEBF7}-- &
  \cellcolor{DDEBF7}-- &
  \cellcolor{DDEBF7}-- &
  \cellcolor{DDEBF7}-- &
  \cellcolor{DDEBF7}\Centerstack{\textbf{1}\\\textbf{1.9\%}} \\ \cdashline{1-15}
 \faCircle & \faCircle & \faCircleO & \faCircle &
  \cellcolor{DDEBF7}-- &
  \cellcolor{DDEBF7}-- &
  \cellcolor{DDEBF7}-- &
  \cellcolor{BED6EB}\Centerstack{S36, S39\\\textbf{(1, 3.8\%)}} &
  \cellcolor{CEE1F1}\Centerstack{S9\\\textbf{(1, 1.9\%)}} &
  \cellcolor{CEE1F1}\Centerstack{S40\\\textbf{(1, 1.9\%)}} &
  \cellcolor{DDEBF7}-- &
  \cellcolor{DDEBF7}-- &
  \cellcolor{DDEBF7}-- &
  \cellcolor{DDEBF7}-- &
  \cellcolor{B2CEE7}\Centerstack{\textbf{4}\\\textbf{7.7\%}} \\ \cdashline{1-15}
 \faCircle & \faCircleO & \faCircle & \faCircle &
  \cellcolor{DDEBF7}-- &
  \cellcolor{DDEBF7}-- &
  \cellcolor{DDEBF7}-- &
  \cellcolor{DDEBF7}-- &
  \cellcolor{DDEBF7}-- &
  \cellcolor{DDEBF7}-- &
  \cellcolor{BED6EB}\Centerstack{S1, S43\\\textbf{(1, 3.8\%)}} &
  \cellcolor{DDEBF7}-- &
  \cellcolor{DDEBF7}-- &
  \cellcolor{DDEBF7}-- &
  \cellcolor{CFE2F2}\Centerstack{\textbf{2}\\\textbf{3.8\%}}\\ \cdashline{1-15}
\faCircleO & \faCircle & \faCircle & \faCircle &
  \cellcolor{DDEBF7}-- &
  \cellcolor{DDEBF7}-- &
  \cellcolor{DDEBF7}-- &
  \cellcolor{CEE1F1}\Centerstack{S26\\\textbf{(1, 1.9\%)}} &
  \cellcolor{DDEBF7}-- &
  \cellcolor{DDEBF7}-- &
  \cellcolor{DDEBF7}-- &
  \cellcolor{DDEBF7}-- &
  \cellcolor{DDEBF7}-- &
  \cellcolor{DDEBF7}-- &
  \cellcolor{DDEBF7}\Centerstack{\textbf{1}\\\textbf{1.9\%}} \\ \cdashline{1-15}
 \faCircle & \faCircle & \faCircle & \faCircle &
  \cellcolor{DDEBF7}-- &
  \cellcolor{CEE1F1}\Centerstack{S3\\\textbf{(1, 1.9\%)}} &
  \cellcolor{CEE1F1}\Centerstack{S23\\\textbf{(1, 1.9\%)}} &
  \cellcolor{CEE1F1}\Centerstack{S30\\\textbf{(1, 1.9\%)}} &
  \cellcolor{DDEBF7}-- &
  \cellcolor{DDEBF7}-- &
  \cellcolor{DDEBF7}-- &
  \cellcolor{DDEBF7}-- &
  \cellcolor{DDEBF7}-- &
  \cellcolor{DDEBF7}-- &
  \cellcolor{C0D8EC}\Centerstack{\textbf{3}\\\textbf{5.8\%}} \\ \hline
\end{tabular}%
}
\end{table*}

\textbf{FT3) Lexical:} Software project source code and log files often contain valuable hidden information, extractable through lexical analysis, which involves two distinct approaches:

\underline{\emph{Content:}} Content-based feature extraction methods are employed in 10 out of 52 studies. Their popularity stems from direct data extraction without computationally intensive processes. This feature is versatile as it is related to text analysis, making it applicable to all programming languages through various textual analysis techniques. These techniques include TF-IDF (Term Frequency–Inverse Document Frequency) (S8, S23, S49), lexical tokenization (S2, S12, S26, S30), Bag-of-Words (S13, S47), and Word Embedding (S13).

\underline{\emph{Syntactical:}} These features are used in two CI studies, specifically for estimating branch coverage in automated testing (S3, S15). Analysis in this category is based on a predefined list of words, often reserved keywords, making them suitable for extracting semantic information from data.

\textbf{FT4) Epochal:} Epochal features, tied to temporal and narrative information, dominate with 35 out of 52 CI studies incorporating them. This prevalence is particularly notable in studies emphasizing test case optimization, underlining the significance of analyzing prior execution history for optimizing test case execution.

\underline{\emph{Temporal:}} While this feature cannot be used solely for decision-making and model training, temporal features critically reveal hidden temporal patterns in data and usually must be used with other features. Encompassing work habits (time, day, and month of actions) (S21, S32), time intervals between the current and the previous event (S21, S43), and the duration of the last event (S37), these features concentrate on events immediately preceding an occurrence. This differs from the broader historical context considered by ``Narrative'' features. They have also contributed to research on order-dependent and non-order-dependent flaky tests (S40).

\underline{\emph{Narrative:}} Narrative features, present in 32 out of 52 studies, are extensively utilized due to their ability to encapsulate valuable insights from past experiences and lessons learned in ML-based CI enhancements~\cite{kim2002history}. For example, in test case prioritization, the likelihood of test cases failing in the future is higher if they have previously failed, justifying their elevated priority in subsequent tests~\cite{haghighatkhah2018test}. Narrative features include attributes like changes in files or text, outcomes of prior test executions, the ratio of failed-to-pass tests, and test execution durations. Notably, features like the Failure Distance attribute (FD - the number of builds since the last failed build) are straightforward to calculate which is used in S21.

\par Table~\ref{Table:Tasks_FeatureTypes} shows that relational features are often combined with other feature types, with only three studies employing all four feature types in their analysis. Additionally, lexical features find extensive application across various tasks.

\subsubsection{Feature Engineering Techniques} 
\par Feature engineering is a crucial step in ML model development~\cite{khurana2016automating}. Its significance lies in two key factors:

Feature engineering techniques significantly impact the accuracy of trained ML models~\cite{bagherzadeh2020reinforcement}. These techniques transform raw data into a format interpretable and meaningful for ML models~\cite{khurana2016automating}. In our analysis of the reviewed papers, we identified five groups of feature engineering (\textbf{FE}) techniques. However, it is worth noting that among the 52 studies in this systematic literature review, 18 did not provide details on their feature engineering techniques.

\begin{table*}
\centering
\caption{Relation between the most frequently used data sets and feature engineering techniques. \textbf{Note:} The \textit{MyBatis}, \textit{Google Closure}, \textit{Google Auto} and \textit{Dspace} datasets did not report any feature engineering techniques. So, we did not present them in this table. The bolded numbers represent the count of studies and their percentage in
total.}
\label{Table:DataandFeatureEng}
\begin{tabular}{l|c|c|c|c|c}
\toprule
\textbf{Datasets} &
  \Centerstack{\textbf{Imputation}\\\textbf{and}\\\textbf{Elimination}} &
  \Centerstack{\textbf{Feature}\\\textbf{Enhancement}\\\textbf{and Labeling}} &
  \Centerstack{\textbf{Outliers}\\\textbf{Handling}} &
  \Centerstack{\textbf{Feature}\\\textbf{Scaling}} &
  \Centerstack{\textbf{Feature}\\\textbf{Tagging and}\\\textbf{Encoding}} \\ \hline

\textbf{IOF/ROL}        
    & \cellcolor{AECBE5}\Centerstack{S11, S37\\ \textbf{(2, 3.8\%)}  }
    & \cellcolor{DDEBF7}-- 
    & \cellcolor{DDEBF7}-- 
    & \cellcolor{84AFD5}\Centerstack{S19, S22, S37\\ \textbf{(3, 5.8\%)}  }
    & \cellcolor{BED6EB}\Centerstack{S19\\ \textbf{(1, 1.9\%)}  }
    \\
\textbf{Paint Control}  
    & \cellcolor{AECBE5}\Centerstack{S11, S37\\ \textbf{(2, 3.8\%)}  }
    & \cellcolor{DDEBF7}-- 
    & \cellcolor{DDEBF7}-- 
    & \cellcolor{84AFD5}\Centerstack{S19, S22, S37\\ \textbf{(3, 5.8\%)}  }
    & \cellcolor{BED6EB}\Centerstack{S19\\ \textbf{(1, 1.9\%)}  }
    \\
\textbf{Apache Commons} 
    & \cellcolor{AECBE5}\Centerstack{S10, S15\\ \textbf{(2, 3.8\%)}  }
    & \cellcolor{BED6EB}\Centerstack{S10\\ \textbf{(1, 1.9\%)}  }
    & \cellcolor{DDEBF7}-- 
    & \cellcolor{6FA0CD}\Centerstack{S3, S15, S19, S22\\ \textbf{(4, 7.7\%)}  }
    & \cellcolor{BED6EB}\Centerstack{S19\\ \textbf{(1, 1.9\%)}  }
    \\
\textbf{GSDTSR}         
    & \cellcolor{BED6EB}\Centerstack{S37\\ \textbf{(1, 1.9\%)}  }
    & \cellcolor{DDEBF7}-- 
    & \cellcolor{DDEBF7}-- 
    & \cellcolor{AECBE5}\Centerstack{S22, S37\\ \textbf{(2, 3.8\%)}  }
    & \cellcolor{DDEBF7}-- 
    \\
\textbf{Google Guava}   
    & \cellcolor{BED6EB}\Centerstack{S15\\ \textbf{(1, 1.9\%)}  }
    & \cellcolor{DDEBF7}-- 
    & \cellcolor{DDEBF7}-- 
    & \cellcolor{AECBE5}\Centerstack{S3, S15\\ \textbf{(2, 3.8\%)}  }
    & \cellcolor{DDEBF7}-- 
    \\
\textbf{Rails}          
    & \cellcolor{DDEBF7}-- 
    & \cellcolor{DDEBF7}-- 
    & \cellcolor{DDEBF7}-- 
    & \cellcolor{BED6EB}\Centerstack{S22\\ \textbf{(1, 1.9\%)}  }
    & \cellcolor{DDEBF7}-- 
    \\
\textbf{TravisTorrent}  
    & \cellcolor{BED6EB}\Centerstack{S5\\ \textbf{(1, 1.9\%)}  }
    & \cellcolor{BED6EB}\Centerstack{S24 \\ \textbf{(1, 1.9\%)} }
    & \cellcolor{BED6EB}\Centerstack{S21\\ \textbf{(1, 1.9\%)}  }
    & \cellcolor{BED6EB}\Centerstack{S24\\ \textbf{(1, 1.9\%)}  }
    & \cellcolor{DDEBF7}-- 
    \\
\textbf{Google Dagger}  
    & \cellcolor{BED6EB}\Centerstack{S15\\ \textbf{(1, 1.9\%)}  }
    & \cellcolor{DDEBF7}-- 
    & \cellcolor{DDEBF7}-- 
    & \cellcolor{AECBE5}\Centerstack{S3, S15\\ \textbf{(2, 3.8\%)}  }
    & \cellcolor{DDEBF7}-- 
    \\ \bottomrule
\end{tabular}%
\end{table*}

\textbf{FE1) Imputation and Elimination:} This technique addresses missing values in a dataset by replacing them with calculated or similar values, preventing data loss due to missing features~\cite{koroglu2016defect}. In CI, it is particularly useful for handling exceptions. For example, S9 calculated the LOC difference when the ``Number of Modified Lines'' was unavailable. It can also derive values by considering interdependencies among data rows, such as finding the start and end times of test cases based on the log of previous execution times. Moreover, to reduce features and lower ML model training costs, elimination methods like the chi-square method can be used to remove closely related features. For instance, authors retained one of two similar features: the number of modified lines or added lines per data entry~\cite{sharif2021deeporder}. Given the large CI data volume, dimensional reduction techniques, as seen in S13, can further reduce computational overhead and input features in ML methods.

\par \textbf{FE2) Feature Enhancement and Labeling:} In CI environments, the presence of diverse data types and their high volume poses a challenge for data analysis. To tackle this challenge, various strategies are employed to enhance data comprehensibility for training ML models and improving their performance. In S10, a new labeling scheme categorized software management profile activities into H(high), M(edium), and L(ow). S46 categorized test cases as entirely or partially redundant based on coverage analysis, augmenting data with additional features. This methodology allows for enriching datasets with contextual information; for instance, S37 labeled commit dates to indicate holidays or regular days. Additionally, S16 introduced a new feature representing the percentage of detected failures relative to the total failures for each test case.

\par \textbf{FE3) Outliers Handling:} In CI environments, unexpected data may arise due to errors or non-repeatable situations, often attributed to human errors or specified limitations. For instance, test case durations may be constrained, and exceeding this time limit results in exceptional values, known as outliers. Outliers can be detected by defining thresholds or cutoff parameters, as demonstrated in S13 and S18.

\par \textbf{FE4) Feature Scaling:} Feature values exhibit diverse distributions and ranges, necessitating normalization through statistical methods. This technique rescales data to fit within a specific range (commonly [0, 1]) or standardizes values to the same magnitude using methods like log transformation, which mitigates the impact of extremely high or low values. For instance, S3 applied log transformation in conjunction with z-score, although they did not specify which feature required this transformation. Feature scaling is widely employed, with 20 out of 34 studies explaining their feature engineering techniques using this method, primarily due to the diversity of feature values and the prevalence of numerical features in CI environments.
 
\par \textbf{FE5) Feature Tagging and Encoding:} CI datasets often involve nominal values, such as test results like ``pass'', ``fail'', or ``canceled''. To facilitate the training of ML models, input data must be made comprehensible, as these models rely on mathematical formulas~\cite{alpaydin2020introduction}. One approach is to assign 0 and 1 values to class labels or represent feature values with binary strings or tags based on predefined rules~\cite{bagherzadeh2020reinforcement, al2020effect}.

\par While encoding applies to various data types, it is particularly common in text-based datasets, involving the assignment of tokens or tags to different text segments. Effective tokenization requires a thorough understanding of the data and the selection of appropriate tokens for each distinct group or category. For example, in S13, authors defined specific tokens (tags) for various groups in source code, such as variables, operands, data types, spaces, and more.

\par In Table~\ref{Table:DataandFeatureEng}, we illustrate the relationship between feature engineering methods and the most commonly used datasets in the literature. It is noteworthy that certain datasets, in comparison with those presented in Table~\ref{Table:DataSourceCI}, are omitted here because the referenced studies did not report their feature engineering methods.


\vspace{1pt}
\noindent\fcolorbox{black}{blue!10}{%
\begin{minipage}{\columnwidth}
\vspace{1pt}
\textbf{Summary:}
\vspace{1pt}

\par $\bullet$ Narrative, Numerical, Content, and Temporal feature types are used more frequently in the selected studies in comparison with other feature types.
\vspace{1pt}

\par $\bullet$ Tools like IntelliJ Idea, RSA, and Aniche can extract features from code with low computational overhead.
\vspace{1pt}

\par $\bullet$ New features can be defined through more complex computations and raw code analysis such as tagging or defining tree-based structures.
\vspace{1pt}

\par $\bullet$ Feature engineering techniques like scaling, outlier handling, and elimination are still useful in CI environments. They cut computational load and boost ML model accuracy and performance.
\end{minipage}}



\subsection{RQ4: Model Training and Tuning}
\label{Sec:Models}
\par The performance and training time of ML methods, as well as their interactions with input data, are influenced by their inherent characteristics and predefined parameters~\cite{treveil2020introducing}. In this section, our emphasis is on improving ML model training in the context of CI and refining algorithms through hyperparameter tuning.

\subsubsection{Types of Learning Algorithms}
Various ML algorithms, including supervised, unsupervised, semi-supervised, and reinforcement learning, play vital roles in ML-based approaches. Brief introductions to these methods are presented in the following.
\par \textit{Supervised Learning:} This approach involves training a model on labelled data~\cite{russell2009artificial}, requiring human effort or pre-existing data labelling before model training~\cite{lee2019classifying}. Supervised learning comprises two primary classes: \textit{classification}, where the model categorizes outputs into fixed or discrete classes, and \textit{regression}, where the model predicts continuous values~\cite{bishop2006pattern}.

\begin{table*}
\centering
\caption{Mapping ML Algorithms to CI Tasks. The table color intensity represents study counts, with lighter shades indicating fewer studies and darker blues representing higher counts. The bolded numbers represent the count of studies and their percentage in total. \textbf{Acronyms:} \textit{\textbf{NN:}} Neural Network, \textit{\textbf{SVM:}} Support Vector Machine, \textit{\textbf{DT:}} Decision Tree, \textit{\textbf{RL:}} Reinforcement Learning, \textit{\textbf{KM:}} K-Means, \textit{\textbf{KNN:}} K-Nearest Neighbors, \textit{\textbf{LR:}} Linear Regression, \textit{\textbf{NB:}} Naive Bayes, \textit{\textbf{TL:}} Transfer learning, \textit{\textbf{UT:}} Unit Test, \textit{\textbf{IT:}} Integration Test, \textit{\textbf{RT:}} Regression Test, \textit{\textbf{BV:}} Build Validation, \textit{\textbf{ST:}} System Test and \textit{\textbf{PM:}} Process Management.}
\label{Table:MLModels}
\resizebox{\textwidth}{!}{%
\begin{tabular}{c|cccccccccc|c}
\hline

 \rotN{\Centerstack{\textbf{Algorithms}}} &
  \rotN{\Centerstack[l]{\textbf{Unit Test}\\\textbf{Prediction}}} &
  \rotN{\Centerstack[l]{\textbf{Branch Coverage}\\\textbf{Prediction}}} &
  \rotN{\Centerstack[l]{\textbf{Integration Test}\\\textbf{Prediction}}} &
  \rotN{\textbf{Test Optimization}} &
  \rotN{\textbf{Defect Prediction}} &
  \rotN{\Centerstack[l]{\textbf{Flaky Test}\\\textbf{Detection}}} &
  \rotN{\textbf{Build Prediction}} &
  \rotN{\Centerstack[l]{\textbf{Installed Software}\\\textbf{Discovery}}} &
  \rotN{\Centerstack[l]{\textbf{Performance Test}\\\textbf{Optimization}}} &
  \rotN{\Centerstack[l]{\textbf{Activity}\\\textbf{Management}}}&
  \rotN{\Centerstack[l]{\textbf{Total number}\\ \textbf{of studies}}} \\ \cline{2-11}
&
\multicolumn{1}{c|}{\textbf{UT}}&
\multicolumn{2}{c|}{\textbf{IT}}&
\multicolumn{3}{c|}{\textbf{RT}}&
\multicolumn{1}{c|}{\textbf{BV}}&
\multicolumn{2}{c|}{\textbf{ST}}&
\multicolumn{1}{c|}{\textbf{PM}}  \\ 
    \hline

\textbf{NN} &
  \cellcolor{CEE1F1}\Centerstack{S2 \\\textbf{(1, 1.9\%)}} &
  \cellcolor{BED6EB}\Centerstack{S3, S15\\\textbf{(2, 3.8\%)}} &
  \cellcolor{CEE1F1}\Centerstack{S18\\\textbf{(1, 1.9\%)}} &
  \cellcolor{7FABD3}\Centerstack{S7, S11, S26, S37,\\S39, S41 \textbf{(6, 11.5\%)}}  &
  \cellcolor{BED6EB}\Centerstack{S13, S32\\\textbf{(2, 3.8\%)}} &
  \cellcolor{DDEBF7}-- &
  \cellcolor{CEE1F1}\Centerstack{S38\\\textbf{(1, 1.9\%)}} &
  \cellcolor{DDEBF7}-- &
  \cellcolor{CEE1F1}\Centerstack{S4\\\textbf{(1, 1.9\%)}} &
  \cellcolor{DDEBF7}-- &
  \cellcolor{7FACD4}\Centerstack{\textbf{14}\\\textbf{26.9\%}} \\
\textbf{Huber} &
  \cellcolor{DDEBF7}-- &
  \cellcolor{CEE1F1}\Centerstack{S3\\\textbf{(1, 1.9\%)}} &
  \cellcolor{DDEBF7}-- &
  \cellcolor{DDEBF7}-- &
  \cellcolor{DDEBF7}-- &
  \cellcolor{DDEBF7}-- &
  \cellcolor{DDEBF7}-- &
  \cellcolor{DDEBF7}-- &
  \cellcolor{DDEBF7}-- &
  \cellcolor{DDEBF7}-- &
  \cellcolor{DDEBF7}\Centerstack{\textbf{1}\\\textbf{1.9\%}} \\
\textbf{SVM} &
  \cellcolor{DDEBF7}-- &
  \cellcolor{BED6EB}\Centerstack{S3, S15\\\textbf{(2, 3.8\%)}} &
  \cellcolor{DDEBF7}-- &
  \cellcolor{AECBE5}\Centerstack{S8, S26, S36\\\textbf{(3, 5.8\%)}} &
  \cellcolor{CEE1F1}\Centerstack{S32\\\textbf{(1, 1.9\%)}} &
  \cellcolor{DDEBF7}-- &
  \cellcolor{DDEBF7}-- &
  \cellcolor{DDEBF7}-- &
  \cellcolor{DDEBF7}-- &
  \cellcolor{CEE1F1}\Centerstack{S49\\\textbf{(1, 1.9\%)}} &
  \cellcolor{B2CEE7}\Centerstack{\textbf{7}\\\textbf{13.5\%}} \\
\textbf{DT} &
  \cellcolor{DDEBF7}-- &
  \cellcolor{CEE1F1}\Centerstack{S15\\\textbf{(1, 1.9\%)}} &
  \cellcolor{BED6EB}\Centerstack{S18, S23\\\textbf{(2, 3.8\%)}} &
  \cellcolor{6FA0CD}\Centerstack{S16, S26, S29, S30,\\S36, S46, S47\\\textbf{(7, 13.5\%)}}  &
  \cellcolor{AECBE5}\Centerstack{S9, S13, S32\\\textbf{(3, 5.7\%)}} &
  \cellcolor{CEE1F1}\Centerstack{S40\\\textbf{(1, 1.9\%)}} &
  \cellcolor{4A88C0}\Centerstack{S1, S5, S21, S24,\\S25, S27, S42,\\S43, S44, S51\\\textbf{(10, 19.2\%)}} &
  \cellcolor{DDEBF7}-- &
  \cellcolor{DDEBF7}-- &
  \cellcolor{CEE1F1}\Centerstack{S49\\\textbf{(1, 1.9\%)}} &
  \cellcolor{4A88C0}\Centerstack{\textbf{25}\\\textbf{48.1\%}} \\
\textbf{RL} &
  \cellcolor{DDEBF7}-- &
  \cellcolor{DDEBF7}-- &
  \cellcolor{DDEBF7}-- &
  \cellcolor{4F8BC1}\Centerstack{S7, S14, S20, S24,\\S27, S28, S31, S33,\\S34, S35, S50, S48\\\textbf{(12, 23.1\%)}} &
  \cellcolor{DDEBF7}-- &
  \cellcolor{DDEBF7}-- &
  \cellcolor{DDEBF7}-- &
  \cellcolor{CEE1F1}\Centerstack{S12\\\textbf{(1, 1.9\%)}} &
  \cellcolor{DDEBF7}-- &
  \cellcolor{DDEBF7}-- &
  \cellcolor{86B0D6}\Centerstack{\textbf{13}\\\textbf{25\%}} \\
\textbf{KM} &
  \cellcolor{DDEBF7}-- &
  \cellcolor{DDEBF7}-- &
  \cellcolor{DDEBF7}-- &
  \cellcolor{CEE1F1}\Centerstack{S45\\\textbf{(1, 1.9\%)}} &
  \cellcolor{DDEBF7}-- &
  \cellcolor{DDEBF7}-- &
  \cellcolor{DDEBF7}-- &
  \cellcolor{DDEBF7}-- &
  \cellcolor{DDEBF7}-- &
  \cellcolor{CEE1F1}\Centerstack{S10\\\textbf{(1, 1.9\%)}} &
  \cellcolor{D6E7F5}\Centerstack{\textbf{2}\\\textbf{3.8\%}} \\
\textbf{KNN} &
  \cellcolor{DDEBF7}-- &
  \cellcolor{DDEBF7}-- &
  \cellcolor{DDEBF7}-- &
  \cellcolor{BED6EB}\Centerstack{S16, S30\\\textbf{(2, 3.8\%)}} &
  \cellcolor{DDEBF7}-- &
  \cellcolor{DDEBF7}-- &
  \cellcolor{CEE1F1}\Centerstack{S5\\\textbf{(1, 1.9\%)}} &
  \cellcolor{DDEBF7}-- &
  \cellcolor{DDEBF7}-- &
  \cellcolor{DDEBF7}-- &
  \cellcolor{CFE2F2}\Centerstack{\textbf{3}\\\textbf{5.8\%}} \\
\textbf{LR} &
  \cellcolor{DDEBF7}-- &
  \cellcolor{DDEBF7}-- &
  \cellcolor{DDEBF7}-- &
  \cellcolor{AECBE5}\Centerstack{S16, S26, S30\\\textbf{(3, 5.7\%)}} &
  \cellcolor{CEE1F1}\Centerstack{S32\\\textbf{(1, 1.9\%)}} &
  \cellcolor{DDEBF7}-- &
  \cellcolor{BED6EB}\Centerstack{S1, S5\\\textbf{(2, 3.8\%)}} &
  \cellcolor{DDEBF7}-- &
  \cellcolor{DDEBF7}-- &
  \cellcolor{DDEBF7}-- &
  \cellcolor{B9D3EA}\Centerstack{\textbf{6}\\\textbf{11.5\%}} \\
\textbf{NB} &
  \cellcolor{DDEBF7}-- &
  \cellcolor{DDEBF7}-- &
  \cellcolor{DDEBF7}-- &
  \cellcolor{CEE1F1}\Centerstack{S16\\\textbf{(1, 1.9\%)}} &
  \cellcolor{CEE1F1}\Centerstack{S32\\\textbf{(1, 1.9\%)}} &
  \cellcolor{DDEBF7}-- &
  \cellcolor{CEE1F1}\Centerstack{S5 \\\textbf{(1, 1.9\%)}}&
  \cellcolor{DDEBF7}-- &
  \cellcolor{DDEBF7}-- &
  \cellcolor{CEE1F1}\Centerstack{S49\\\textbf{(1, 1.9\%)}} &
  \cellcolor{C8DDEF}\Centerstack{\textbf{4}\\\textbf{7.7\%}} \\
\textbf{TL} &
  \cellcolor{DDEBF7}-- &
  \cellcolor{DDEBF7}-- &
  \cellcolor{DDEBF7}-- &
  \cellcolor{CEE1F1}\Centerstack{S52\\\textbf{(1, 1.9\%)}} &
  \cellcolor{CEE1F1}\Centerstack{S17\\\textbf{(1, 1.9\%)}} &
  \cellcolor{DDEBF7}-- &
  \cellcolor{DDEBF7}-- &
  \cellcolor{DDEBF7}-- &
  \cellcolor{DDEBF7}-- &
  \cellcolor{DDEBF7}-- &
  \cellcolor{D6E7F5}\Centerstack{\textbf{2}\\\textbf{7.7\%}} \\ \hline
\end{tabular}
}
\end{table*}

\par \textit{Unsupervised learning:} Unsupervised learning algorithms, unlike supervised learning, do not require data labeling before training. These models uncover data relationships and cluster data points~\cite{russell2009artificial}. Unsupervised \textit{clustering} algorithms are particularly suited for large datasets where manual data labeling efforts are impractical~\cite{russell2009artificial}. 
\textit{Semi-supervised learning:} Semi-supervised learning algorithms leverage labeled data to classify unlabeled data by identifying underlying data relationships~\cite{chapelle2009semi}, making them more practical when dealing with imprecise or noisy datasets containing both labeled and unlabeled data~\cite{ratner2019weak}. 
Lastly, \textit{Reinforcement Learning} (RL) algorithms learn through trial and error, receiving rewards or penalties at each step until they achieve the desired output or accuracy~\cite{bishop2006pattern}. RL algorithms benefit from continuous model updates, making them suitable for dynamically changing CI environments and data.

\par Among the primary studies, supervised learning is the most commonly employed approach, with 44 out of 52 studies utilizing it. Unsupervised learning is the second most used, with five out of 52 studies employing this approach, and one study using both supervised and unsupervised ML methods. Additionally, four studies applied semi-supervised learning algorithms, while 12 studies employed RL algorithms. Our observation also indicates a significant increase in the usage of RL algorithms in ``Test Optimization'' tasks since 2020. Notably, two out of the three state-of-the-art methods in this task (COLEMAN and RETECS, see Table~\ref{table:TestOpt_SOTA}) are RL-based methods, underscoring the growing applicability of RL-based algorithms in the realm of test optimization.


\par Our analysis also reveals that classification algorithms are the most widely employed methods in the CI context, with 32 out of 52 studies utilizing them. This preference can be attributed to the characteristics of testing, where ML-based binary classifiers excel in predicting test outcomes and build results without the need for explicit execution. Additionally, one study (S49) in the CI environment employed multi-class classification methods to categorize reported issues into five classes, aiding developers in identifying and addressing bug-related issues based on the most frequent labels in the training dataset.

\par Table~\ref{Table:MLModels} demonstrates that Decision Tree (DT) algorithms are the prevailing classification method in the CI environment, with 25 out of 52 studies employing them. The popularity of DT algorithms can be attributed to several factors, two of which are derived from the selected studies, while the third is drawn from the literature.

\par First, CI environments continuously generate vast amounts of data, and ML models require frequent updates~\cite{yaraghi2022scalable, porres2020automatic}. Training and updating DT algorithms demand low computational resources, making them a feasible choice in CI settings.
\par Second, DT algorithms have high performance in classifying unseen data~\cite{finlay2014data}.
\par Third, DT algorithms are interpretable and easily comprehensible for human users~\cite{witten2002data}.

\par Table~\ref{Table:MLModels} shows that, except for one study (S12), RL algorithms have primarily been used in regression testing (RT) tasks within CI environments. In RT, a predefined set of test cases is established, and new test cases are incrementally added as new features are developed for the final software product. The objective of RT is to identify test cases with failure outcomes and prioritize their execution before those with passing outcomes. RL algorithms are well-suited for RT because this CI phase can be formulated as a sequential decision-making problem, a key characteristic of RL-based solutions~\cite{pan2020dynamic}. Moreover, RL algorithms adapt to new data more effectively with frequent updates compared to other supervised and unsupervised ML algorithms~\cite{bagherzadeh2020reinforcement}.

\par However, researchers and practitioners should consider the advantages and disadvantages of RL-based models. For instance, the ROCKET solution~\cite{marijan2013test} faces scalability issues in long runtime, while the RETECS method~\cite{spieker2018reinforcement}, an RL-based solution, requires a substantial amount of time for training.

\par Alongside the widespread adoption of DT and RL algorithms in regression testing, Table~\ref{Table:MLModels} highlights the utilization of Neural Network (NN) algorithms in 14 studies. Remarkably, NN algorithms have been applied in 7 out of 10 CI tasks, underscoring their flexibility in addressing various challenges within CI. The appeal of NN algorithms, despite their need for significant computational resources during training, lies in their capacity to automatically extract features from datasets and their high predictive accuracy~\cite{porres2020automatic}.

\subsubsection{Hyper-Parameter Tuning} 
\par In general, ML models comprise a basic formula that requires configuration by identifying the optimal values for their hyper-parameters~\cite{treveil2020introducing}. Selected studies present varying perspectives on hyperparameter tuning. Some emphasize the importance of hyperparameter tuning in achieving the optimal performance of trained models~\cite{elsner2021empirically}, considering skipping this process as a potential threat that can affect model accuracy~\cite{abdelkarim2022tcp, lima2020learning}. However, other studies, like~\cite{mondal2019exploratory}, found that tuning hyperparameters did not significantly impact model accuracy. Additionally, Al-Sabbagh et al.~\cite{al2019predicting} reported that automatic hyperparameter tuning tools can be time-consuming, prompting them to manually tune hyperparameters to save time.
\par It is worth noting that hyperparameter tuning can affect model training speed. For instance, adjusting the `training rate' hyperparameter can increase model instability while reducing training time or decrease training speed while enhancing overall model performance stability~\cite{alpaydin2020introduction}. Hence, our investigation delves into how the selected studies in the application of ML algorithms within CI environments performed hyperparameter tuning.

\par Based on the data extracted from the selected studies, we categorized the hyper-parameter tuning strategies \textbf{(HT)} into five groups.

\par \textbf{HT1)} The first group used \underline{searching methods} to find the best hyperparameters. These studies employed methods such as \textit{Grid Search (GS)}, \textit{Bayesian Search (BS)}, and \textit{Genetic Algorithm (GA)}.

\par In GS, researchers fix the domain of hyperparameters, and the algorithm finds the best combination from these fixed values~\cite{bergstra2011algorithms}. In BS, the next hyperparameter values are determined based on the evaluation results of the previous values, avoiding unnecessary evaluations~\cite{eggensperger2013towards}. In GA, hyperparameters are adjusted iteratively by evaluating the model's performance and making slight changes~\cite{foo2007efficient}.

\par In S38, a unique Genetic Algorithms-based hyperparameter tuning method is utilized. Random values were assigned to each hyperparameter, and a single-point crossover operator generated two new offspring in the Crossover step. The best solutions were retained using the elitism method and fitness function calculation. In the Mutation step, hyperparameter values were slightly modified, and the Crossover step was repeated. This process continued until it reached the predetermined stopping criteria. Table~\ref{Table_old:Hyper} details the hyperparameter tuning methods used for ML models, acknowledging that one tuning method might be applied to multiple ML models in studies utilizing more than one ML method.

\begin{table*}

\centering
\caption{Summary of employed hyper-parameter strategies and mapping to the ML methods. color intensity corresponds to study counts, with lighter shades indicating lower and darker blues representing higher counts. The bolded numbers represent the count of studies and their percentage in total. \textbf{Acronyms:} \textit{\textbf{NN:}} Neural Network, \textit{\textbf{SVM:}} Support Vector Machine, \textit{\textbf{DT:}} Decision Tree, \textit{\textbf{RL:}} Reinforcement Learning, \textit{\textbf{KM:}} K-Means, \textit{\textbf{KNN:}} K-Nearest Neighbors, \textit{\textbf{LR:}} Linear Regression, \textit{\textbf{NB:}} Naive Bayes and \textit{\textbf{TL:}} Transfer learning, \textit{\textbf{GS:}} Grid Search, \textit{\textbf{GA:}} Genetic Algorithm and \textit{\textbf{BS:}} Bayesian Search}
\label{Table_old:Hyper}
\resizebox{\textwidth}{!}{%
\begin{tabular}{l|cccccc}
\hline
                   \textbf{Alg.}    &\textbf{Methods}& \textbf{Default} & \textbf{Literature} & \textbf{Test-and-Trial} & \textbf{Formula} & \textbf{Did not report} \\ \hline
\textbf{NN}    & \cellcolor{A3C4E1}\Centerstack{S15(GS), S38(GA),\\ S39(BS)\\\textbf{(3, 5.8\%)}} & \cellcolor{CADEF0}\Centerstack{S26\\\textbf{(1, 1.9\%)}} & \cellcolor{CADEF0}\Centerstack{S7\\\textbf{(1, 1.9\%)}} & \cellcolor{699DCB}\Centerstack{S2, S3, S4, S13,\\S18, S37, S41\\\textbf{(7, 13.4\%)}} & \cellcolor{CADEF0}\Centerstack{S11\\\textbf{(1, 1.9\%)}} & \cellcolor{CADEF0}\Centerstack{S32\\\textbf{(1, 1.9\%)}} \\
\textbf{Huber} & \cellcolor{DDEBF7}--          & \cellcolor{DDEBF7}-- & \cellcolor{DDEBF7}-- & \cellcolor{CADEF0}\Centerstack{S3\\\textbf{(1, 1.9\%)}} & \cellcolor{DDEBF7}-- & \cellcolor{DDEBF7}-- \\
\textbf{SVM}   & \cellcolor{CADEF0}\Centerstack{S15(GS)\\\textbf{(1, 1.9\%)}}         & \cellcolor{CADEF0}\Centerstack{S26\\\textbf{(1, 1.9\%)}} & \cellcolor{DDEBF7}-- & \cellcolor{B7D1E9}\Centerstack{S3, S49\\\textbf{(2, 3.8\%)}} & \cellcolor{DDEBF7}-- & \cellcolor{A3C4E1}\Centerstack{S8, S32, S36\\\textbf{(3, 5.7\%)}} \\
\textbf{DT}    & \cellcolor{A3C4E1}\Centerstack{S15(GS), S29(GS),\\S30(BS)\\\textbf{(3, 5.7\%)}}     & \cellcolor{7DAAD3}\Centerstack{S17, S23, S25,\\S26, S47, S51\\\textbf{(6, 11.5\%)}} & \cellcolor{DDEBF7}-- & \cellcolor{7DAAD3}\Centerstack{S9, S13, S49,\\S18, S46, S16\\\textbf{(6, 11.5\%)}} & \cellcolor{B7D1E9}\Centerstack{S24, S44\\\textbf{(2, 3.8\%)}}  & \cellcolor{3F80BB}\Centerstack{S1, S5, S21, S27, S32,\\S36, S40, S42, S43\\\textbf{(9, 17.3\%)}} \\
\textbf{RL}    & \cellcolor{DDEBF7}--          & \cellcolor{B7D1E9}\Centerstack{S19, S35\\\textbf{(2, 3.8\%)}} & \cellcolor{7DAAD3}\Centerstack{S20, S48, S7, S28,\\S34 \textbf{(5, 9.6\%)}} & \cellcolor{B7D1E9} \Centerstack{S14, S22\\\textbf{(2, 3.8\%)}} & \cellcolor{DDEBF7}-- & \cellcolor{A3C4E1}\Centerstack{S12, S31, S33\\\textbf{(3, 5.8\%)}} \\
\textbf{KM}    & \cellcolor{DDEBF7}--          & \cellcolor{CADEF0}\Centerstack{S10\\\textbf{(1, 1.9\%)}} & \cellcolor{DDEBF7}-- & \cellcolor{DDEBF7}-- & \cellcolor{DDEBF7}-- & \cellcolor{CADEF0}\Centerstack{S45\\\textbf{(1, 1.9\%)}} \\
\textbf{KNN}   & \cellcolor{CADEF0}\Centerstack{S30(BS)\\\textbf{(1, 1.9\%)}}         & \cellcolor{DDEBF7}-- & \cellcolor{DDEBF7}-- & \cellcolor{CADEF0}\Centerstack{S16\\\textbf{(1, 1.9\%)}} & \cellcolor{DDEBF7}-- & \cellcolor{CADEF0}\Centerstack{S5\\\textbf{(1, 1.9\%)}} \\
\textbf{LR}    & \cellcolor{CADEF0}\Centerstack{S30(BS)\\\textbf{(1, 1.9\%)}}         & \cellcolor{CADEF0}\Centerstack{S26\\\textbf{(1, 1.9\%)}} & \cellcolor{DDEBF7}-- & \cellcolor{CADEF0}\Centerstack{S16\\\textbf{(1, 1.9\%)}} & \cellcolor{DDEBF7}-- & \cellcolor{A3C4E1}\Centerstack{S1, S5, S32\\\textbf{(3, 5.8\%)}} \\
\textbf{NB}    & \cellcolor{DDEBF7}--          & \cellcolor{DDEBF7}-- & \cellcolor{DDEBF7}-- & \cellcolor{CADEF0}\Centerstack{S49\\\textbf{(1, 1.9\%)}} & \cellcolor{DDEBF7}-- & \cellcolor{B7D1E9}\Centerstack{S5, S32\\\textbf{(2, 3.8\%)}} \\
\textbf{TL}    & \cellcolor{DDEBF7}--          & \cellcolor{B7D1E9}\Centerstack{S17, S52\\\textbf{(2, 3.8\%)}} & \cellcolor{DDEBF7}-- & \cellcolor{DDEBF7}-- & \cellcolor{DDEBF7}-- & \cellcolor{DDEBF7}--  \\ \hline
\end{tabular}
}
\end{table*}

\par In Table~\ref{Table_old:Hyper}, the hyperparameter tuning methods used for ML models are detailed. Considering that several studies employed more than one ML method, a single tuning method might be applied to multiple ML models.

\par \textbf{HT2)} Among the 52 reviewed studies, 10 papers opted for the \underline{Default} hyperparameter values, determined by ML training libraries like scikit-learn. This approach allows efficient training without investing time in hyperparameter tuning. Default hyperparameters are commonly established through expert knowledge, serving as a dependable initial configuration and aiding in the prevention of overfitting issues~\cite{mantovani2020rethinking}.

\par \textbf{HT3)} Four studies reduced ML model training time by utilizing hyperparameters tuned from \underline{Literature}. Given the frequent retraining needs of ML models in CI environments, both HT2 and HT3 methods contribute to significant time savings.

\par \textbf{HT4)} The majority of studies (11 out of 52) manually explored hyperparameter values for ML models, determining optimal settings through the \underline{Test and Trial} method. Similar to Random Search (RS), this method allows parallel and independent evaluation of candidate hyperparameters, facilitating quicker identification of suitable solutions in the agile CI environment, even before the arrival of new data~\cite{yang2020hyperparameter}.

\par \textbf{HT5)} The fifth strategy involves defining a \underline{Formula} for hyperparameter tuning, observed in three studies: S44 calculated the Hoeffding bound for navigating Hoeffding decision tree nodes based on observations and confidence parameters; S11 determined memory units for LSTM, Sigmoid, and Tanh using a specified formula; and S24 adjusted the K value for clustering build logs with the K-Means algorithm as $\sqrt{n/2}$ during data preparation, where n is the number of build logs, without providing evidence for this assumption.

\par Notably, 14 studies \textit{did not report} how they adjusted hyperparameters in their studies.

\par While hyperparameter tuning enhances model accuracy, it may lead to overfitting~\cite{cheng2021towards}. Overfitting occurs when a model learns excessively from the training data, such as by increasing the number of layers in neural networks. This includes capturing noise or random patterns, which hinders its performance with new data and reduces its generalizability~\cite{montesinos2022overfitting}. To address this issue, researchers commonly employ K-fold cross-validation~\cite{martins2021supervised}. This technique involves iterative training and evaluation on different data subsets, allowing hyperparameter adjustment for improved model performance. Notably, the selected studies do not address the impact of hyperparameter tuning techniques on ML model performance in CI tasks.


\vspace{1pt}
\noindent\fcolorbox{black}{blue!10}{%
\begin{minipage}{\columnwidth}
\vspace{1pt}
\textbf{Summary:}
\vspace{1pt}
\par $\bullet$ Decision tree algorithms are favored in agile CI due to high accuracy and low computational overhead.
\vspace{1pt}
\par $\bullet$ Reinforcement learning suits agile settings for sequential decision-making challenges.
\vspace{1pt}
\par $\bullet$ Neural network algorithms, with high accuracy, find broad application across five CI tasks.

\vspace{1pt}
\par $\bullet$ Hyperparameter tuning, despite increasing training time, enhances ML model accuracy.
\vspace{1pt}


\par $\bullet$ Five strategies, including search methods, default settings, literature-tuned parameters, manual tuning, and formulas, are identified for hyperparameter tuning.
\vspace{1pt}
\par $\bullet$ 14 studies in the literature did not report their hyperparameter tuning methods. 

\end{minipage}}


\subsection{RQ5: Evaluation Methods}
\label{Sec:MethodEval}

\par In practical ML applications, performance evaluation is crucial. In this section, we summarize the validation techniques and performance metrics used in the selected studies. For assessing supervised ML methods, predictions were compared with actual labels from untrained data, revealing seven distinct evaluation methods \textbf{(EM)} in the literature as presented in Table~\ref{Table_old:EvalTech}.

\begin{table*}
\caption{Relationships between seven evaluation methods and CI phases are illustrated in the table. The color intensity in this table corresponds to the number of studies, with lighter shades indicating lower counts and darker blues representing higher counts. The bolded numbers represent the count of studies and their percentage in total.}
\label{Table_old:EvalTech}
\resizebox{\textwidth}{!}{%
\begin{tabular}{l|l|c|cccccc}
\toprule
\Centerstack{\textbf{Evaluation}\\\textbf{technique}}             & \Centerstack{\textbf{Selection}\\\textbf{method}} &
\Centerstack{\textbf{Total}\\\textbf{papers}} &
\Centerstack[c]{\textbf{Unit}\\\textbf{Test}} &
\Centerstack[c]{\textbf{Integration}\\\textbf{Test}} &
\Centerstack[c]{\textbf{Regression}\\\textbf{Test}} &
\Centerstack[c]{\textbf{Build}\\\textbf{Validation}} &
\Centerstack[c]{\textbf{System}\\\textbf{Test}} &
\Centerstack[c]{\textbf{Process}\\\textbf{Management}} \\ 
\toprule
\multirow{2}{*}{\textbf{K-Fold}}          & \textbf{Sorted}           & \cellcolor{AAC9E4}\Centerstack{\textbf{5}\\\textbf{9.6\%}} &
  \cellcolor{DDEBF7}-- &
  \cellcolor{DDEBF7}-- &
  \cellcolor{B5D0E8}\Centerstack{S26, S30, S32,\\S40 \textbf{(4, 7.7\%)}} &
  \cellcolor{D3E5F4}\Centerstack{S38 \textbf{(1, 1.9\%)}} &
  \cellcolor{DDEBF7}-- &
  \cellcolor{DDEBF7}-- \\ \cdashline{2-9} 
                                          & \textbf{Random}           & \cellcolor{4F8BC1}\Centerstack{\textbf{16}\\\textbf{30.8\%}} &
  \cellcolor{D3E5F4}\Centerstack{S2\\\textbf{(1, 1.9\%)}} &
  \cellcolor{BFD7EC}\Centerstack{S3, S15,\\S23 \textbf{(3, 5.8\%)}} &
  \cellcolor{AAC9E4}\Centerstack{S13, S32, S36, S46,\\S47 \textbf{(5, 9.6\%)}} &
  \cellcolor{AAC9E4}\Centerstack{S21, S27, S43, S44,\\S51 \textbf{(5, 9.6\%)}} &
  \cellcolor{D3E5F4}\Centerstack{S12\\\textbf{(1, 1.9\%)}} &
  \cellcolor{D3E5F4}\Centerstack{S49\\\textbf{(1, 1.9\%)}}           \\ \hline

\multirow{2}{*}{\textbf{Percentage}}      & \textbf{Sorted}           & \cellcolor{B5D0E8}\Centerstack{\textbf{4}\\\textbf{7.7\%}} &
  \cellcolor{DDEBF7}-- &
  \cellcolor{DDEBF7}-- &
  \cellcolor{BFD7EC}\Centerstack{S8, S16, S29\\\textbf{(3, 5.8\%)}} &
  \cellcolor{D3E5F4}\Centerstack{S25\\\textbf{(1, 1.9\%)}} &
  \cellcolor{DDEBF7}-- &
  \cellcolor{DDEBF7}-- \\ \cdashline{2-9} 
                                          & \textbf{Random}           & \cellcolor{B5D0E8}\Centerstack{\textbf{4}\\\textbf{7.7\%}} &
  \cellcolor{DDEBF7}-- &
  \cellcolor{D3E5F4}\Centerstack{S18\\\textbf{(1, 1.9\%)}} &
  \cellcolor{D3E5F4}\Centerstack{S39\\\textbf{(1, 1.9\%)}} &
  \cellcolor{C9DEF0}\Centerstack{S24, S42\\\textbf{(2, 3.8\%)}} &
  \cellcolor{DDEBF7}-- &
  \cellcolor{DDEBF7}-- \\ \hline
\textbf{Constant Number}                  & \textbf{Sorted}           & \cellcolor{BFD7EC}\Centerstack{\textbf{3}\\\textbf{5.8\%}} &
  \cellcolor{DDEBF7}-- &
  \cellcolor{DDEBF7}-- &
  \cellcolor{C9DEF0}\Centerstack{S11, S41\\\textbf{(2, 3.8\%)}} &
  \cellcolor{D3E5F4}\Centerstack{S6\\\textbf{(1, 1.9\%)}} &
  \cellcolor{DDEBF7}-- &
  \cellcolor{DDEBF7}-- \\ \hline
\textbf{Time}                             & \textbf{Sorted}           & \cellcolor{B5D0E8}\Centerstack{\textbf{4}\\\textbf{7.7\%}} &
  \cellcolor{DDEBF7}-- &
  \cellcolor{DDEBF7}-- &
  \cellcolor{BFD7EC}\Centerstack{S17, S37, S52\\\textbf{(3, 5.8\%)}} &
  \cellcolor{D3E5F4}\Centerstack{S1\\\textbf{(1, 1.9\%)}} &
  \cellcolor{DDEBF7}-- &
  \cellcolor{DDEBF7}-- \\ \hline
\textbf{Version}                          & \textbf{Sorted}           & \cellcolor{D3E5F4}\Centerstack{\textbf{1}\\\textbf{1.9\%}} &
  \cellcolor{DDEBF7}-- &
  \cellcolor{DDEBF7}-- &
  \cellcolor{D3E5F4}\Centerstack{S9\\\textbf{(1, 1.9\%)}} &
  \cellcolor{DDEBF7}-- &
  \cellcolor{DDEBF7}-- &
  \cellcolor{DDEBF7}-- \\ \hline
\textbf{Gradually Evaluation}             & \textbf{Incremental}      & \cellcolor{5891C5}\Centerstack{\textbf{13}\\\textbf{25\%}} &
  \cellcolor{DDEBF7}-- &
  \cellcolor{DDEBF7}-- &
  \cellcolor{6398C9}\Centerstack{S7, S14, S19, S20,\\S22, S28, S31, S33,\\S34, S35, S48, S50\\\textbf{(12, 23.1\%)}} &
  \cellcolor{DDEBF7}-- &
  \cellcolor{D3E5F4}\Centerstack{S4\\\textbf{(1, 1.9\%)}} &
  \cellcolor{DDEBF7}-- \\ \hline
\textbf{Baseline}                         & \textbf{Train=Test}       & \cellcolor{C9DEF0}\Centerstack{\textbf{2}\\\textbf{3.8\%}} &
  \cellcolor{DDEBF7}-- &
  \cellcolor{DDEBF7}-- &
  \cellcolor{D3E5F4}\Centerstack{S10\\\textbf{(1, 1.9\%)}} &
  \cellcolor{DDEBF7}-- &
  \cellcolor{DDEBF7}-- &
  \cellcolor{D3E5F4}\Centerstack{S45\\\textbf{(1, 1.9\%)}} \\ \bottomrule
\end{tabular}
}
\end{table*}

\par \textbf{EM1) K-Fold:} Generally, data is divided into K equal parts, with ML models trained on K-1 segments and evaluated on the unseen segment.

\par \textbf{EM2) Percentage:} Data is split into training and testing sets based on a specified percentage (e.g., 80\% training and 20\% testing).

\par \textbf{EM3) Constant number:} N samples are designated as the test set, while the model is trained on the remaining samples.

\par \textbf{EM4) Time:} Dataset division is based on specific dates or time spans, mirroring real-world CI data dynamics.

\par \textbf{EM5) Version:} Utilized in version-based releases, where old software versions serve as the training data and the earliest versions as the test data.

\par \textbf{EM6) Gradual Evaluation:} Common in RL research, it assesses model performance incrementally through a reward-based approach, reflecting real-world CI conditions.

\par \textbf{EM7) Baseline:} Unsupervised ML models use the same dataset for training and testing, with evaluation based on comparing model outcomes to the expected solution.

\par Notably, besides gradual evaluation for RL-based methods, S4 employed an online supervised learning approach and assessed model performance by iterative training and evaluating positive predictive values (PPV) as the main effectiveness metric for a test suite.

\par Table~\ref{Table_old:EvalTech} classifies K-Fold and Percentage methods into Sorted and Random types. In Sorted types, models train on older data and evaluate earlier unseen segments, prioritizing reliability in real-world CI scenarios. Random types, randomly segment data, enhancing solution generalizability but may be less robust for real-world problem-solving~\cite{ashok2023remediating, nougnanke2022ml}.

\par According to Table~\ref{Table_old:EvalTech}, K-Fold methods are widely employed for evaluating ML techniques. This prevalence may be attributed to the availability of numerous packages capable of calculating evaluation metrics with minimal intervention. However, the reliance on random selection methods for training and testing data, as observed in many studies, can compromise the reliability of the presented methods in real-world scenarios. It is worth mentioning that in CI environments, ML models should always be applied to newly produced, unseen data.

\par In contrast to random selection in K-Fold methods, as shown in Table~\ref{Table_old:EvalTech}, a majority (30 out of 36) of the studies utilize sorted selection methods to align their approaches with real-world Continuous Integration (CI) environments.

\par Additionally, Gradual Evaluation emerges as another commonly utilized evaluation method, particularly notable due to the extensive use of Reinforcement Learning methods in CI's Regression Testing step.

\par
\begin{figure*}
    \centering
    \includegraphics[width=\textwidth]{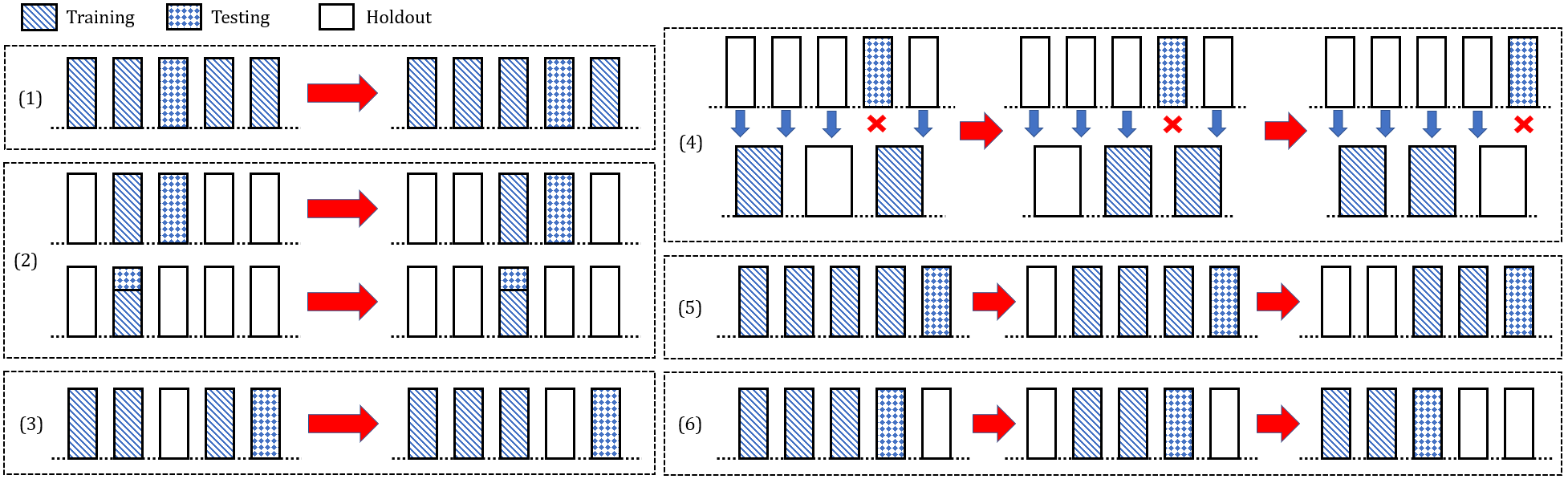}
    \caption{Six types of employed K-Fold evaluation techniques in the selected studies. \textbf{Note:} In this picture red arrows present the next evaluation round, blue arrows show the new data arrangement, and the X mark shows the neglected data parts in the new data arrangement.}
    \label{Fig:KFoldTypes}
\end{figure*}

 Our analysis identified six types of K-Fold cross-validation methods (Figure~\ref{Fig:KFoldTypes}), each denoted by a number. K-Fold validation typically divides the dataset into three groups: \textit{Training} data for model training, \textit{Testing} data for model evaluation, and a \textit{Holdout} subset unused for training or evaluation (in some techniques, Testing and Holdout sets are the same).

\par Figure~\ref{Fig:KFoldTypes} presents six K-Fold cross-validation types, each addressing specific challenges in CI environments:

\par Classical K-Fold uses the same testing and holdout sets, risking overfitting in sequential CI data~\cite{grano2019branch}.

\par The second type of K-Fold is ideal for environments with extensive data, addressing the need for training models on large volumes of data in CI settings~\cite{pan2021continuous}. Here, testing data is consistently more recent than the training data. In this approach, after dividing the data into K folds, we have two options: we can select one fold as the training set and the next fold as the testing set (first row), or we can divide one fold into both training and testing datasets (second row).

\par The third K-Fold type, similar to the classical type, uses the most recent data as testing while altering the holdout data. This method lowers the risk of overfitting compared to the first K-Fold type.

\par The fourth type of K-Fold Cross-Validation, known as Nested Cross-Validation (Nested CV), utilizes two K-Fold procedures to conduct hyperparameter tuning and increase the number of iterations. However, Nested CV incorporates futuristic data during training~\cite{grano2019branch}. This method resembles the third type of K-Fold, with the key difference being that the testing dataset varies in each iteration. Additionally, Nested CV allows for the selection of a new K value for the training dataset. Specifically, the process involves initially reserving one random fold as the testing dataset, followed by model training using a new K-Fold method on the remaining dataset.

\par The fifth K-Fold type is the modified version of the third type, uses the most recent data for testing, and varies the number of holdout folds to determine the data required for acceptable model accuracy~\cite{elsner2021empirically}.

\par The last K-Fold type employs the last consecutive data fold for testing while adjusting the number of holdout folds, allowing the examination of model stability over time~\cite{elsner2021empirically}.

\par The chronological order is vital in CI data, rendering many K-Fold techniques employed in the selected studies inadequate for CI environments due to their lack of consideration for the latest data in evaluations~\cite{chen2020buildfast, hassan2006using}.


\subsubsection{Performance Measures}
\par Choosing appropriate performance measures allows us to identify weaknesses in a solution and compare our results with others’ methods detached from the research settings. 

\begin{table*}
\centering
\caption{Performance measurements description and formulas.}
\label{Table:PerfMeas}
\begin{tabular}{{p{0.1\linewidth} p{0.55\linewidth} p{0.25\linewidth}}}
\hline
\multicolumn{1}{c}{\textbf{Measure}} &
  \multicolumn{1}{c}{\textbf{Description}} &
  \multicolumn{1}{c}{\textbf{Formula}} \\ \hline
Precision & The percentage of the detected positive instances that were correct               & $\frac{TP}{TP+FP}$ \\[6pt]
Recall    & The proportion of positive instances that were correctly identified               & $\frac{TP}{TP+FN}$    \\[6pt]
F1-score  & The harmonic mean of recall and precision                                           & $\frac{2 \times Precision \times Recall}{Precision+ Recall}$  \\[6pt]
Accuracy  & Percentage of correctly classified instances                                        & $\frac{TP+TN}{TP+FP+FN+TN}$  \\[6pt]
APFD      & Ratio between detected and detectable instances in classes                          & $1 - \frac{\sum_1^mTF_{i}}{m \times n} + \frac{1}{2n}$      \\[6pt]
NAPFD     & Normalized $APFD$                                                                     & $p - \frac{\sum_1^mTF_{i}}{m \times n} + \frac{p}{2n}$     \\[6pt]
MAE       & Average of errors between predicted and actual value                                & $\frac{\sum_1^n(y_{i}-\widehat{y}_{i})}{n}$       \\[6pt]
RMSE &
  The square root of the average of squared differences between prediction and actual observation &
  $\sqrt{\frac{1}{n}\sum_1^n(y_{i}-\widehat{y}_{i})^{2}}$\\
PP        & The ratio of the detected positive instances and all instances                      & $\frac{TP + FP}{TP + FP + TN + FN}$        \\[6pt]
MCC       & Correlation coefficient between the observed and predicted binary classifications & $\frac{TP\times TN - FP\times FN}{\sqrt{(TP+FP)(TP+FN)(TN+FP)(TN+FN)}}$       \\ 
G-Measure  & The Harmonic mean of TPR ($recall$) and true negative rate ($TNR$)                    & $\frac{2 \times TPR \times TNR}{TPR+ TNR}$  \\[6pt]
NRPA      & Measures how close a predicted ranking of items (${s_e}$) is to the optimal ranking (${s_o}$)         &  $\frac{RPA(s_e)}{RPA(s_o)}$      \\
AUC       & Two-dimensional area under Receiver Operating Characteristic ($ROC$).               &        \\
CCOFF    & Measures the closeness of the predicted values and the testing set                 & $\rho_{xy} = \frac{Cov(x,y)}{\sigma_x \sigma_y}$    \\
T-test    & Determining mean difference between two sets                                         &     \\\hline

\end{tabular}

\end{table*}

\par In the 52 selected studies, recall, precision, and F1-score are the most frequently used performance metrics, appearing in 30, 24, and 21 studies, respectively. Table~\ref{Table:PerfMeas} provides a summary of performance measurements, including formulas and brief descriptions. Key variables encompass $n$ (number of test cases), $m$ (number of faults), $p$ (ratio of detected faults to total faults), $TF_{i}$ (position of the first test exposing fault $i$), $y_{i}$ and $\widehat{y}_{i}$ (actual and predicted values), $\rho$ (Pearson product-moment correlation coefficient), and $\sigma_x$ (standard deviation of variable $x$).

\par In the reviewed studies, frequently used performance measurements include Average Percentage of Fault Detection (APFD) in 16 studies, Normalized APFD (NAPFD) in 10 studies, Accuracy in 9 studies, and AUC (Area under the ROC Curve) in 9 studies. AUC quantifies performance by plotting the True Positive Rate (TPR or Recall) against the False Positive Rate ($FPR = \frac{FP}{FP+TN}$) in a ROC curve, where $T$, $F$, $P$, and $N$ stand for True, Negative, Positive, and Negative in the confusion matrix.

\par It is crucial to note that while APFD is a valuable performance metric, researchers are encouraged to prefer Normalized APFD (NAPFD) for enhanced consistency~\cite{qu2007combinatorial}. APFD measures the area under the curve with the y-axis indicating the percentage of faults found and the x-axis showing the percentage of test cases. However, reporting the average APFD value can be misleading, as it equals 1 when no failures occur in a test cycle and may yield high values in imbalanced datasets, a common characteristic in CI datasets~\cite{bagherzadeh2020reinforcement}.

\begin{table*}
\caption{Ratio of ML types to performance metrics (left) and CI phases (right). Numbers under `ML Types' indicate study counts. Study IDs are omitted to prevent table dimension expansion. color intensity corresponds to percentages, with lighter shades indicating lower and darker blues representing higher percentages. The bolded numbers represent the count of studies and their percentage in each ML type. The percentage in the ML types column is in total.}
\label{Table:MLMeasureCI}
\resizebox{\textwidth}{!}{%
\begin{tabular}{ccccccc|c|cccccc}
\hline
\multicolumn{1}{c}{\textbf{NAPFD}} &
  \multicolumn{1}{c}{\textbf{Accuracy}} &
  \multicolumn{1}{c}{\textbf{AUC}} &
  \multicolumn{1}{c}{\textbf{APFD}} &
  \multicolumn{1}{c}{\textbf{F1-Score}} &
  \multicolumn{1}{c}{\textbf{Precision}} &
  \multicolumn{1}{c|}{\textbf{Recall}} &
  \multicolumn{1}{c|}{\textbf{ML types}} &

  \Centerstack{\textbf{Unit}\\\textbf{Test}} &
  \Centerstack{\textbf{Integration}\\\textbf{Test}} &
  \Centerstack{\textbf{Regression}\\\textbf{Test}} &
  \Centerstack{\textbf{Build}\\\textbf{Validation}} &
  \Centerstack{\textbf{System}\\\textbf{Test}} &
  \Centerstack{\textbf{Process}\\\textbf{Management}} \\
  \hline
\cellcolor{D8E8F5}\Centerstack{\textbf{1}\\\textbf{3.0\%}} &
  \cellcolor{BED6EB}\Centerstack{\textbf{6}\\\textbf{18.2\%}} &
  \cellcolor{AECBE5}\Centerstack{\textbf{9}\\\textbf{27.3\%}} &
  \cellcolor{C3DAED}\Centerstack{\textbf{5}\\\textbf{15.2\%}} &
  \cellcolor{84AFD5}\Centerstack{\textbf{17}\\\textbf{51.5\%}} &
  \cellcolor{79A8D1}\Centerstack{\textbf{19}\\\textbf{57.6\%}} &
  \cellcolor{79A8D1}\Centerstack{\textbf{19}\\\textbf{57.6\%}} &
  \Centerstack{\textbf{Classification}\\\textbf{(33, 63.5\%)}} &
  \cellcolor{D8E8F5}\Centerstack{\textbf{1}\\\textbf{3.0\%}} &
  \cellcolor{CEE1F1}\Centerstack{\textbf{3}\\\textbf{9.1\%}} &
  \cellcolor{8EB6D9}\Centerstack{\textbf{15}\\\textbf{45.5\%}} &
  \cellcolor{A4C4E2}\Centerstack{\textbf{11}\\\textbf{33.3\%}} &
  \cellcolor{D3E4F3}\Centerstack{\textbf{2}\\\textbf{6.1\%}} &
  \cellcolor{D8E8F5}\Centerstack{\textbf{1}\\\textbf{3.0\%}} \\
\cellcolor{DDEBF7}-- &
  \cellcolor{C0D8EC}\Centerstack{\textbf{1}\\\textbf{16.7\%}} &
  \cellcolor{DDEBF7}-- &
  \cellcolor{A3C4E1}\Centerstack{\textbf{2}\\\textbf{33.3\%}} &
  \cellcolor{DDEBF7}-- &
  \cellcolor{C0D8EC}\Centerstack{\textbf{1}\\\textbf{16.7\%}} &
  \cellcolor{C0D8EC}\Centerstack{\textbf{1}\\\textbf{16.7\%}} &
  \Centerstack{\textbf{Regression}\\\textbf{(6, 11.5\%)}} &
  \cellcolor{DDEBF7}-- &
  \cellcolor{A4C4E2}\Centerstack{\textbf{2}\\\textbf{33.3\%}} &
  \cellcolor{86B0D6}\Centerstack{\textbf{3}\\\textbf{50.0\%}} &
  \cellcolor{C1D8ED}\Centerstack{\textbf{1}\\\textbf{16.7\%}} &
  \cellcolor{DDEBF7}-- &
  \cellcolor{DDEBF7}-- \\
\cellcolor{A3C4E1}\Centerstack{\textbf{4}\\\textbf{33.3\%}} &
  \cellcolor{DDEBF7}-- &
  \cellcolor{DDEBF7}-- &
  \cellcolor{95BADC}\Centerstack{\textbf{5}\\\textbf{41.7\%}} &
  \cellcolor{DDEBF7}-- &
  \cellcolor{DDEBF7}-- &
  \cellcolor{B2CEE7}\Centerstack{\textbf{3}\\\textbf{25.0\%}} &
  \Centerstack{\textbf{Reinforcement}\\ \textbf{Learning (12, 23.1\%)}} &
  \cellcolor{DDEBF7}-- & 
  \cellcolor{DDEBF7}-- &
  \cellcolor{3F80BB}\Centerstack{\textbf{12}\\\textbf{100\%}} &
  \cellcolor{DDEBF7}-- &
  \cellcolor{DDEBF7}-- &
  \cellcolor{DDEBF7}-- \\
\cellcolor{DDEBF7}-- &
  \cellcolor{86B0D6}\Centerstack{\textbf{1}\\\textbf{50.0\%}} &
  \cellcolor{DDEBF7}-- &
  \cellcolor{86B0D6}\Centerstack{\textbf{1}\\\textbf{50.0\%}} &
  \cellcolor{86B0D6}\Centerstack{\textbf{1}\\\textbf{50.0\%}} &
  \cellcolor{3F80BB}\Centerstack{\textbf{2}\\\textbf{100\%}} &
  \cellcolor{3F80BB}\Centerstack{\textbf{2}\\\textbf{100\%}} &
  \Centerstack{\textbf{Clustering}\\\textbf{(2, 3.8\%)}} &
  \cellcolor{DDEBF7}-- &
  \cellcolor{DDEBF7}-- &
  \cellcolor{86B0D6}\Centerstack{\textbf{1}\\\textbf{50.0\%}} &
  \cellcolor{DDEBF7}-- &
  \cellcolor{DDEBF7}-- &
  \cellcolor{86B0D6}\Centerstack{\textbf{1}\\\textbf{50.0\%}} \\ \hline
\end{tabular}%
}
\end{table*}

\par Table~\ref{Table:MLMeasureCI} indicates that classification ML methods often used Recall, Precision, and F1-Score, while RL methods predominantly utilized APFD and NAPFD. The strong association between RL algorithms and APFD metrics is due to RL's compatibility with Test Case Prioritization (TCP) in CI, leveraging action-reward policies. APFD and NAPFD quantify the weighted mean of fault detection percentages over the test suite's lifecycle, making them suitable for evaluating test case quality. Notably, studies using APFD or NAPFD mainly focused on TCP, except for S29, which aimed to enhance mutant test precision within their MuDelta approach.

\par Several studies reported additional performance metrics, including MAE, RMSE, pairwise t-tests, PP, NRPA, MCC, TTF, and G-Measure. Nine studies used simple rate values like misclassification rates and failure detection rates. Combining these simple rate values with more comprehensive measures like F-score and Accuracy is recommended for a more thorough assessment of results. This strategy aids in preventing resource misallocation, as relying solely on the evaluation of results based on True Positive (failing tests) values could lead to the execution of numerous small failing test cases, thereby increasing resource consumption, particularly by resource-intensive passing test cases.

\subsubsection{Connection Between CI Tasks and Evaluation Metrics}

\begin{table*}
\caption{Heat map displays the percentage distribution of evaluation metrics across CI tasks. Study IDs are omitted to prevent table dimension expansion. color intensity corresponds to percentages, with lighter shades indicating lower and darker blues representing higher percentages. The bolded numbers represent the count of studies and their percentage in each CI task.}
\label{Table:CITasksandMeasure}
\resizebox{\textwidth}{!}{%
\begin{tabular}{l|l|c|ccccccccccccccccc}
\hline
\rotN{\textbf{CI Phases}} &
\multicolumn{1}{c|}{\rotN{\textbf{CI Tasks}}} &
  \rotN{\Centerstack{\textbf{Num of}\\\textbf{papers}}} &
  \rotN{\textbf{Recall}} &
  \rotN{\textbf{Precision}} &
  \rotN{\textbf{F1-Score}} &
  \rotN{\textbf{APFD}} &
  \rotN{\textbf{NAPFD}} &
  \rotN{\textbf{Accuracy}} &
  \rotN{\textbf{AUC}} &
  \rotN{\textbf{MCC}} &
  \rotN{\textbf{TTF}} &
  \rotN{\textbf{MAE}} &
  \rotN{\textbf{RMSE}} &
  \rotN{\textbf{NRPA}} &
  \rotN{\textbf{T-test}} &
  \rotN{\textbf{PP}} &
  \rotN{\textbf{G-Measure}} &
  \rotN{\textbf{Time Cost}} &
  \rotN{\textbf{Simple}} \\
  \hline
  \textbf{UT} &
\Centerstack[l]{\textbf{Unit Test}\\\textbf{Prediction}} &
    1 &
  \cellcolor{3F80BB}\Centerstack{\textbf{1}\\\textbf{100\%}} &
  \cellcolor{3F80BB}\Centerstack{\textbf{1}\\\textbf{100\%}} &
  \cellcolor{3F80BB}\Centerstack{\textbf{1}\\\textbf{100\%}} &
  \cellcolor{DDEBF7}-- &
  \cellcolor{DDEBF7}-- &
  \cellcolor{3F80BB}\Centerstack{\textbf{1}\\\textbf{100\%}} &
  \cellcolor{DDEBF7}-- &
  \cellcolor{DDEBF7}-- &
  \cellcolor{DDEBF7}-- &
  \cellcolor{DDEBF7}-- &
  \cellcolor{DDEBF7}-- &
  \cellcolor{DDEBF7}-- &
  \cellcolor{DDEBF7}-- &
  \cellcolor{DDEBF7}-- &
  \cellcolor{DDEBF7}-- &
  \cellcolor{DDEBF7}-- &
  \cellcolor{DDEBF7}-- \\\cdashline{1-20}
\multirow{2}{*}{\textbf{IT}} &
\Centerstack[l]{\textbf{Branch Coverage}\\\textbf{Prediction}} &
    2&
  \cellcolor{DDEBF7}-- &
  \cellcolor{DDEBF7}-- &
  \cellcolor{DDEBF7}-- &
  \cellcolor{DDEBF7}-- &
  \cellcolor{DDEBF7}-- &
  \cellcolor{DDEBF7}-- &
  \cellcolor{DDEBF7}-- &
  \cellcolor{DDEBF7}-- &
  \cellcolor{DDEBF7}-- &
  \cellcolor{3F80BB}\Centerstack{\textbf{2}\\\textbf{100\%}} &
  \cellcolor{DDEBF7}-- &
  \cellcolor{DDEBF7}-- &
  \cellcolor{DDEBF7}-- &
  \cellcolor{DDEBF7}-- &
  \cellcolor{DDEBF7}-- &
  \cellcolor{DDEBF7}-- &
  \cellcolor{DDEBF7}-- \\\cdashline{2-20}
&
\Centerstack[l]{\textbf{Integration Test}\\\textbf{Prediction}} &
    2&
  \cellcolor{3F80BB}\Centerstack{\textbf{2}\\\textbf{100\%}} &
  \cellcolor{3F80BB}\Centerstack{\textbf{2}\\\textbf{100\%}} &
  \cellcolor{86B0D6}\Centerstack{\textbf{1}\\\textbf{50.0\%}} &
  \cellcolor{DDEBF7}-- &
  \cellcolor{DDEBF7}-- &
  \cellcolor{DDEBF7}-- &
  \cellcolor{86B0D6}\Centerstack{\textbf{1}\\\textbf{50.0\%}} &
  \cellcolor{DDEBF7}-- &
  \cellcolor{DDEBF7}-- &
  \cellcolor{DDEBF7}-- &
  \cellcolor{DDEBF7}-- &
  \cellcolor{DDEBF7}-- &
  \cellcolor{DDEBF7}-- &
  \cellcolor{DDEBF7}-- &
  \cellcolor{DDEBF7}-- &
  \cellcolor{DDEBF7}-- &
  \cellcolor{DDEBF7}-- \\\cdashline{1-20}
\multirow{3}{*}{\textbf{RT}} &
\Centerstack[l]{\textbf{Test}\\\textbf{Optimization}} &
    26&
  \cellcolor{80ACD4}\Centerstack{\textbf{14}\\\textbf{53.8\%}} &
  \cellcolor{A8C7E3}\Centerstack{\textbf{8}\\\textbf{30.8\%}} &
  \cellcolor{B5D0E8}\Centerstack{\textbf{6}\\\textbf{23.1\%}} &
  \cellcolor{79A7D1}\Centerstack{\textbf{15}\\\textbf{57.7\%}} &
  \cellcolor{9BBEDE}\Centerstack{\textbf{10}\\\textbf{38.5\%}} &
  \cellcolor{B5D0E8}\Centerstack{\textbf{6}\\\textbf{23.1\%}} &
  \cellcolor{D0E2F2}\Centerstack{\textbf{2}\\\textbf{7.7\%}} &
  \cellcolor{D7E7F5}\Centerstack{\textbf{1}\\\textbf{3.8\%}} &
  \cellcolor{C9DEF0}\Centerstack{\textbf{3}\\\textbf{11.5\%}} &
  \cellcolor{DDEBF7}-- &
  \cellcolor{D0E2F2}\Centerstack{\textbf{2}\\\textbf{7.7\%}} &
  \cellcolor{DDEBF7}-- &
  \cellcolor{DDEBF7}-- &
  \cellcolor{DDEBF7}-- &
  \cellcolor{DDEBF7}-- &
  \cellcolor{A1C3E1}\Centerstack{\textbf{9}\\\textbf{34.6\%}} &
  \cellcolor{C3D9ED}\Centerstack{\textbf{4}\\\textbf{15.4\%}} \\\cdashline{2-20}
 &
\Centerstack[l]{\textbf{Defect}\\\textbf{Prediction}} &
    4&
  \cellcolor{5B93C6}\Centerstack{\textbf{3}\\\textbf{75\%}} &
  \cellcolor{86B0D6}\Centerstack{\textbf{2}\\\textbf{50\%}} &
  \cellcolor{86B0D6}\Centerstack{\textbf{2}\\\textbf{50\%}} &
  \cellcolor{DDEBF7}-- &
  \cellcolor{DDEBF7}-- &
  \cellcolor{B2CEE7}\Centerstack{\textbf{1}\\\textbf{25\%}} &
  \cellcolor{B2CEE7}\Centerstack{\textbf{1}\\\textbf{25\%}} &
  \cellcolor{B2CEE7}\Centerstack{\textbf{1}\\\textbf{25\%}} &
  \cellcolor{DDEBF7}-- &
  \cellcolor{DDEBF7}-- &
  \cellcolor{DDEBF7}-- &
  \cellcolor{DDEBF7}-- &
  \cellcolor{DDEBF7}-- &
  \cellcolor{B2CEE7}\Centerstack{\textbf{1}\\\textbf{25\%}} &
  \cellcolor{B2CEE7}\Centerstack{\textbf{1}\\\textbf{25\%}} &
  \cellcolor{DDEBF7}-- &
  \cellcolor{DDEBF7}--  \\\cdashline{2-20}
 &
\Centerstack[l]{\textbf{Flaky Test}\\\textbf{Detection}} &
    1&
  \cellcolor{3F80BB}\Centerstack{\textbf{1}\\\textbf{100\%}} &
  \cellcolor{3F80BB}\Centerstack{\textbf{1}\\\textbf{100\%}} &
  \cellcolor{3F80BB}\Centerstack{\textbf{1}\\\textbf{100\%}} &
  \cellcolor{DDEBF7}-- &
  \cellcolor{DDEBF7}-- &
  \cellcolor{DDEBF7}-- &
  \cellcolor{DDEBF7}-- &
  \cellcolor{DDEBF7}-- &
  \cellcolor{DDEBF7}-- &
  \cellcolor{DDEBF7}-- &
  \cellcolor{DDEBF7}-- &
  \cellcolor{DDEBF7}-- &
  \cellcolor{DDEBF7}-- &
  \cellcolor{DDEBF7}-- &
  \cellcolor{DDEBF7}-- &
  \cellcolor{DDEBF7}-- &
  \cellcolor{3F80BB}\Centerstack{\textbf{1}\\\textbf{100\%}} \\\cdashline{1-20}
\textbf{BV} &
\Centerstack[l]{\textbf{Build}\\\textbf{Prediction}} &
    12&
  \cellcolor{86B0D6}\Centerstack{\textbf{6}\\\textbf{50\%}} &
  \cellcolor{78A7D1}\Centerstack{\textbf{7}\\\textbf{58.3\%}} &
  \cellcolor{78A7D1}\Centerstack{\textbf{7}\\\textbf{58.3\%}} &
  \cellcolor{DDEBF7}-- &
  \cellcolor{DDEBF7}-- &
  \cellcolor{CFE2F2}\Centerstack{\textbf{1}\\\textbf{8.3\%}} &
  \cellcolor{A3C4E1}\Centerstack{\textbf{4}\\\textbf{33.3\%}} &
  \cellcolor{CFE2F2}\Centerstack{\textbf{1}\\\textbf{8.3\%}} &
  \cellcolor{DDEBF7}-- &
  \cellcolor{DDEBF7}-- &
  \cellcolor{DDEBF7}-- &
  \cellcolor{DDEBF7}-- &
  \cellcolor{CFE2F2}\Centerstack{\textbf{1}\\\textbf{8.3\%}} &
  \cellcolor{DDEBF7}-- &
  \cellcolor{DDEBF7}-- &
  \cellcolor{DDEBF7}-- &
  \cellcolor{A3C4E1}\Centerstack{\textbf{4}\\\textbf{33.3\%}} 
  \\\cdashline{1-20}
\multirow{2}{*}{\textbf{ST}} &
\Centerstack[l]{\textbf{Installed Software}\\\textbf{Discovery}} &
    1&
  \cellcolor{3F80BB}\Centerstack{\textbf{1}\\\textbf{100\%}} &
  \cellcolor{3F80BB}\Centerstack{\textbf{1}\\\textbf{100\%}} &
  \cellcolor{3F80BB}\Centerstack{\textbf{1}\\\textbf{100\%}} &
  \cellcolor{DDEBF7}-- &
  \cellcolor{DDEBF7}-- &
  \cellcolor{DDEBF7}-- &
  \cellcolor{DDEBF7}-- &
  \cellcolor{DDEBF7}-- &
  \cellcolor{DDEBF7}-- &
  \cellcolor{DDEBF7}-- &
  \cellcolor{DDEBF7}-- &
  \cellcolor{DDEBF7}-- &
  \cellcolor{DDEBF7}-- &
  \cellcolor{DDEBF7}-- &
  \cellcolor{DDEBF7}-- &
  \cellcolor{DDEBF7}-- &
  \cellcolor{DDEBF7}-- \\\cdashline{2-20}
 &
\Centerstack[l]{\textbf{Performance Test}\\\textbf{Optimization}} &
    1&
  \cellcolor{DDEBF7}-- &
  \cellcolor{DDEBF7}-- &
  \cellcolor{DDEBF7}-- &
  \cellcolor{DDEBF7}-- &
  \cellcolor{DDEBF7}-- &
  \cellcolor{DDEBF7}-- &
  \cellcolor{DDEBF7}-- &
  \cellcolor{DDEBF7}-- &
  \cellcolor{DDEBF7}-- &
  \cellcolor{DDEBF7}-- &
  \cellcolor{DDEBF7}-- &
  \cellcolor{DDEBF7}-- &
  \cellcolor{DDEBF7}-- &
  \cellcolor{DDEBF7}-- &
  \cellcolor{DDEBF7}-- &
  \cellcolor{DDEBF7}-- &
  \cellcolor{3F80BB}\Centerstack{\textbf{1}\\\textbf{100\%}} \\\cdashline{1-20}
\textbf{PM} &
\Centerstack[l]{\textbf{Activity} \\\textbf{Management}} &
    2&
  \cellcolor{3F80BB}\Centerstack{\textbf{2}\\\textbf{100\%}} &
  \cellcolor{3F80BB}\Centerstack{\textbf{2}\\\textbf{100\%}} &
  \cellcolor{3F80BB}\Centerstack{\textbf{2}\\\textbf{100\%}} &
  \cellcolor{DDEBF7}-- &
  \cellcolor{DDEBF7}-- &
  \cellcolor{DDEBF7}-- &
  \cellcolor{86B0D6}\Centerstack{\textbf{1}\\\textbf{50\%}} &
  \cellcolor{DDEBF7}-- &
  \cellcolor{DDEBF7}-- &
  \cellcolor{DDEBF7}-- &
  \cellcolor{DDEBF7}-- &
  \cellcolor{DDEBF7}-- &
  \cellcolor{DDEBF7}-- &
  \cellcolor{DDEBF7}-- &
  \cellcolor{DDEBF7}-- &
  \cellcolor{DDEBF7}-- &
  \cellcolor{DDEBF7}-- \\ \hline
\end{tabular}%
}
\end{table*}

\par The selection of evaluation metrics to assess tasks' performance, depends on the objectives of the tasks. Table~\ref{Table:CITasksandMeasure} illustrates the correlation between CI tasks and the metrics applied in the selected studies. Recall, Precision, and F1-Score emerge as the most frequently employed metrics across various tasks. Notably, accuracy is often avoided in CI environments due to data imbalances, which can introduce bias into the results~\cite{mortaz2020imbalance}.

\par Furthermore, APFD and NAPFD metrics find primary application in Test Optimization solutions, where ML methods aim to optimize the execution order of test cases. These metrics are specifically designed for the evaluation of test case prioritization and selection techniques~\cite{mor2014evaluate}.

\par AUC measurements play a crucial role in various CI tasks, offering robust performance assessment in imbalanced datasets. Their application is vital for addressing data imbalances, ensuring fairness, and enhancing the effectiveness of ML-based methodologies in such datasets~\cite{yang2021learning}.

\par Additionally, other metrics are sporadically used across the literature. MAE, for instance, is consistently employed in studies related to the Branch Coverage Prediction task. This metric, measuring the average error between predicted and actual values, provides a more intuitive understanding compared to other error measures like MSE or RMSE, as it utilizes absolute values instead of squared values~\cite{hodson2022root}.

\par While evaluating ML methods' performance, authors often neglected reporting associated time and resource costs and the reduced costs in real-world projects. However, S41 stood out by emphasizing time-related metrics and introducing Normalized Time Reduction (NTR) and Prioritization Time (PT). NTR measures time saved by using ML models to detect the first failure instead of executing all tests, while PT denotes the time required for ML models to prioritize test cases.

\par Table~\ref{Table:CITasksandMeasure} illustrates that classification methods find application across all steps of CI, notably in regression testing and build validation. Given the structural similarities between regression testing and unit or integration testing, the underutilization of classification methods in these steps highlights a significant gap in the literature.

\par Additionally, Table~\ref{Table:CITasksandMeasure} indicates that RL methods are predominantly utilized in RT. Despite their suitability for continuous adjustment and analysis of high-volume data in RT~\cite{yang2021adaptive}, RL methods hold the potential for providing just-in-time predictions in other CI steps.

\par Moreover, Table~\ref{Table:CITasksandMeasure} reveals a notable neglect of various ML types across different CI steps. This underscores an important opportunity for researchers to explore and assess the performance of ML models in these overlooked areas of CI.

\vspace{1pt}
\noindent\fcolorbox{black}{blue!10}{%
\begin{minipage}{\columnwidth}
\vspace{1pt}
\textbf{Summary:}
\vspace{1pt}
\par $\bullet$ The sorted data selection method in the evaluation phase provides higher reliability than random selection, simulating real data streams in CI environments.
\vspace{1pt}

\par $\bullet$ Researchers should choose K-fold types based on specific ML method requirements, considering the pros and cons of each type and the characteristics of CI data.
\vspace{1pt}

\par $\bullet$ F1-score, Precision, and Recall, detailed in Table~\ref{Table:PerfMeas}, are frequently employed performance metrics, aiding comparisons with other ML-based CI solutions.
\vspace{1pt}

\par $\bullet$ RL and consequently Gradual Evaluation methods are commonly employed in CI, primarily because of their capability for continuous evaluation.
\vspace{1pt}


\par $\bullet$ Designed metrics like APFD and NAPFD for Test Optimization and MAE for Branch Coverage Prediction underscore the importance of selecting appropriate metrics for assessing ML-based CI solution performance.

\end{minipage}}


\section{Limitations}
\label{sec:limit}
\par While adhering to the guidelines outlined by Kitchenham, and Charters~\cite{kitchenham2007guidelines}, and Braun and Clarke~\cite{braun2006using}, and taking into account the documented threats in systematic literature reviews (SLRs) within the software engineering domain~\cite{ampatzoglou2020guidelines, zhou2016map}, efforts have been made to minimize potential defects. However, despite these efforts, there remain persistent threats to the validity of this SLR.

\par One such threat relates to the completeness and inclusiveness of relevant studies. We conducted a preliminary search to identify papers already known to us that have been published. However, it's important to note that our coverage may be impacted by potential inconsistencies in terminology found within paper titles and abstracts.

\par We executed our search string on ACM, IEEE, and Scopus indexing systems, renowned for their comprehensive coverage in software engineering and computer science, including well-known venues in these fields~\cite{kitchenham2010systematic}. To address potential omissions, we implemented backward and forward snowballing techniques, manually exploring references of selected studies and scrutinizing cited papers. However, limitations may still exist with this approach.

\par Another validity threat may arise from potential bias in the selection, analysis, and synthesis of data by researchers. To mitigate this threat at every step, multiple researchers were involved, and all steps were supervised by the fourth author.

\par We presented the CI pipeline and its six identified phases based on reviewing the selected studies in Figure~\ref{Fig:CIPhases}. However, the risk remains that uninvestigated CI phases could have been overlooked.
 

\section{Discussion}
\label{sec:Disc}
\par This study primarily focused on breaking down CI phases into automatable tasks optimized by ML and providing detailed insights into ML method preparation in this emerging domain. The rapid software release cycle associated with CI has sparked substantial interest among software companies and researchers, driving the need for efficient CI practices.

\par Our study focused on the development steps of ML methods, to assist researchers and practitioners in handling the ever-increasing volume of data in the CI field by reviewing the published studies over the last two decades. The surge in research studies on fault prediction within CI necessitates more structured research to achieve reliable and comprehensive results. The paper discusses key assumptions and areas for future research.

\par In this study, we conducted a systematic literature review (SLR) to give a comprehensive overview of the use of ML approaches in the context of CI. Our findings focused on numerous elements of the utilization of ML in CI, such as the phases, methodologies, performance indicators, and data sources that are often used. The CI process, with its continuous and quick software release cycle, has presented unique problems to software developers and companies. As a result, researchers are increasingly turning to ML techniques to improve CI procedures. Our evaluation identified major trends and gaps in the existing literature, which we will now examine in depth to provide insights into future research areas and practical implications.

\par \textbf{Usage of Large Language Models:} Despite high expectations regarding Large Language Models (LLMs), our analysis revealed a significant research gap in this area within the CI domain. Future research avenues might include the generation of new test cases for defect detection, leveraging various data types and historical features. Researchers could also explore the identification of test cases closely related to software requirements. Consideration of automation, agility, and code commit frequency is crucial in CI research to align with CI's core principles. It is noteworthy that our initial pilot study employed the search query ``\textit{((`Large Language Model' OR `LLM') AND `Continuous Integration')}'', which, regrettably, did not return any papers suitable for inclusion in the final list of the selected papers

\par \textbf{Security of Proposed Methods:} Most studies reviewed here overlooked the security aspect of their ML methods except S12. ML models are vulnerable to causative and exploratory attacks, particularly in open-source projects~\cite{wang2019security}. Researchers should evaluate their methods' security, especially when employing Neural Network (NN) techniques, which lack interpretability. Comparative evaluations that account for security concerns are essential in addressing these identified gaps.

\par \textbf{Performance metrics:} Our analysis of Research Question 5 (RQ5) in Section~\ref{Sec:MethodEval} reveals diverse performance measures employed across the studies. In particular, 12 studies (S3, S4, S6, S7, S15, S22, S27, S36, S42, S43, S44 and S52) solely utilized a single metric, indicating the importance of adopting a comprehensive set of metrics like precision, recall, F1-score, and accuracy for classification methods, RMSE for regression methods, and APFD and NAPFD for test case prioritization and selection tasks. Researchers are encouraged to report multiple metrics to enable better comparisons and informed decisions, alongside qualitative evaluations to enhance method comprehension. Given that the primary objective of automating CI tasks is to enhance the overall efficiency of software developers, it becomes necessary for researchers to not only rely on the metrics provided in Table~\ref{Table:CITasksandMeasure} for assessing the performance of ML models but also to embark on the development of novel metrics designed to quantify the changes in developers' performance.

\par \textbf{Cost benchmark:} We emphasize the significance of reporting benchmark performance measures in addition to performance metrics to assess the efficiency of ML models. Metrics like time to train and perform provide insights into model effectiveness. By integrating both types of measures, studies can perform cost-benefit analyses. Future research can consider metrics like Average Percentage of Fault Detected with Cost ($APFD_{c}$), accounting for hardware costs, for more thorough evaluations. The $APFD_{c}$ metric is calculated using equation (\ref{Eq:APFDC}), where $n$ is the number of test cases in test suite $T$, $t_{i}$ is the cost of executing test case $i$, $f_{i}$ is the severity of faults, and $TF_{i}$ is the first test case that reveals fault $i$.

\begin{equation}
\label{Eq:APFDC}
\centering
APFD_{c} = \frac{\sum_{i=1}^m(f_{i} \times (\sum_{j=TF_{i}}^n t_{j}-\frac{1}{2}t_{TF_{i}}))}{\sum_{i=1}^n t_{i} \times \sum_{i=1}^m f_{i}}
\end{equation}

\par \textbf{Diversity of data:} Diverse input data is critical, as it increases the generality of solutions~\cite{koroglu2016defect}. Studies should explore different datasets with various sizes and characteristics. Data sources like version control and issue-tracking tools or using user requirement files automating the test designing issues are valuable for ML-based solutions in CI. Evaluating ML methods on a range of projects enhances their generalizability.

\par \textbf{New strategies for solutions:} Researchers can enhance their proposed methods by exploring innovative strategies, such as hybrid models that optimize performance for different data segments. Notably, there is an underutilization of unsupervised ML methods, emphasizing the need to consider the learning type distribution in ML algorithms. Novelty can be introduced through solution refinements, like defining new reward functions for reinforcement learning methods (e.g., S33) or following the divide-and-conquer solutions (e.g. separate models for predicting pass and fail test cases).

\par \textbf{Human interactions:} ML methods significantly improve CI tasks, but human involvement remains crucial, as evidenced by 30 out of 34 selected papers in this study from 2020 to 2023 employing supervised ML algorithms. Human effort cannot be entirely replaced, making professional guidance and expertise vital in ML-based solutions for validating the ML models, guiding the data analyzer in the training phase, and helping with feature extraction and data collection.

\par \textbf{Personalized solutions:} Out of the 52 reviewed studies, only 5 addressed individual features for modeling (see Table~\ref{Table:Tasks_FeatureTypes}). However, these models cannot be considered personalized as they focus on individual factors within a comparative context, revealing a significant gap in truly personalized ML-based methods for CI tasks.

\par \textbf{Data drifting challenge:} Agile CI environments introduce data drifting challenges, where new data may differ from training data, affecting various phases of ML model development~\cite{ackerman2020detection}. While most selected studies considered concept drift, data drifting received little attention, except in studies S4 and S36. These studies mitigated data drifting by retraining ML models at specific intervals in the CI cycle, a crucial consideration for industry partners in determining retraining frequencies.


\section{Conclusion}
\label{sec:conc}
\par In this paper, we conducted a systematic literature review (SLR) to investigate the current state of the application of ML methods in CI. We analyzed a total of 52 primary studies published between 2000 and 2023 and examined the use of ML methods and their properties in this context. While many studies have reported successful results, a few studies suffer from imperfections in certain dimensions and lack detailed information about the presented solutions. These issues can be considered significant research gaps in the application of ML methods in CI. 
\par In summary, our work comprises the following findings:
\par \textbf{(1)} A comprehensive depiction of the CI pipeline, which begins with the code commitment of the developer and progresses through various testing phases until the final release of a product that is ready for customers.
\par \textbf{(2)} An examination of different aspects of data engineering, including data sources, data types, and data preparation.
\par \textbf{(3)} Utilizing thematic analysis, we classify four types of features used in ML-based solutions for CI, along with their subclasses, and examine five feature engineering techniques employed in the selected studies.
\par \textbf{(4)} Presentation of statistics related to the ML techniques implemented and their association with different phases of the CI pipeline, as well as an investigation into the methods of hyper-parameter tuning.
\par \textbf{(5)} Finally, in this study, we provide a thorough description of the evaluation methods and metrics that were commonly used in the primary studies we selected. By demonstrating the relationship between the evaluation methods and ML algorithms used in these studies, our work enables researchers to select the most suitable evaluation methods when comparing their findings with those of other studies in the literature.

\par To summarize, this SLR has presented an overview of ML-based solutions for improving the CI pipeline concerning speed and resource consumption. However, we believe that further research is necessary for various stages of the CI cycle, as discussed in the Discussion section (see section~\ref{sec:Disc}). Given that a considerable proportion of the studies we reviewed were conducted in industrial settings, we encourage practitioners to adapt and implement the proposed approaches and solutions in actual CI environments. Furthermore, to make additional progress in this research domain, we suggest the use of standardized research methods and more general approaches, as well as consideration of the security aspects of the studies presented.

\par In the future, we plan to update this SLR with new studies and extend our research scope to other libraries.

\section*{Declaration of competing interest}
The authors declare that they have no known competing financial interests or personal relationships that could have appeared to influence the work reported in this paper.

\section*{Acknowledgement}
We acknowledge the contribution of Dr Zohaib Md. Jan and Roshan Namal Rajapakse during the first phase of collecting and analyzing data in this study.

\bibliographystyle{unsrt}  
\bibliography{references}

\begin{thebibliography}{10}

\bibitem{shahin2017continuous}
Mojtaba Shahin, Muhammad~Ali Babar, and Liming Zhu.
\newblock Continuous integration, delivery and deployment: a systematic review on approaches, tools, challenges and practices.
\newblock {\em IEEE Access}, 5:3909--3943, 2017.

\bibitem{jin2020cost}
Xianhao Jin and Francisco Servant.
\newblock A cost-efficient approach to building in continuous integration.
\newblock In {\em 2020 IEEE/ACM 42nd International Conference on Software Engineering (ICSE)}, pages 13--25. IEEE, 2020.

\bibitem{xie2018cutting}
Zheng Xie and Ming Li.
\newblock Cutting the software building efforts in continuous integration by semi-supervised online auc optimization.
\newblock In {\em Proceedings of the 27th International Joint Conference on Artificial Intelligence}, pages 2875--2881, 2018.

\bibitem{mamata2022failure}
Rezwana Mamata, Kevin Smith, Akramul Azim, Yee-Kang Chang, Qasim Taiseef, Ramiro Liscano, and Gkerta Seferi.
\newblock Failure prediction using transfer learning in large-scale continuous integration environments.
\newblock In {\em Proceedings of the 32nd Annual International Conference on Computer Science and Software Engineering}, pages 193--198, 2022.

\bibitem{hilton2016usage}
Michael Hilton, Timothy Tunnell, Kai Huang, Darko Marinov, and Danny Dig.
\newblock Usage, costs, and benefits of continuous integration in open-source projects.
\newblock In {\em Proceedings of the 31st IEEE/ACM international conference on automated software engineering}, pages 426--437, 2016.

\bibitem{abdelkarim2022tcp}
Mohamed Abdelkarim and Reem ElAdawi.
\newblock Tcp-net: Test case prioritization using end-to-end deep neural networks.
\newblock In {\em 2022 IEEE International Conference on Software Testing, Verification and Validation Workshops (ICSTW)}, pages 122--129. IEEE, 2022.

\bibitem{figalist2019supporting}
Iris Figalist, Andreas Biesdorf, Christoph Brand, Sebastian Feld, and Marie Kiermeier.
\newblock Supporting the devops feedback loop using unsupervised machine learning.
\newblock In {\em 2019 IEEE International Symposium on INnovations in Intelligent SysTems and Applications (INISTA)}, pages 1--6. IEEE, 2019.

\bibitem{shafiq2021literature}
Saad Shafiq, Atif Mashkoor, Christoph Mayr-Dorn, and Alexander Egyed.
\newblock A literature review of machine learning and software development life cycle stages.
\newblock {\em IEEE Access}, 2021.

\bibitem{lee2019classifying}
Seongmin Lee, Shin Hong, Jungbae Yi, Taeksu Kim, Chul-Joo Kim, and Shin Yoo.
\newblock Classifying false positive static checker alarms in continuous integration using convolutional neural networks.
\newblock In {\em 2019 12th IEEE Conference on Software Testing, Validation and Verification (ICST)}, pages 391--401. IEEE, 2019.

\bibitem{yang2021sparse}
Yang Yang, Zheng Li, Ying Shang, and Qianyu Li.
\newblock Sparse reward for reinforcement learning-based continuous integration testing.
\newblock {\em Journal of Software: Evolution and Process}, page e2409, 2021.

\bibitem{brandtner2015sqa}
Martin Brandtner, Sebastian~C M{\"u}ller, Philipp Leitner, and Harald~C Gall.
\newblock Sqa-profiles: Rule-based activity profiles for continuous integration environments.
\newblock In {\em 2015 IEEE 22nd International Conference on Software Analysis, Evolution, and Reengineering (SANER)}, pages 301--310. IEEE, 2015.

\bibitem{arani2023sok}
Ali~Kazemi Arani, Mansooreh Zahedi, Triet Huynh~Minh Le, and Muhammad~Ali Babar.
\newblock Sok: Machine learning for continuous integration.
\newblock In {\em 2023 IEEE/ACM International Workshop on Cloud Intelligence \& AIOps (AIOps)}, pages 8--13. IEEE, 2023.

\bibitem{fitzgerald2017continuous}
Brian Fitzgerald and Klaas-Jan Stol.
\newblock Continuous software engineering: A roadmap and agenda.
\newblock {\em Journal of Systems and Software}, 123:176--189, 2017.

\bibitem{debois2011devops}
Patrick Debois et~al.
\newblock Devops: A software revolution in the making.
\newblock {\em Journal of Information Technology Management}, 24(8):3--39, 2011.

\bibitem{mitchell2006discipline}
Tom~Michael Mitchell.
\newblock {\em The discipline of machine learning}, volume~9.
\newblock Carnegie Mellon University, School of Computer Science, Machine Learning~…, 2006.

\bibitem{bacstanlar2014introduction}
Yalin Ba{\c{s}}tanlar and Mustafa {\"O}zuysal.
\newblock Introduction to machine learning.
\newblock In {\em miRNomics: MicroRNA Biology and Computational Analysis}, pages 105--128. Springer, 2014.

\bibitem{russell2009artificial}
J~Russell~Stuart and Peter Norvig.
\newblock {\em Artificial intelligence: a modern approach}.
\newblock Prentice Hall, 2009.

\bibitem{andrew1998reinforcement}
Alex~M Andrew.
\newblock Reinforcement learning.
\newblock {\em Kybernetes}, 1998.

\bibitem{hassan2017change}
Foyzul Hassan and Xiaoyin Wang.
\newblock Change-aware build prediction model for stall avoidance in continuous integration.
\newblock In {\em 2017 ACM/IEEE International Symposium on Empirical Software Engineering and Measurement (ESEM)}, pages 157--162. IEEE, 2017.

\bibitem{pan2022test}
Rongqi Pan, Mojtaba Bagherzadeh, Taher~A Ghaleb, and Lionel Briand.
\newblock Test case selection and prioritization using machine learning: a systematic literature review.
\newblock {\em Empirical Software Engineering}, 27(2):1--43, 2022.

\bibitem{durelli2019machine}
Vinicius~HS Durelli, Rafael~S Durelli, Simone~S Borges, Andre~T Endo, Marcelo~M Eler, Diego~RC Dias, and Marcelo~P Guimar{\~a}es.
\newblock Machine learning applied to software testing: A systematic mapping study.
\newblock {\em IEEE Transactions on Reliability}, 68(3):1189--1212, 2019.

\bibitem{zhang2003machine}
Du~Zhang and Jeffrey~JP Tsai.
\newblock Machine learning and software engineering.
\newblock {\em Software Quality Journal}, 11(2):87--119, 2003.

\bibitem{ali2019systematic}
Asad Ali and Carmine Gravino.
\newblock A systematic literature review of software effort prediction using machine learning methods.
\newblock {\em Journal of Software: Evolution and Process}, 31(10):e2211, 2019.

\bibitem{gada2021automated}
Mihir Gada, Zenil Haria, Arnav Mankad, Kaustubh Damania, and Smita Sankhe.
\newblock Automated feature engineering and hyperparameter optimization for machine learning.
\newblock In {\em 2021 7th International Conference on Advanced Computing and Communication Systems (ICACCS)}, volume~1, pages 981--986. IEEE, 2021.

\bibitem{zhang2003data}
Shichao Zhang, Chengqi Zhang, and Qiang Yang.
\newblock Data preparation for data mining.
\newblock {\em Applied artificial intelligence}, 17(5-6):375--381, 2003.

\bibitem{khurana2016automating}
Udayan Khurana, Fatemeh Nargesian, Horst Samulowitz, Elias Khalil, and Deepak Turaga.
\newblock Automating feature engineering.
\newblock {\em Transformation}, 10(10):10, 2016.

\bibitem{alpaydin2020introduction}
Ethem Alpaydin.
\newblock {\em Introduction to machine learning}.
\newblock MIT press, 2020.

\bibitem{kitchenham2007guidelines}
Barbara~A Kitchenham and Stuart Charters.
\newblock Guidelines for performing systematic literature reviews in software engineering technical report.
\newblock {\em Software Engineering Group, EBSE Technical Report, Keele University and Department of Computer Science University of Durham}, 2, 2007.

\bibitem{kitchenham2010systematic}
Barbara Kitchenham, Rialette Pretorius, David Budgen, O~Pearl Brereton, Mark Turner, Mahmood Niazi, and Stephen Linkman.
\newblock Systematic literature reviews in software engineering--a tertiary study.
\newblock {\em Information and software technology}, 52(8):792--805, 2010.

\bibitem{van2020crop}
Thomas Van~Klompenburg, Ayalew Kassahun, and Cagatay Catal.
\newblock Crop yield prediction using machine learning: A systematic literature review.
\newblock {\em Computers and Electronics in Agriculture}, 177:105709, 2020.

\bibitem{wang2022machine}
Simin Wang, Liguo Huang, Amiao Gao, Jidong Ge, Tengfei Zhang, Haitao Feng, Ishna Satyarth, Ming Li, He~Zhang, and Vincent Ng.
\newblock Machine/deep learning for software engineering: A systematic literature review.
\newblock {\em IEEE Transactions on Software Engineering}, 49(3):1188--1231, 2022.

\bibitem{malhotra2015systematic}
Ruchika Malhotra.
\newblock A systematic review of machine learning techniques for software fault prediction.
\newblock {\em Applied Soft Computing}, 27:504--518, 2015.

\bibitem{wohlin2014guidelines}
Claes Wohlin.
\newblock Guidelines for snowballing in systematic literature studies and a replication in software engineering.
\newblock In {\em Proceedings of the 18th international conference on evaluation and assessment in software engineering}, pages 1--10, 2014.

\bibitem{braun2006using}
Virginia Braun and Victoria Clarke.
\newblock Using thematic analysis in psychology.
\newblock {\em Qualitative research in psychology}, 3(2):77--101, 2006.

\bibitem{cruzes2011recommended}
Daniela~S Cruzes and Tore Dyba.
\newblock Recommended steps for thematic synthesis in software engineering.
\newblock In {\em 2011 international symposium on empirical software engineering and measurement}, pages 275--284. IEEE, 2011.

\bibitem{hassan2006using}
Ahmed~E Hassan and Ken Zhang.
\newblock Using decision trees to predict the certification result of a build.
\newblock In {\em 21st IEEE/ACM International Conference on Automated Software Engineering (ASE'06)}, pages 189--198. IEEE, 2006.

\bibitem{philip2019fastlane}
Adithya~Abraham Philip, Ranjita Bhagwan, Rahul Kumar, Chandra~Sekhar Maddila, and Nachiappan Nagppan.
\newblock Fastlane: test minimization for rapidly deployed large-scale online services.
\newblock In {\em 2019 IEEE/ACM 41st International Conference on Software Engineering (ICSE)}, pages 408--418. IEEE, 2019.

\bibitem{firestone1987meaning}
William~A Firestone.
\newblock Meaning in method: The rhetoric of quantitative and qualitative research.
\newblock {\em Educational researcher}, 16(7):16--21, 1987.

\bibitem{dixon2005synthesising}
Mary Dixon-Woods, Shona Agarwal, David Jones, Bridget Young, and Alex Sutton.
\newblock Synthesising qualitative and quantitative evidence: a review of possible methods.
\newblock {\em Journal of health services research \& policy}, 10(1):45--53, 2005.

\bibitem{stolberg2009enabling}
Sean Stolberg.
\newblock Enabling agile testing through continuous integration.
\newblock In {\em 2009 agile conference}, pages 369--374. IEEE, 2009.

\bibitem{al2019predicting}
Khaled~Walid Al-Sabbagh, Miroslaw Staron, Regina Hebig, Wilhelm Meding, AK~Tarhan, and A~Coskun{\c{c}}ay.
\newblock Predicting test case verdicts using textual analysis of committed code churns.
\newblock In {\em IWSM-Mensura}, pages 138--153, 2019.

\bibitem{ali2019enhanced}
Sadia Ali, Yaser Hafeez, Shariq Hussain, and Shunkun Yang.
\newblock Enhanced regression testing technique for agile software development and continuous integration strategies.
\newblock {\em Software Quality Journal}, pages 1--27, 2019.

\bibitem{elsner2021empirically}
Daniel Elsner, Florian Hauer, Alexander Pretschner, and Silke Reimer.
\newblock Empirically evaluating readily available information for regression test optimization in continuous integration.
\newblock In {\em Proceedings of the 30th ACM SIGSOFT International Symposium on Software Testing and Analysis}, pages 491--504, 2021.

\bibitem{pan2020dynamic}
Chaoyue Pan, Yang Yang, Zheng Li, and Junxia Guo.
\newblock Dynamic time window based reward for reinforcement learning in continuous integration testing.
\newblock In {\em 12th Asia-Pacific Symposium on Internetware}, pages 189--198, 2020.

\bibitem{lima2020learning}
Jackson A~Prado Lima, Willian~DF Mendon{\c{c}}a, Silvia~R Vergilio, and Wesley~KG Assun{\c{c}}{\~a}o.
\newblock Learning-based prioritization of test cases in continuous integration of highly-configurable software.
\newblock In {\em Proceedings of the 24th ACM Conference on Systems and Software Product Line: Volume A-Volume A}, pages 1--11, 2020.

\bibitem{martins2021supervised}
Ricardo Martins, Rui Abreu, Manuel Lopes, and Jo{\~a}o Nadkarni.
\newblock Supervised learning for test suit selection in continuous integration.
\newblock In {\em 2021 IEEE International Conference on Software Testing, Verification and Validation Workshops (ICSTW)}, pages 239--246. IEEE, 2021.

\bibitem{finlay2014data}
Jacqui Finlay, Russel Pears, and Andy~M Connor.
\newblock Data stream mining for predicting software build outcomes using source code metrics.
\newblock {\em Information and Software Technology}, 56(2):183--198, 2014.

\bibitem{xia2017empirical}
Jing Xia, Yanhui Li, and Chuanqi Wang.
\newblock An empirical study on the cross-project predictability of continuous integration outcomes.
\newblock In {\em 2017 14th Web Information Systems and Applications Conference (WISA)}, pages 234--239. IEEE, 2017.

\bibitem{schwabacher2005survey}
Mark Schwabacher.
\newblock A survey of data-driven prognostics.
\newblock {\em Infotech@ Aerospace}, page 7002, 2005.

\bibitem{vassallo2019enabling}
Carmine Vassallo.
\newblock Enabling continuous improvement of a continuous integration process.
\newblock In {\em 2019 34th IEEE/ACM International Conference on Automated Software Engineering (ASE)}, pages 1246--1249. IEEE, 2019.

\bibitem{chen2022focus}
Fanliang Chen, Zheng Li, Ying Shang, and Yang Yang.
\newblock Focus on new test cases in continuous integration testing based on reinforcement learning.
\newblock In {\em 2022 IEEE 22nd International Conference on Software Quality, Reliability and Security (QRS)}, pages 830--841. IEEE, 2022.

\bibitem{kaur2019systematic}
Harsurinder Kaur, Husanbir~Singh Pannu, and Avleen~Kaur Malhi.
\newblock A systematic review on imbalanced data challenges in machine learning: Applications and solutions.
\newblock {\em ACM Computing Surveys (CSUR)}, 52(4):1--36, 2019.

\bibitem{sharif2021deeporder}
Aizaz Sharif, Dusica Marijan, and Marius Liaaen.
\newblock Deeporder: Deep learning for test case prioritization in continuous integration testing.
\newblock In {\em 2021 IEEE International Conference on Software Maintenance and Evolution (ICSME)}, pages 525--534. IEEE, 2021.

\bibitem{xiao2020lstm}
Lei Xiao, Huaikou Miao, Tingting Shi, and Yu~Hong.
\newblock Lstm-based deep learning for spatial--temporal software testing.
\newblock {\em DISTRIBUTED AND PARALLEL DATABASES}, 2020.

\bibitem{wu2019time}
Zhaolin Wu, Yang Yang, Zheng Li, and Ruilian Zhao.
\newblock A time window based reinforcement learning reward for test case prioritization in continuous integration.
\newblock In {\em Proceedings of the 11th Asia-Pacific Symposium on Internetware}, pages 1--6, 2019.

\bibitem{vig2018test}
Vidhi Vig and Arvinder Kaur.
\newblock Test effort estimation and prediction of traditional and rapid release models using machine learning algorithms.
\newblock {\em Journal of Intelligent \& Fuzzy Systems}, 35(2):1657--1669, 2018.

\bibitem{grano2018high}
Giovanni Grano, Timofey~V Titov, Sebastiano Panichella, and Harald~C Gall.
\newblock How high will it be? using machine learning models to predict branch coverage in automated testing.
\newblock In {\em 2018 IEEE Workshop on Machine Learning Techniques for Software Quality Evaluation (MaLTeSQuE)}, pages 19--24. IEEE, 2018.

\bibitem{chidamber1994metrics}
Shyam~R Chidamber and Chris~F Kemerer.
\newblock A metrics suite for object oriented design.
\newblock {\em IEEE Transactions on software engineering}, 20(6):476--493, 1994.

\bibitem{halstead1977elements}
Maurice~H Halstead.
\newblock {\em Elements of Software Science (Operating and programming systems series)}.
\newblock Elsevier Science Inc., 1977.

\bibitem{kim2002history}
Jung-Min Kim and Adam Porter.
\newblock A history-based test prioritization technique for regression testing in resource constrained environments.
\newblock In {\em Proceedings of the 24th international conference on software engineering}, pages 119--129, 2002.

\bibitem{haghighatkhah2018test}
Alireza Haghighatkhah, Mika M{\"a}ntyl{\"a}, Markku Oivo, and Pasi Kuvaja.
\newblock Test prioritization in continuous integration environments.
\newblock {\em Journal of Systems and Software}, 146:80--98, 2018.

\bibitem{bagherzadeh2020reinforcement}
Mojtaba Bagherzadeh, Nafiseh Kahani, and Lionel Briand.
\newblock Reinforcement learning for test case prioritization.
\newblock {\em arXiv preprint arXiv:2011.01834}, 2020.

\bibitem{koroglu2016defect}
Yavuz Koroglu, Alper Sen, Doruk Kutluay, Akin Bayraktar, Yalcin Tosun, Murat Cinar, and Hasan Kaya.
\newblock Defect prediction on a legacy industrial software: A case study on software with few defects.
\newblock In {\em 2016 IEEE/ACM 4th International Workshop on Conducting Empirical Studies in Industry (CESI)}, pages 14--20. IEEE, 2016.

\bibitem{al2020effect}
Khaled~Walid Al-Sabbagh, Regina Hebig, and Miroslaw Staron.
\newblock The effect of class noise on continuous test case selection: A controlled experiment on industrial data.
\newblock In {\em International Conference on Product-Focused Software Process Improvement}, pages 287--303. Springer, 2020.

\bibitem{treveil2020introducing}
Mark Treveil, Nicolas Omont, Cl{\'e}ment Stenac, Kenji Lefevre, Du~Phan, Joachim Zentici, Adrien Lavoillotte, Makoto Miyazaki, and Lynn Heidmann.
\newblock {\em Introducing MLOps}.
\newblock O'Reilly Media, 2020.

\bibitem{bishop2006pattern}
Christopher~M Bishop.
\newblock {\em Pattern recognition and machine learning}.
\newblock springer, 2006.

\bibitem{chapelle2009semi}
Olivier Chapelle, Bernhard Scholkopf, and Alexander Zien.
\newblock Semi-supervised learning (chapelle, o. et al., eds.; 2006)[book reviews].
\newblock {\em IEEE Transactions on Neural Networks}, 20(3):542--542, 2009.

\bibitem{ratner2019weak}
Alex Ratner, Stephen Bach, Paroma Varma, and Chris R{\'e}.
\newblock Weak supervision: the new programming paradigm for machine learning.
\newblock {\em Hazy Research. Available via https://dawn. cs. stanford. edu//2017/07/16/weak-supervision/. Accessed}, pages 05--09, 2019.

\bibitem{yaraghi2022scalable}
Ahmadreza~Saboor Yaraghi, Mojtaba Bagherzadeh, Nafiseh Kahani, and Lionel Briand.
\newblock Scalable and accurate test case prioritization in continuous integration contexts.
\newblock {\em IEEE Transactions on Software Engineering}, 2022.

\bibitem{porres2020automatic}
Ivan Porres, Tanwir Ahmad, Hergys Rexha, S{\'e}bastien Lafond, and Dragos Truscan.
\newblock Automatic exploratory performance testing using a discriminator neural network.
\newblock In {\em 2020 IEEE International Conference on Software Testing, Verification and Validation Workshops (ICSTW)}, pages 105--113. IEEE, 2020.

\bibitem{witten2002data}
Ian~H Witten and Eibe Frank.
\newblock Data mining: practical machine learning tools and techniques with java implementations.
\newblock {\em Acm Sigmod Record}, 31(1):76--77, 2002.

\bibitem{marijan2013test}
Dusica Marijan, Arnaud Gotlieb, and Sagar Sen.
\newblock Test case prioritization for continuous regression testing: An industrial case study.
\newblock In {\em 2013 IEEE International Conference on Software Maintenance}, pages 540--543. IEEE, 2013.

\bibitem{spieker2018reinforcement}
Helge Spieker, Arnaud Gotlieb, Dusica Marijan, and Morten Mossige.
\newblock Reinforcement learning for automatic test case prioritization and selection in continuous integration.
\newblock {\em arXiv preprint arXiv:1811.04122}, 2018.

\bibitem{mondal2019exploratory}
Amit~Kumar Mondal, Banani Roy, and Kevin~A Schneider.
\newblock An exploratory study on automatic architectural change analysis using natural language processing techniques.
\newblock In {\em 2019 19th International Working Conference on Source Code Analysis and Manipulation (SCAM)}, pages 62--73. IEEE, 2019.

\bibitem{bergstra2011algorithms}
James Bergstra, R{\'e}mi Bardenet, Yoshua Bengio, and Bal{\'a}zs K{\'e}gl.
\newblock Algorithms for hyper-parameter optimization.
\newblock {\em Advances in neural information processing systems}, 24, 2011.

\bibitem{eggensperger2013towards}
Katharina Eggensperger, Matthias Feurer, Frank Hutter, James Bergstra, Jasper Snoek, Holger Hoos, Kevin Leyton-Brown, et~al.
\newblock Towards an empirical foundation for assessing bayesian optimization of hyperparameters.
\newblock In {\em NIPS workshop on Bayesian Optimization in Theory and Practice}, volume~10, pages 1--5, 2013.

\bibitem{foo2007efficient}
Chuan-sheng Foo, Andrew Ng, et~al.
\newblock Efficient multiple hyperparameter learning for log-linear models.
\newblock {\em Advances in neural information processing systems}, 20, 2007.

\bibitem{mantovani2020rethinking}
Rafael~Gomes Mantovani, Andr{\'e} Luis~Debiaso Rossi, Edesio Alcoba{\c{c}}a, Jadson~Castro Gertrudes, Sylvio~Barbon Junior, and Andr{\'e} Carlos Ponce de Leon~Ferreira de~Carvalho.
\newblock Rethinking default values: a low cost and efficient strategy to define hyperparameters.
\newblock {\em arXiv preprint arXiv:2008.00025}, 2020.

\bibitem{yang2020hyperparameter}
Li~Yang and Abdallah Shami.
\newblock On hyperparameter optimization of machine learning algorithms: Theory and practice.
\newblock {\em Neurocomputing}, 415:295--316, 2020.

\bibitem{cheng2021towards}
Lei Cheng and Qingjiang Shi.
\newblock Towards overfitting avoidance: Tuning-free tensor-aided multi-user channel estimation for 3d massive mimo communications.
\newblock {\em IEEE Journal of Selected Topics in Signal Processing}, 15(3):832--846, 2021.

\bibitem{montesinos2022overfitting}
Osval~Antonio Montesinos~L{\'o}pez, Abelardo Montesinos~L{\'o}pez, and Jose Crossa.
\newblock Overfitting, model tuning, and evaluation of prediction performance.
\newblock In {\em Multivariate statistical machine learning methods for genomic prediction}, pages 109--139. Springer, 2022.

\bibitem{ashok2023remediating}
Sreeja Ashok, Sangeetha Ezhumalai, and Tanvi Patwa.
\newblock Remediating data drifts and re-establishing ml models.
\newblock {\em Procedia Computer Science}, 218:799--809, 2023.

\bibitem{nougnanke2022ml}
Benoit Nougnanke, Yann Labit, Marc Bruyere, Ulrich Aivodji, and Simone Ferlin.
\newblock Ml-based performance modeling in sdn-enabled data center networks.
\newblock {\em IEEE Transactions on Network and Service Management}, 20(1):815--829, 2022.

\bibitem{grano2019branch}
Giovanni Grano, Timofey~V Titov, Sebastiano Panichella, and Harald~C Gall.
\newblock Branch coverage prediction in automated testing.
\newblock {\em Journal of Software: Evolution and Process}, 31(9):e2158, 2019.

\bibitem{pan2021continuous}
Cong Pan and Michael Pradel.
\newblock Continuous test suite failure prediction.
\newblock In {\em Proceedings of the 30th ACM SIGSOFT International Symposium on Software Testing and Analysis}, pages 553--565, 2021.

\bibitem{chen2020buildfast}
Bihuan Chen, Linlin Chen, Chen Zhang, and Xin Peng.
\newblock Buildfast: History-aware build outcome prediction for fast feedback and reduced cost in continuous integration.
\newblock In {\em 2020 35th IEEE/ACM International Conference on Automated Software Engineering (ASE)}, pages 42--53. IEEE, 2020.

\bibitem{qu2007combinatorial}
Xiao Qu, Myra~B Cohen, and Katherine~M Woolf.
\newblock Combinatorial interaction regression testing: A study of test case generation and prioritization.
\newblock In {\em 2007 IEEE International Conference on Software Maintenance}, pages 255--264. IEEE, 2007.

\bibitem{mortaz2020imbalance}
Ebrahim Mortaz.
\newblock Imbalance accuracy metric for model selection in multi-class imbalance classification problems.
\newblock {\em Knowledge-Based Systems}, 210:106490, 2020.

\bibitem{mor2014evaluate}
Anil Mor.
\newblock Evaluate the effectiveness of test suite prioritization techniques using apfd metric.
\newblock {\em IOSR Journal of Computer}, 16(4):47--51, 2014.

\bibitem{yang2021learning}
Zhiyong Yang, Qianqian Xu, Shilong Bao, Xiaochun Cao, and Qingming Huang.
\newblock Learning with multiclass auc: Theory and algorithms.
\newblock {\em IEEE Transactions on Pattern Analysis and Machine Intelligence}, 44(11):7747--7763, 2021.

\bibitem{hodson2022root}
Timothy~O Hodson.
\newblock Root-mean-square error (rmse) or mean absolute error (mae): When to use them or not.
\newblock {\em Geoscientific Model Development}, 15(14):5481--5487, 2022.

\bibitem{yang2021adaptive}
Yang Yang, Chaoyue Pan, Zheng Li, and Ruilian Zhao.
\newblock Adaptive reward computation in reinforcement learning-based continuous integration testing.
\newblock {\em IEEE Access}, 9:36674--36688, 2021.

\bibitem{ampatzoglou2020guidelines}
Apostolos Ampatzoglou, Stamatia Bibi, Paris Avgeriou, and Alexander Chatzigeorgiou.
\newblock Guidelines for managing threats to validity of secondary studies in software engineering.
\newblock {\em Contemporary empirical methods in software engineering}, pages 415--441, 2020.

\bibitem{zhou2016map}
Xin Zhou, Yuqin Jin, He~Zhang, Shanshan Li, and Xin Huang.
\newblock A map of threats to validity of systematic literature reviews in software engineering.
\newblock In {\em 2016 23rd Asia-Pacific Software Engineering Conference (APSEC)}, pages 153--160. IEEE, 2016.

\bibitem{wang2019security}
Xianmin Wang, Jing Li, Xiaohui Kuang, Yu-an Tan, and Jin Li.
\newblock The security of machine learning in an adversarial setting: A survey.
\newblock {\em Journal of Parallel and Distributed Computing}, 130:12--23, 2019.

\bibitem{ackerman2020detection}
Samuel Ackerman, Eitan Farchi, Orna Raz, Marcel Zalmanovici, and Parijat Dube.
\newblock Detection of data drift and outliers affecting machine learning model performance over time.
\newblock {\em arXiv preprint arXiv:2012.09258}, 2020.

\end{thebibliography}

\newpage

\section*{Appendix}
\appendix
\section{Selected Studies}
See Table~\ref{Table_old:PaprID}.

 \begin{table*}
 
    \tablefirsthead{%
        \hline
        
        \textbf{ID} & \textbf{Title} & \textbf{Authors} & \textbf{Venue} & \textbf{Cite} & \textbf{Year}  \\
        \hline
        }
    \tablehead{%
        \hline
        \textbf{ID} & \textbf{Title} & \textbf{Authors} & \textbf{Venue} & \textbf{Cite} & \textbf{Year} \\
        }
    \tabletail{%
        \hline 
    }
    \tablelasttail{\hline}
    \tablecaption{List of selected studies in this review. Here, ID denotes the study identification number. \textbf{Note:} The number of citations for each study was gathered on 14th August 2023. \label{Table_old:PaprID}}
    \fontsize{8}{9}\selectfont
    
    \begin{supertabular}{l p{4cm} p{4cm} p{5cm} l l }
        \toprule
        S1 & FastLane: Test Minimization for Rapidly Deployed Large-Scale Online Services & Philip, A. A.,   Bhagwan, R., Kumar, R., Maddila, C. S., \& Nagppan, N. & IEEE/ACM   International Conference on Software Engineering (ICSE) & 24 & 2019 \\ \hline
	S2 & Classifying false positive static checker alarms in continuous integration using convolutional neural networks & Lee, S., Hong, S.,   Yi, J., Kim, T., Kim, C. J., \& Yoo, S. & IEEE Conference on   Software Testing, Validation and Verification (ICST) & 22 & 2019 \\ \hline
	S3 & How high will it be? Using machine learning models to predict branch coverage in automated testing & Grano, G., Titov, T.   V., Panichella, S., \& Gall, H. C. & IEEE workshop on   machine learning techniques for software quality evaluation (MaLTeSQuE) & 34 & 2018 \\ \hline
	S4 & Automatic exploratory performance testing using a discriminator neural network & Porres, I., Ahmad,   T., Rexha, H., Lafond, S., \& Truscan, D. & IEEE international   conference on software testing, verification and validation workshops (ICSTW) & 14 & 2020 \\ \hline
	S5 & An empirical study on the cross-project predictability of continuous integration outcomes & Xia, J., Li, Y., \& Wang, C. & Web Information   Systems and Applications Conference (WISA) & 13 & 2017 \\ \hline
	S6 & Cutting the software building efforts in continuous integration by semi-supervised online AUC optimization & Xie, Z., \& Li, M. & International Joint   Conference on Artificial Intelligence (IJCAI) & 20 & 2018 \\ \hline
	S7 & Reinforcement learning for automatic test case prioritization and selection in continuous integration & Spieker, H., Gotlieb,   A., Marijan, D., \& Mossige, M. & Proceedings of the   ACM SIGSOFT International Symposium on Software Testing and Analysis & 225 & 2017 \\ \hline
	S8 & Learning for test prioritization: An industrial case study & Busjaeger, B., \&   Xie, T. & Proceedings of the   ACM SIGSOFT International symposium on foundations of software engineering & 108 & 2016 \\ \hline
	S9 & Defect prediction on   a legacy industrial software: A case study on software with few defects & Koroglu, Y., Sen, A.,   Kutluay, D., Bayraktar, A., Tosun, Y., Cinar, M., \& Kaya, H. & Proceedings of the   International Workshop on Conducting Empirical Studies in Industry & 17 & 2016 \\ \hline
	S10 & SQA-Profiles:   Rule-based activity profiles for Continuous Integration environments & Brandtner, M.,   Müller, S. C., Leitner, P., \& Gall, H. C. & IEEE International   Conference on Software Analysis, Evolution, and Reengineering (SANER) & 13 & 2015 \\ \hline
	S11 & LSTM-based deep   learning for spatial–temporal software testing & Xiao, L., Miao, H.,   Shi, T., \& Hong, Y. & Distributed and   Parallel Databases & 9 & 2020 \\ \hline
	S12 & Praxi: Cloud Software   Discovery That Learns From Practice & Byrne, A., Allen, S.   L., Nadgowda, S., \& Coskun, A. K. & Proceedings of the   International Middleware Conference Demos and Posters & 6 & 2019 \\ \hline
	S13 & Early prediction of   test case verdict with bag-of-words vs. word embeddings & Meding, W. & CEUR Workshop   Proceedings & 1 & 2020 \\ \hline
	S14 & A time window based   reinforcement learning reward for test case prioritization in continuous   integration & Wu, Z., Yang, Y., Li,   Z., \& Zhao, R. & Proceedings of the   Asia-Pacific Symposium on Internetware & 21 & 2019 \\ \hline
	S15 & Branch coverage   prediction in automated testing & Grano, G., Titov, T.   V., Panichella, S., \& Gall, H. C. & Journal of Software:   Evolution and Process & 21 & 2019 \\ \hline
	S16 & Failure Prediction   Using Machine Learning in IBM WebSphere Liberty Continuous Integration   Environment & Khan, M. A., Azim,   A., Liscano, R., Smith, K., Chang, Y. K., Garcon, S., \& Tauseef, Q. & Proceedings of the   Annual International Conference on Computer Science and Software Engineering & 88 & 2021 \\ \hline
	S17 & Failure Prediction   Using Transfer Learning in Large-Scale Continuous Integration Environments & Mamata, R., Smith,   K., Azim, A., Chang, Y. K., Taiseef, Q., Liscano, R., \& Seferi, G. & Proceedings of the   Annual International Conference on Computer Science and Software Engineering & 47 & 2022 \\ \hline
	S18 & Predicting Test Case   Verdicts Using Textual Analysis of Committed Code Churns & Al Sabbagh, K.,   Staron, M., Hebig, R., \& Meding, W. & CEUR Workshop   Proceedings & 11 & 2019 \\ \hline

 \hline
 
    \end{supertabular}
\end{table*}

 \begin{table*}
 
    \tablefirsthead{%
         \multicolumn{6}{l}{\textit{(Continue) List of selected studies in this review.}}\\\hline
        
        \textbf{ID} & \textbf{Title} & \textbf{Authors} & \textbf{Venue} & \textbf{Cite} & \textbf{Year}  \\
        \hline
        }
    \tablehead{%
        \hline
        }
    \tabletail{%
        \hline 
    }
    \tablelasttail{\hline}
    \fontsize{8}{9}\selectfont
    
    \begin{supertabular}{l p{4cm} p{4cm} p{5cm} l l }
        \toprule
 
	S19 & Reinforcement   Learning for Test Case Prioritization & Bagherzadeh, M.,   Kahani, N., \& Briand, L. & IEEE Transactions on   Software Engineering & 49 & 2021 \\ \hline 
	S20 & Focus on New Test   Cases in Continuous Integration Testing based on Reinforcement Learning & Chen, F., Li, Z.,   Shang, Y., \& Yang, Y. & IEEE International   Conference on Software Quality, Reliability and Security (QRS) & 8 & 2022 \\ \hline
	S21 & A cost-efficient   approach to building in continuous integration & Jin, X., \&   Servant, F. & Proceedings of the   ACM/IEEE International Conference on Software Engineering & 25 & 2020 \\ \hline 
    S22 & Occurrence Frequency   and All Historical Failure Information Based Method for TCP in CI & Shang, Y., Li, Q.,   Yang, Y., \& Li, Z. & Proceedings of the   International Conference on Software and System Processes & 2 & 2020 \\ \hline
	S23 & A Machine Learning   Approach to Improve the Detection of CI Skip Commits & Abdalkareem, R.,   Mujahid, S., \& Shihab, E. & IEEE Transactions on   Software Engineering & 34 & 2020 \\ \hline
	S24 & Change-Aware Build   Prediction Model for Stall Avoidance in Continuous Integration & Hassan, F., \&   Wang, X. & ACM/IEEE   International Symposium on Empirical Software Engineering and Measurement   (ESEM) & 38 & 2017 \\ \hline
	S25 & BuildFast:   History-Aware Build Outcome Prediction for Fast Feedback and Reduced Cost in   Continuous Integration & Chen, B., Chen, L.,   Zhang, C., \& Peng, X. & Proceedings of the   IEEE/ACM International Conference on Automated Software Engineering & 1 & 2020 \\ \hline
	S26 & Empirically   Evaluating Readily Available Information for Regression Test Optimization in   Continuous Integration & Elsner, D., Hauer,   F., Pretschner, A., \& Reimer, S. & Proceedings of the   ACM SIGSOFT International Symposium on Software Testing and Analysis & 28 & 2021 \\ \hline
	S27 & Continuous Build   Outcome Prediction: A Small-N Experiment in Settings of a Real Software   Project & Kawalerowicz, M.,   \& Madeyski, L. & Advances and Trends   in Artificial Intelligence. From Theory to Practice: International Conference   on Industrial, Engineering and Other Applications of Applied Intelligent   Systems, IEA/AIE & 5 & 2021 \\ \hline
	S28 & Multi-Armed Bandit   Test Case Prioritization in Continuous Integration Environments: A Trade-off   Analysis. & Lima, J. A. P., \&   Vergilio, S. R. & Proceedings of the   Brazilian symposium on systematic and automated software testing & 25 & 2020 \\ \hline
	S29 & MuDelta:   Delta-Oriented Mutation Testing at Commit Time & Ma, W., Chekam, T.   T., Papadakis, M., \& Harman, M. & IEEE/ACM   International Conference on Software Engineering (ICSE) & 11 & 2021 \\ \hline
	S30 & Supervised Learning   for Test Suit Selection in Continuous Integration & Martins, R., Abreu,   R., Lopes, M., \& Nadkarni, J. & IEEE International   Conference on Software Testing, Verification and Validation Workshops (ICSTW) & 5 & 2021 \\ \hline
	S31 & Reinforcement   learning based test case prioritization for enhancing the security of   software & Shi, T., Xiao, L.,   \& Wu, K. & IEEE International   Conference on Data Science and Advanced Analytics (DSAA) & 4 & 2020 \\ \hline
	S32 & Continuous Test Suite   Failure Prediction & Pan, C., \&   Pradel, M. & Proceedings of the   ACM SIGSOFT International Symposium on Software Testing and Analysis & 13 & 2021 \\ \hline
	S33 & Weighted Reward for   Reinforcement Learning based Test Case Prioritization in Continuous   Integration Testing & Li, G., Yang, Y., Wu,   Z., Cao, T., Liu, Y., \& Li, Z. & IEEE Annual   Computers, Software, and Applications Conference (COMPSAC) & 1 & 2021 \\ \hline
	S34 & Learning-based   Prioritization of Test Cases in Continuous Integration of Highly-Configurable   Software & Lima, J. A. P.,   Mendonça, W. D., Vergilio, S. R., \& Assunção, W. K. & Proceedings of the   ACM conference on systems and software product line & 14 & 2020 \\ \hline
	S35 & Dynamic Time Window   based Reward for Reinforcement Learning in Continuous Integration Testing & Pan, C., Yang, Y.,   Li, Z., \& Guo, J. & Proceedings of the   Asia-Pacific Symposium on Internetware & 5 & 2020 \\ \hline

\hline
 
    \end{supertabular}
\end{table*}

 \begin{table*}
 
    \tablefirsthead{%
        
         \multicolumn{6}{l}{\textit{(Continue) List of selected studies in this review.}}\\\hline
        \textbf{ID} & \textbf{Title} & \textbf{Authors} & \textbf{Venue} & \textbf{Cite} & \textbf{Year}  \\
        \hline
        }
    \tablehead{%
        \hline
        }
    \tabletail{%
        \hline 
    }
    \tablelasttail{\hline}
    \fontsize{8}{9}\selectfont
    
    \begin{supertabular}{l p{4cm} p{4cm} p{5cm} l l }
        \toprule

	S36 & Scalable and Accurate   Test Case Prioritization in Continuous Integration Contexts & Yaraghi, A. S.,   Bagherzadeh, M., Kahani, N., \& Briand, L. C. & IEEE Transactions on   Software Engineering & 14 & 2022 \\ \hline

	S37 & DeepOrder: Deep   Learning for Test Case Prioritization in Continuous Integration Testing & Sharif, A., Marijan,   D., \& Liaaen, M. & IEEE International   Conference on Software Maintenance and Evolution (ICSME) & 18 & 2021 \\ \hline
	S38 & Improving the   prediction of continuous integration build failures using deep learning & Saidani, I., Ouni,   A., \& Mkaouer, M. W. & Automated Software   Engineering & 27 & 2022 \\ \hline
	S39 & TCP-Net: Test Case   Prioritization using End-to-End Deep Neural Networks & Abdelkarim, M., \&   ElAdawi, R. & International   Conference on Software Testing, Verification and Validation Workshops (ICSTW) & 5 & 2022 \\ \hline
 
	S40 & Evaluating Features   for Machine Learning Detection of Order- and Non-Order-Dependent Flaky Tests & Parry, O.,   Kapfhammer, G. M., Hilton, M., \& McMinn, P. & IEEE Conference on   Software Testing, Verification and Validation (ICST) & 11 & 2022 \\ \hline

	S41 & Machine Learning   Regression Techniques for Test Case Prioritization in Continuous Integration   Environment & Da Roza, E. A., Lima,   J. A. P., Silva, R. C., \& Vergilio, S. R. & IEEE International   Conference on Software Analysis, Evolution and Reengineering (SANER) & 5 & 2022 \\ \hline
    S42 & Jaskier: A Supporting   Software Tool for Continuous Build Outcome Prediction Practice & Kawalerowicz, M.,   \& Madeyski, L. & Advances and Trends   in Artificial Intelligence. From Theory to Practice: International Conference   on Industrial, Engineering and Other Applications of Applied Intelligent   Systems, IEA/AIE & 2 & 2021 \\ \hline
	S43 & Using decision trees   to predict the certification result of a build & Hassan, A. E., \&   Zhang, K. & IEEE/ACM   International Conference on Automated Software Engineering (ASE) & 11 & 2006 \\ \hline
	S44 & Data stream mining   for predicting software build outcomes using source code metrics & Finlay, J., Pears,   R., \& Connor, A. M. & Information and   Software Technology & 50 & 2014 \\ \hline
	S45 & Enhanced regression   testing technique for agile software development and continuous integration   strategies & Ali, S., Hafeez, Y.,   Hussain, S., \& Yang, S. & Software Quality   Journal & 29 & 2020 \\ \hline
	S46 & A learning algorithm   for optimizing continuous integration development and testing practice & Marijan, D., Gotlieb,   A., \& Liaaen, M. & Software: Practice   and Experience & 37 & 2019 \\ \hline
	S47 & The Effect of Class   Noise on Continuous Test Case Selection: A Controlled Experiment on   Industrial Data & Al-Sabbagh, K. W.,   Hebig, R., \& Staron, M. & International   Conference on Product-Focused Software Process Improvement & 13 & 2020 \\ \hline
	S48 & Adaptive Reward   Computation in Reinforcement Learning-Based Continuous Integration Testing & Yang, Y., Pan, C.,   Li, Z., \& Zhao, R. & IEEE Access & 3 & 2021 \\ \hline
	S49 & Towards   auto-labelling issue reports for pull-based software development using text   mining approach & Fazayeli, H.,   Syed-Mohamad, S. M., \& Akhir, N. S. M. & Procedia Computer   Science & 21 & 2019 \\ \hline
	S50 & Sparse reward for   reinforcement learning‐based continuous integration testing & Yang, Y., Li, Z.,   Shang, Y., \& Li, Q. & Journal of Software:   Evolution and Process & 15 & 2023 \\ \hline
	S51 & Predicting Build   Outcomes in Continuous Integration using Textual Analysis of Source Code   Commits & Al-Sabbagh, K.,   Staron, M., \& Hebig, R. & Proceedings of the   International Conference on Predictive Models and Data Analytics in Software   Engineering & 0 & 2022 \\ \hline
	S52 & Test Case   Prioritization using Transfer Learning in Continuous Integration Environments & Mamata, R., Azim, A.,   Liscano, R., Smith, K., Chang, Y. K., Seferi, G., \& Tauseef, Q. & IEEE/ACM   International Conference on Automation of Software Test (AST) & 0 & 2023 \\
 \hline
 
    \end{supertabular}
\end{table*}

\section{Data Extraction From}
See Table~\ref{Table:ExtractedData}.

\begin{table*}
\centering
\caption{The extracted data items from each study and their relationship with research questions.}
\label{Table:ExtractedData}
\resizebox{\textwidth}{!}{%
\begin{tabular}{lll}
\hline
\textbf{ID}  & \textbf{Data Item}        & \textbf{Description}                                                     \\ \hline
             & \multicolumn{2}{l}{\textbf{Demographic data}}                                                        \\ \hline
\textbf{D1}  & Paper Title               & The title of the study                                                   \\
\textbf{D2}  & Authors                   & Name(s) of the author(s)                                                 \\
\textbf{D3}  & Publication Year          & Publication year of the paper                                            \\
\textbf{D4}  & Publication Venue         & Name of the conference or journal where the paper   is published         \\
\textbf{D5}  & Publication Type          & Publication type i.e., workshop, conference,   journal                   \\
\textbf{D6}  & Number of Citation        & How many citations does the paper have according to   GoogleScholar      \\
\textbf{D7}  & Keywords                  & List of keywords of the paper                                            \\
\textbf{D8}  & Context of the Study                      & The study contexts are categorized into industry   and non-industry cases            \\ \hline
             & \multicolumn{2}{l}{\textbf{RQ1: CI Tasks}}                                                           \\ \hline
\textbf{D9}  & CI Phases Addressed by ML & The definition of the CI phases and their role in   CI                   \\
\textbf{D10} & CI Phase Mediator         & Abstraction of the outcome and required input of   each CI phases        \\
\textbf{D11} & CI Tasks Enhanced by ML   & The definition of the enhanced CI tasks by ML   methods                  \\
\textbf{D12} & Insights on Underexplored CI Phases/Tasks & We summarized if any underexplored CI phases/tasks   are identified by the study     \\ \hline
    & \multicolumn{2}{l}{\textbf{RQ2: Data Engineering}}                                                   \\ \hline
\textbf{D13} & Commonly Used Datasets    & The most employed datasets in each CI tasks                              \\
\textbf{D14} & Data Engineering Methods  & The required data preparation and engineering   methods                \\
\textbf{D15} & Data Types and Tasks                      & Correlation between employed data types and the   enhanced CI tasks                  \\
\textbf{D16} & Data Quality Impact       & The impact of data quality on the performance of ML   models             \\ \hline
    & \multicolumn{2}{l}{\textbf{RQ3: Feature Engineering}}                                                \\ \hline
\textbf{D17} & Feature Types             & The features that have been used in studies for   training the ML models \\
\textbf{D18} & Feature Engineering Techniques            & The employed techniques for preparing the data as   an input of ML models            \\
\textbf{D19} & Feature and Tasks         & The correlation between feature types and CI tasks                       \\ \hline
    & \multicolumn{2}{l}{\textbf{RQ4: Model Tuning}}                                                       \\ \hline
\textbf{D20} & ML Learning Types         & Learning types of ML models based on the nature of   input data          \\
\textbf{D21} & ML Model Types            & Types of ML models based on nature of the problem                        \\
\textbf{D22} & ML Types and CI Phases    & Correlation between ML model types and CI phases                          \\
\textbf{D23} & Model Tuning              & The employed techniques for tuning the   hyper-parameters                \\ \hline
             & \multicolumn{2}{l}{\textbf{RQ5: Evaluation}}                                                         \\ \hline
\textbf{D24} & Evaluation Metrics                        & Which evaluation metrics   have been used for measuring the performance of ML models \\
\textbf{D25} & Evaluation Techniques     & Data division and evaluation technique                                   \\
\textbf{D26} & Metrics and ML Models     & Correlation between employed evaluation metrics and   ML models          \\ \hline
\end{tabular}%
}

\end{table*}

\end{document}